\documentclass[letterpaper,12pt]{article}
\usepackage[hyperindex,breaklinks]{hyperref}
\usepackage{amsfonts}
\usepackage{amsmath, amsthm, amssymb, bbm}
\usepackage{xcolor,multirow,graphicx}
\usepackage[disable]{todonotes}
\usepackage{enumerate,pgfplots}
\usepackage{natbib} 
\usepackage{comment}

\usepackage{pstricks}
\newrgbcolor{dblue}{0 0 0.5}
\newrgbcolor{dgray}{0.3 0.3 0.3}
\newrgbcolor{dred}{0.5 0 0}
\newrgbcolor{dgreen}{0 0.5 0}
\newrgbcolor{lgrey}{0.7 0.7 0.7}
\newrgbcolor{dblue2}{0 0 0.7}
\newrgbcolor{dred2}{0.7 0 0}
\definecolor{lyellow}{rgb}{1,1,0.7}

\newtheorem{assumption}{Assumption} 
\newtheorem{theorem}{Theorem}

\newtheorem{proposition}{Proposition}

\newtheorem{definition}{Definition}
\newtheorem{lemma}{Lemma}
\newtheorem{corollary}{Corollary}
\newtheorem{remark}{Remark}

\newcommand{\plim}{\operatorname*{plim}}

\newcommand{\argmin}{\operatorname*{argmin}}

\renewcommand{\baselinestretch}{1.22}

\usepackage{geometry}
\begin{document}

\title{Bounding Treatment Effects by \\ Pooling Limited Information across Observations%
\thanks{We are grateful for useful comments from an associate editor, two anonymous referees, Tim Armstrong, Michal Koles\'{a}r, Myunghyun Song,
and from seminar/conference participants at 
the 2023 ASSA meeting, Arizona, Chicago, Cornell, Cowles Foundation, Harvard/MIT, Maryland,  Monash, Montreal, 
 NYU, Princeton,
UC Davis, UPenn,  Vanderbilt, Virginia, Wisconsin, and Zurich.
This research was
supported by   the Economic and Social Research Council through the ESRC Centre for
Microdata Methods and Practice (grant numbers RES-589-28-0001, RES-589-28-0002 and ES/P008909/1),
and by the European Research Council grants ERC-2014-CoG-646917-ROMIA and
ERC-2018-CoG-819086-PANEDA.
}
}
\author{\setcounter{footnote}{2}Sokbae Lee\thanks{%
Department of Economics, Columbia University.  Email:  {\tt sl3841@columbia.edu.}  }
 \and 
 Martin Weidner%
\thanks{%
Dept.\ of Economics \& Nuffield College, Univ.\ of Oxford. Email:~{\tt martin.weidner@economics.ox.ac.uk}.
}
 }
\date{May 2026}

\maketitle

\begin{abstract}
\noindent
We provide novel bounds on average treatment effects (on the treated) that are valid under an unconfoundedness assumption. Our bounds are designed to be robust in challenging situations, for example, when the conditioning variables take on a large number of different values in the observed sample, or when the overlap condition is violated. This robustness is achieved by only using limited ``pooling'' of information across observations. Namely, the bounds are constructed as sample averages over functions of the observed outcomes such that the contribution of each outcome only depends on the treatment status of a limited number of observations. No information pooling across observations leads to so-called ``Manski bounds'', while unlimited information pooling leads to standard inverse propensity score weighting. We explore the intermediate range between these two extremes and provide corresponding inference methods. We show in Monte Carlo experiments and through two empirical applications that our bounds are indeed robust and informative in practice.
\\[0.3cm]
\end{abstract}

\newpage

\section{Introduction} 

\todo[size=tiny]{The page number starts from the title page as required; The latex option `baselinestretch' is modified to fit the main text within 45 pages.}

In many applications, causal inference hinges on strong ignorability, namely unconfoundedness and overlap \citep[see, e.g.,][for a monograph]{imbens2015causal}. 
The former condition is non-testable but requires that  all confounders be used as covariates;
the latter is a testable condition that may not be satisfied in practice.

The overlap condition has received increasing attention in the literature.
In applications, it is not uncommon to have a situation where the estimated propensity scores are close to zero or one. This problem is referred to as limited overlap \citep[e.g.,][]{crump2009dealing}. 
The existence of limited overlap may change the asymptotic behavior of the estimators 
\citep[e.g.,][]{Khan:Tamer:2010,Hong:et:al:2019}
and may necessitate using a more robust inference method 
\citep[e.g.,][]{Rothe:2017,sasaki_ura_2021}.
\citet{DAMOUR2020} provide a cautionary tale on the overlap condition when high-dimensional covariates are adopted to make unconfoundedness more plausible.

There are several approaches in the literature to estimate treatment effects when facing limited overlap. Arguably, the most popular method is to focus on a subpopulation where the overlap condition holds \citep[e.g.,][]{crump2009dealing,Yang:Ding:18}.
For example, 
\citet{crump2009dealing} recommend a simple rule of thumb to drop all observations with estimated propensity scores outside the range $[\alpha,1-\alpha]$ for some predetermined constant $\alpha$, say $\alpha = 0.1$.  
Alternatively, 
\citet{Li-et-al:2018} advocate the use of the so-called `overlap weights' to define the average treatment effect. This amounts to assigning 
weights equal to
one minus the propensity score for the treated units
and equal to the propensity score for the control units.
 If the treatment effects are heterogeneous, both trimming and overlap weighting change the parameter of interest from the population average treatment effect.  
Without changing it, 
\citet{nethery2019} develop a Bayesian framework by extrapolating estimates from the overlap region to the non-overlap region via a spline model. 
 However, identification by extrapolation is subject to model misspecification.

In this paper, we start with the observation that none of the aforementioned papers would work well if the overlap condition is not satisfied at the population level and it is a priori unknown where it fails. In that case, the population average treatment effect is not point-identified and one may 
resort to \citet{manski1989anatomy,manski1990nonparametric}'s bounds, provided that the support of outcome is bounded and known. 
However, it may not yield tight bounds if unconfoundedness assumption is plausible, while the overlap condition being the only source of identification failure. This paper provides a systematic method to explore this possibility.

Our contributions are two-fold. First, we provide novel \todo[size=tiny]{wording change: ``population bounds'' to ``bounds''} %
bounds on both
average treatment effects (ATE) and 
average treatment effects on the treated (ATT) that are valid under an unconfoundedness assumption. Our bounds are applicable if the conditioning variables do not satisfy the overlap condition and take on a large number of different values in the observed sample. This robustness is achieved by only using limited ``pooling'' of information across observations. Namely, the bounds are constructed as the expectations of functions of the observed outcomes such that the contribution of each outcome only depends on the treatment status of a limited number of observations. No information pooling across observations leads to \citet{manski1989anatomy,manski1990nonparametric}'s bounds, which we call ``first-order bounds'', while unlimited information pooling leads to standard inverse propensity score weighting. 
We explore the intermediate range between these two extremes by considering the setup where an applied researcher provides a reference propensity score.
Our bounds are valid independent of the value of this reference propensity score, but if it happens to be close to the true propensity score, then our bounds 
are optimal in terms of expected width within the class of limited pooling bounds considered in this paper.
The reference propensity score is therefore crucial to construct our novel treatment effect bounds uniquely, and it also allows to incorporate prior knowledge on the propensity score
in a robust way.

Second, we develop estimation and inference methods for the bounds we have established under the unconfoundedness assumption. 
 Our formal theory assumes that the observed covariates are discrete, so that multiple observations can share the same covariate value. A leading data scenario we analyze assumes that the number of distinct covariate values is large relative to the sample size, implying that for each possible covariate value only a small number of observations are available.
In this scenario, it is a statistically challenging problem to provide a valid confidence interval for the treatment effects, which we tackle in this paper.

In many empirical applications, however, some or all covariates are continuous rather than discrete. To apply our method in such settings, one must first discretize the covariates, for example by binning or clustering observations with similar covariate values. This is a practical solution that we discuss in Section~\ref{sec:Cluster} and employ in our Monte Carlo experiments and empirical applications. However, discretization introduces an approximation error for which we do not provide formal theory. Controlling this error would require additional smoothness assumptions on the conditional mean of the outcome variable, which we do not impose. This is a limitation of our approach in its current form, and we leave the formal analysis of discretization bias for future work.

 An alternative approach to robust inference for treatment effects under unconfoundedness is provided
 by   \cite{Armstrong:Kolesar:21}. In particular, 
 their confidence intervals are 
asymptotically valid  under a violation of the overlap condition, as long as the researcher specifies a 
 Lipschitz bound on the conditional mean of the outcome variable. Their approach
  is distinct from and complementary to ours.
The approach of \cite{Armstrong:Kolesar:21} reduces to a matching estimator for the average treatment effect
(e.g., \citealt{abadie2006large,abadie2008failure,abadie2011bias}) if the Lipschitz bound is chosen to be very large. 
Those matching estimators crucially require that for every observation we can find other observations with similar
covariate values but opposite treatment status. This is not required in our approach. Crucially, we only pool information across
observations with similar covariate values, but in contrast to \cite{Armstrong:Kolesar:21} and matching estimators, we do so completely
independent of the treatment status of the observations involved. This is the key difference compared to those existing methods.

The remainder of the paper is organized as follows. In Section~\ref{sec:Setup}, we
describe the setup and intuition behind our approach. 
Section~\ref{sec:MainIdea} illustrates our key ideas through the simple two-unit example and introduces the main ideas, including a formal characterization of our ``second-order bounds''. In Section~\ref{sec:PopulationBounds}, we extend the framework to the general bounds of arbitrary order, and in Section~\ref{sec:implementation}, we construct sample analogs and develop corresponding inference methods.
Using those bounds we then
provide asymptotically valid confidence intervals.
We discuss how to cluster the covariate observations in Section~\ref{sec:Cluster}.
The results of Monte Carlo experiments are reported in Section~\ref{sec:MC}.
In Section~\ref{sec:EE}, we present two empirical applications. The first uses the well-known dataset from \citet{Connors1996}'s study of the efficacy of right heart catheterization (RHC), which has been extensively analyzed in the context of limited overlap (see, e.g., \citealp{crump2009dealing}, \citealp{Rothe:2017}, \citealp{Li-et-al:2018}). The second application uses the dataset from \citet{DehejiaWahba-JASA}, which exhibits limited overlap and serves as a useful complement. These applications illustrate the practical relevance and robustness of our method. The appendices contain all proofs, technical derivations omitted from the main text, and additional results from our Monte Carlo experiments. An accompanying R package is available on the Comprehensive R Archive Network (CRAN) at \url{https://CRAN.R-project.org/package=ATbounds}.

\section{Setup}
\label{sec:Setup}

For units $i=1,\ldots,n$, we observe treatment status $D_i \in \{0,1\}$, regressors $X_i \in {\cal X}$,  where ${\cal X}$ is a discrete set, and outcome $Y_i = (1-D_i) \, Y_i(0) + D_i \, Y_i(1)$, where $Y_i(0)$ and $Y_i(1)$ are potential outcomes. While we observe the realized outcome $Y_i$, we never observe both potential outcomes for the same unit.
Our main objective is to conduct inference on the average treatment effect (ATE) and average treatment effect on the treated (ATT), conditional on the covariates:
\begin{align}
   {\rm ATE} & := \frac 1 n \sum_{i=1}^n  \tau(X_i),
   & \tau(x)  &:= \mathbb{E}\left[  Y_i(1) -  Y_i(0) \, \big| \, X_i=x   \right]   ,
 \nonumber  \\
   {\rm ATT} &:= \frac{  \frac 1 n \sum_{i=1}^n  \pi(X_i)  }
     {\frac 1 n \sum_{i=1}^n \mathbb{E}(D_i|X_i)}  ,
     &
      \pi(x) &:= \mathbb{E}\left\{  D_i \left[ Y_i(1) -  Y_i(0)  \right] \,  \big| \, X_i=x  \right\} .
     \label{ATestimands}
\end{align}
We do not assume i.i.d.\ sampling. Instead, we allow the distribution of covariates $X^{(n)} := (X_1,\ldots,X_n)$ to vary with $n$. This leads us to define estimands conditional on $X^{(n)}$, but under our assumptions, the relevant expectations depend only on $X_i$.
\begin{assumption}~
     \label{ass:MAIN}
    \begin{itemize}
         \item[(i)]  $\forall i \in \{1,\ldots,n\}: \; \; \left[Y_i(0), Y_i(1)\right] \; \perp \; D_i \; \big| \; X_i$. \qquad (unconfoundedness)
         
         \item[(ii)]  There are known constants $a_{\min}, a_{\max} \in \mathbb{R}$ such that $a_{\min} \leq Y_i(d) \leq a_{\max}$, for all
         $d \in \{0,1\}$ and  $i \in \{1,\ldots,n\}$.

         \item[(iii)] Let $f(\cdot|\cdot)$ be the probability density  or probability mass function of 
         the distribution of  $[ D_i,Y_i(0),Y_i(1)]_{i=1,\ldots,n}$  conditional
         on  $X^{(n)}=(X_1,\ldots,X_n)$. Then, there exists a function $g : \{0,1\} \times \mathbb{R}^2 \times {\cal X} \rightarrow \mathbb{R}$ such that, almost surely,\footnote{
      From     Assumption~\ref{ass:MAIN}(i)  we know that {there exist functions $g^*$ and $g^{**}$ such that} $g\left( D_i,Y_i(0),Y_i(1) \, |\, X_i \right)  = g^*\left( D_i \, |\, X_i \right) \, \cdot \, g^{**}\left( Y_i(0),Y_i(1)  \, |\, X_i \right)$.
         }
            $$
            f\left(  \big[D_i,Y_i(0),Y_i(1) \big]_{i=1,\ldots,n}  \, \Big| \,   X^{(n)} \right) = 
            \prod_{i=1}^n  g\left( D_i,Y_i(0),Y_i(1) \, |\, X_i \right) .
         $$         
    \end{itemize}
\end{assumption}

Assumption~\ref{ass:MAIN}(i) imposes unconfoundedness, meaning that treatment assignment is effectively randomized conditional on covariates $X_i$. While we could weaken this to mean independence (i.e.\ $ \mathbb{E}\left[ Y_i(d) \, \big| \, D_i , \, X_i  \right] = \mathbb{E}\left[ Y_i(d) \, \big| \, X_i  \right] $) for most of our results, scenarios that justify mean independence typically also support full conditional independence. Similarly, Assumption~\ref{ass:MAIN}(ii) could be relaxed to bounds on conditional expectations (i.e.\ $a_{\min} \leq  \mathbb{E}\left[ Y_i(d) \, \big| \, X_i  \right]  \leq a_{\max}$), but in practice, known bounds usually apply directly to the potential outcomes themselves.

Assumption~\ref{ass:MAIN}(iii) has two key implications for the sampling scheme. First, it ensures that $(D_i,Y_i(0),Y_i(1))$ is independently distributed across units conditional on $X^{(n)}$. Second, it requires that the conditional distribution of $(D_i,Y_i(0),Y_i(1))$ depends only on $X_i$, not on other covariates $X_j$ or the unit index $i$. This implies that both treatment effects and propensity scores depend only on individual covariates:
$ \mathbb{E}\left[  Y_i(1) -  Y_i(0) \, \big| \, X^{(n)}  \right]  =  \mathbb{E}\left[  Y_i(1) -  Y_i(0) \, \big| \, X_i   \right] $, and
$\mathbb{E}\left( D_i  \, \big| \, X^{(n)} \right)  = \mathbb{E}\left( D_i  \, \big| \, X_i \right) =: p(x)$.
If the propensity score $p(x)$ were known and satisfied $0<p(x)<1$ (overlap condition), the treatment effects would be point-identified through inverse propensity score weighting:
\begin{align}
      \tau(x)  &= \mathbb{E}\left[ \left. \frac{D_i \, Y_i} {p(X_i)} -  \frac{(1-D_i) \, Y_i} {1-p(X_i)}  \, \right| \, X_i=x \right] ,
 \nonumber \\
\pi(x)  &=  
       \mathbb{E}\left[  \left. D_i \, Y_i  -  \frac{p(X_i) \, (1-D_i) \, Y_i} {1-p(X_i)} \, \right| \, X_i=x \right] .
       \label{PSW}
\end{align}
However, since $p(x)$ is unknown in practice and the overlap condition may fail, we can generally only obtain partial identification of ATE and ATT. This means we can construct valid large-sample confidence intervals, but these may not converge to a point as $n \rightarrow \infty$.

Let $D_{-i}$ denote the treatment statuses of all units $j \neq i$ sharing the same covariate value as unit $i$. Our approach to constructing valid confidence intervals for ATE relies on finding functions $L(Y_i, D_i, D_{-i},X_i)$ and $U(Y_i, D_i, D_{-i},X_i)$ that provide bounds on the conditional treatment effect:
\begin{align}
   \mathbb{E}\left[ L(Y_i, D_i, D_{-i},X_i)  \, \Big| \, X^{(n)}  \right] \leq  \tau(X_i) \leq    \mathbb{E}\left[ U(Y_i, D_i, D_{-i},X_i)    \, \Big| \, X^{(n)}  \right]  .
   \label{BoundUL}
\end{align}
Let ${\cal X}_*$ denote the set of distinct observed covariate values with cardinality $m=|{\cal X}_*|$. In asymptotic sequences where $m \rightarrow \infty$ as $n \rightarrow \infty$, equation \eqref{BoundUL} implies:
\begin{align}
    \plim_{n \rightarrow \infty}   \frac 1 n \sum_{i=1}^n L(Y_i, D_i, D_{-i},X_i )  \leq  {\rm ATE}  \leq    \plim_{n \rightarrow \infty}   \frac 1 n \sum_{i=1}^n U(Y_i, D_i, D_{-i},X_i) ,
       \label{BoundULplim}
\end{align}
where the probability limit is taken conditional on $X^{(n)}$. These cross-sectional averages of the bounds form the basis for constructing asymptotically valid confidence intervals for ATE.

For ATT, we similarly construct bounds using functions of the form $L(Y_i, D_i, D_{-i}, X_i)$ and $U(Y_i, D_i, D_{-i}, X_i)$ that satisfy \eqref{BoundUL} with $\pi(X_i)$ in place of $\tau(X_i)$. This allows us to bound the numerator $\frac{1}{n} \sum_{i=1}^n \pi(X_i)$ from \eqref{ATestimands}, while the denominator $\frac{1}{n} \sum_{i=1}^n \mathbb{E}(D_i|X_i)$ can be consistently estimated by $\frac{1}{n} \sum_{i=1}^n D_i$. This approach to ATT estimation requires only that $\frac{1}{n} \sum_{i=1}^n p(X_i) > 0$, a mild condition that permits $p(X_i)=0$ for many units.

The key advantage of our approach is that the asymptotic validity of the confidence intervals for ATE and ATT relies only on Assumption~\ref{ass:MAIN}, with the sole additional requirement of $\frac{1}{n} \sum_{i=1}^n p(X_i) > 0$ for ATT inference. Notably, we require neither assumptions on $X^{(n)}$, nor the overlap condition ($0 < p(x) < 1$), nor knowledge or consistent estimation of $p(x)$.

For each covariate value $X_i$, the treatment assignments $(D_i,D_{-i})$ are
assumed to be
independent Bernoulli draws with the same mean $p(X_i)$. However, since the number of units sharing any given covariate value may be small and non-increasing asymptotically, $p(X_i)$ may not be consistently estimable under our assumptions.

\subsection{Manski bounds}

Manski bounds (\citealt{manski1989anatomy,manski1990nonparametric}) represent a simplified version of \eqref{BoundUL} and \eqref{BoundULplim} where $L(Y_i, D_i, D_{-i}, X_i)$ and $U(Y_i, D_i, D_{-i}, X_i)$ reduce to functions $L^{(1)}(Y_i, D_i)$ and $U^{(1)}(Y_i, D_i)$ that depend only on individual outcomes and treatment status. These bounds are particularly robust as they do not require unconfoundedness (Assumption~\ref{ass:MAIN}(i)) and apply even when covariate values are unique to each unit.
Using the outcome bounds $a_{\min}$ and $a_{\max}$, we can establish:
\begin{align}
    B^{(1)}_{0,a_{\min}}(Y_i, D_i)   \leq   Y_i(0)  \leq  B^{(1)}_{0,a_{\max}}(Y_i, D_i) ,
    \nonumber \\
     B^{(1)}_{1,a_{\min}}(Y_i, D_i)   \leq Y_i(1)  \leq  B^{(1)}_{1,a_{\max}}(Y_i, D_i)  ,
     \label{ManskiBounds}
\end{align}
where
\begin{align*}
     B^{(1)}_{0,a}(Y_i, D_i) &:= a + (1-D_i) \, (Y_i-a)  ,
     &
       B^{(1)}_{1,a}(Y_i, D_i) &:= a + D_i \, (Y_i-a) .
\end{align*}
These lead to bounds for ATE
\begin{align*}
    L^{(1)}(Y_i, D_i) &:=  B^{(1)}_{1,a_{\min}}(Y_i, D_i) - B^{(1)}_{0,a_{\max}}(Y_i, D_i)  ,
    \\
     U^{(1)}(Y_i, D_i) &:= B^{(1)}_{1,a_{\max}}(Y_i, D_i) - B^{(1)}_{0,a_{\min}}(Y_i, D_i) ,
\end{align*}
satisfying
\begin{align}
   \mathbb{E}\left[  L^{(1)}(Y_i, D_i)  \, \Big| \, X^{(n)}  \right] \leq  \tau(X_i) \leq    \mathbb{E}\left[   U^{(1)}(Y_i, D_i)   \, \Big| \, X^{(n)}  \right] . 
      \label{FirstBoundsATE}
\end{align}
For ATT, defining $C^{(1)}_a(Y_i, D_i)  := D_i \, (Y_i   - a)$, we have:
\begin{align}
   \mathbb{E}\left[  C^{(1)}_{a_{\max}}(Y_i, D_i)  \, \Big| \, X^{(n)}  \right] \leq  \pi(X_i) \leq    \mathbb{E}\left[   C^{(1)}_{a_{\min}}(Y_i, D_i)     \, \Big| \, X^{(n)}  \right] . 
      \label{FirstBoundsATT}
\end{align}
The bounds in \eqref{FirstBoundsATE} and \eqref{FirstBoundsATT} are well-known,
and we  denote those bounds on $ \tau(X_i)$ and $\pi(X_i) $ as either Manski bounds (\citealt{manski1989anatomy,manski1990nonparametric}) or as ``first-order bounds''.

\subsection{Pooling information across observations}

The Manski bounds in \eqref{FirstBoundsATE} and \eqref{FirstBoundsATT} are sharp when we only impose outcome boundedness (Assumption~\ref{ass:MAIN}(ii)). However, under unconfoundedness (Assumption~\ref{ass:MAIN}(i)) and overlap ($0<p(x)<1$), both $\tau(x)$ and $\pi(x)$ become point-identified via inverse propensity score weighting as shown in \eqref{PSW}.

This presents us with two extremes. The Manski bounds require minimal assumptions but pool no information across observations, leading to potentially wide bounds. In contrast, point identification through propensity score methods requires extensive pooling of information across observations to estimate $p(x)$ consistently. This latter approach demands strong data requirements and faces the curse of dimensionality as the dimension of $X_i$ increases.

This paper explores a middle ground between these extremes. Our approach pools some information across observations to tighten the bounds on average treatment effects, but requires much less pooling than needed for consistent nonparametric estimation of $p(x)$. The key idea is to use information from pairs or small groups of observations with similar or identical covariate values to construct tighter bounds.

For example, when two observations share the same covariate value ($X_i = X_j$), we can construct ``second-order" bounds that use information from both observations jointly. These bounds improve upon Manski bounds by leveraging unconfoundedness and using the treatment status of both observations with the same covariate value. Similar principles can be extended to construct higher-order bounds that pool information across larger groups of observations.

\subsection{Lack of overlap and curse of dimensionality}

\begin{figure}[tb]
\begin{center}
\includegraphics[scale=0.35]{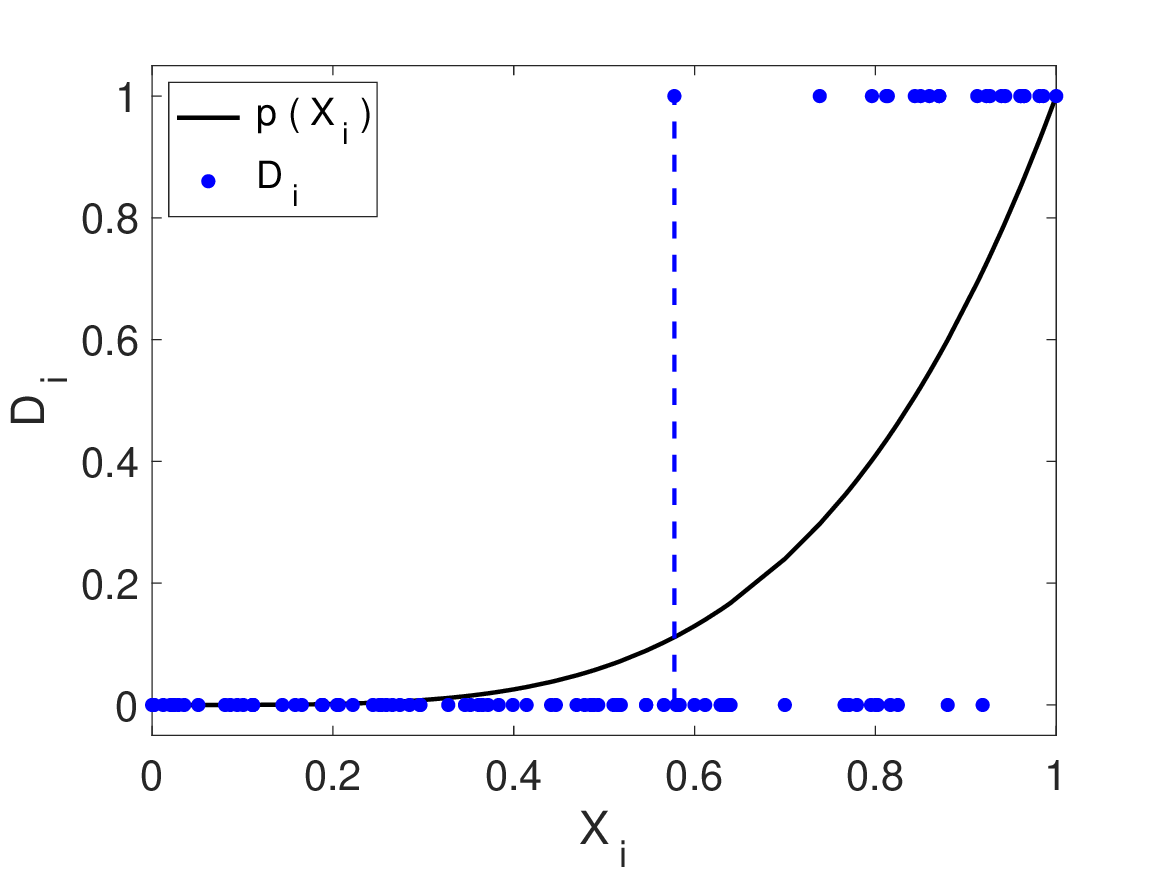} \quad
\includegraphics[scale=0.35]{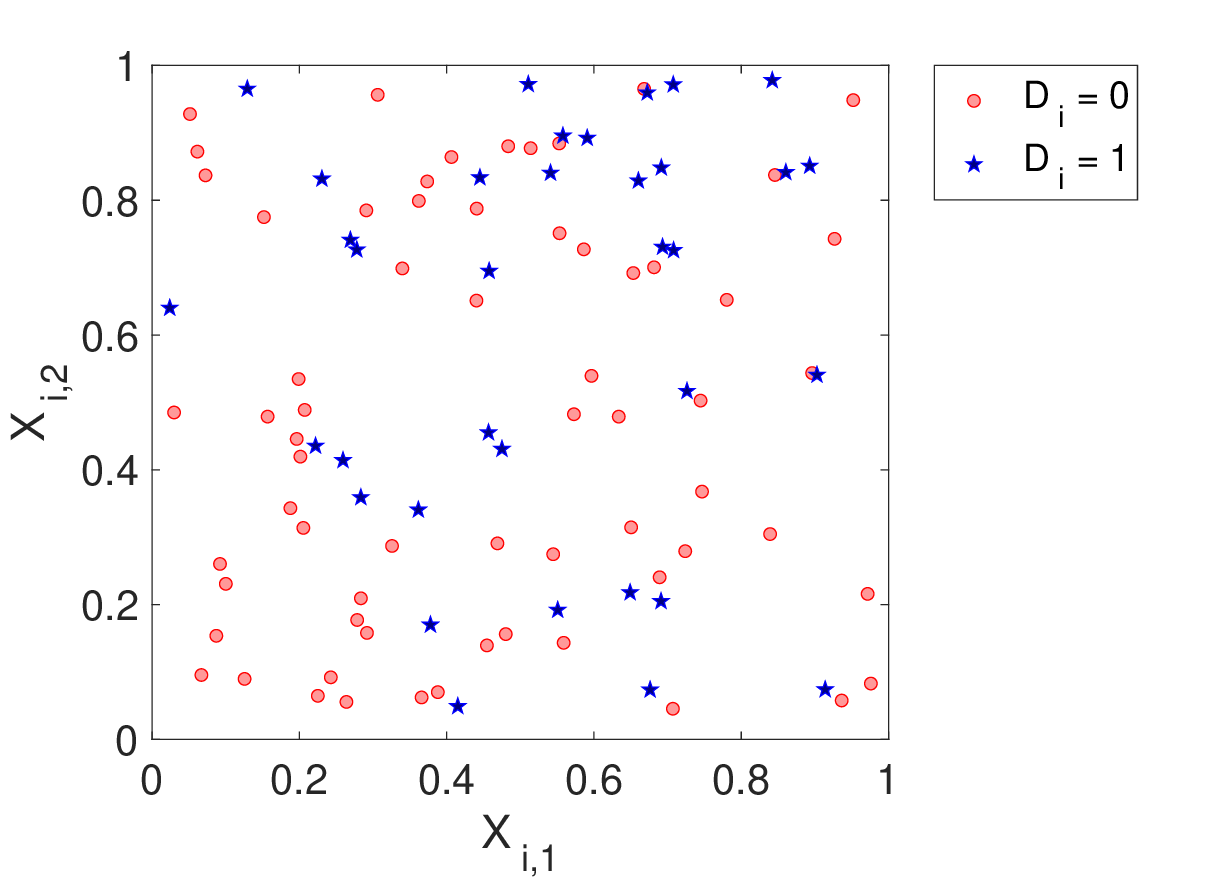}
\end{center}
\vspace{-0.5cm}
\caption{\label{fig_prop_score_example} Two simple examples for samples of $(X_i,D_i)$, $i=1,\ldots,n$, with $n=100$.
For the example on the left we have one-dimensional $X_i \sim U[0,1]$ and $p(x)=x^4$. For the example on the right
we have two-dimensional $X_i \sim U[0,1]^2$, and $p(x)=0.3$.
}
\end{figure}

While equation \eqref{PSW} shows that ATE and ATT are point-identified under Assumption~\ref{ass:MAIN} and overlap ($0<p(x)<1$), estimating $p(x)= \mathbb{E}\left( D  \, \big| \, X=x  \right)$ in finite samples presents significant challenges, particularly due to the curse of dimensionality for multi-dimensional covariates. Consider two illustrative examples in Figure~\ref{fig_prop_score_example}, each with sample size $n=100$.

In the left example, with $X_i \sim U[0,1]$ and $p(x)=x^4$, the overlap condition is theoretically satisfied for all $x \in (0,1)$. However, we observe no treated units ($D_i=1$) for $X_i < 0.58$, making precise point estimation of ATE infeasible without strong parametric assumptions. A natural approach here would combine Manski bounds for $X_i<0.58$ with matching or inverse propensity score weighting for $X_i \geq 0.58$.

The right example, with $X_i \sim U[0,1]^2$ and constant $p(x)=0.3$, illustrates a more complex challenge. Despite perfect theoretical overlap, large regions of the covariate space (e.g., near $x=(0,0)$) contain no treated observations. From the finite sample alone, we cannot determine whether this reflects true lack of overlap or merely finite-sample variation, as these regions contain few observations overall. This challenge intensifies with higher-dimensional covariates.

In this paper, we provide asymptotically valid inference on ATE or ATT under Assumption~\ref{ass:MAIN} that remains robust to any form of the unknown propensity score $p(x)$, including $p(x)=0$ and $p(x)=1$
for some $x$. While some observations share identical (or similar) covariate values and thus propensity scores, each observation typically has few such neighbors. This scenario can be modeled asymptotically by having each covariate value appear exactly $k$ times ($k=2,3,4,\ldots$) in the sample, with $m=|{\cal X}_*|=n/k$ distinct values.

Our bounds solve this inference problem where $n/m$ may stay bounded as $n \rightarrow \infty$, addressing situations where limited local sample sizes prevent reliable propensity score estimation, as in the right panel of Figure~\ref{fig_prop_score_example}. Moreover, our method can incorporate prior information about the propensity score, potentially achieving point identification when this information is correct and overlap holds, while maintaining robust confidence intervals otherwise.

\section{Main idea for second order bounds on $\mathbb{E}[ Y_i(1) ]$}
\label{sec:MainIdea}

 To illustrate the fundamental concept of our new bounds, we begin with the simplest non-trivial case. Consider two units $i \in \{1,2\}$ sharing identical covariate values $X_1=X_2$. For each unit, we observe an outcome $Y_i \in \mathbb{R}$ and a treatment indicator $D_i \in \{0,1\}$. As introduced in Section~\ref{sec:Setup}, each unit has potential outcomes $Y_i(0)$ and $Y_i(1)$, with the observed outcome being $Y_i=Y_i(D_i)$.

All stochastic statements in this section are implicitly conditional on $(X_1,X_2)$ and $X_1=X_2$.
Let $P$ denote the corresponding conditional distribution of $[(D_i,Y_i(0),Y_i(1)) \, : i \in \{1,2\}]$, 
and let ${\cal P}$ be the set of distributions $P$ satisfying:
\begin{itemize}
  \item[(i)]     $\left[Y_i(0), Y_i(1)\right] \; \perp \; D_i$, for $i\in \{1,2\}$.
  
   \item[(ii)]    $a_{\min} \leq Y_i(d) \leq a_{\max}$, for 
         $d \in \{0,1\}$ and  $i \in \{1,2\}$.
         
    \item[(iii)] $(D_i,Y_i(0),Y_i(1))$  are independent
         and identically distributed across $i \in \{1,2\}$.
\end{itemize}
These conditions are  restatements of Assumptions~\ref{ass:MAIN} for the case where $X_1=X_2$. We denote expectations under $P$ by $\mathbb{E}_P$.

Our goal in this section is to construct a valid upper bound function $B(Y_1,D_1,D_2) \in \mathbb{R}$ such that, for all $P \in {\cal P}$,
\begin{align}
  \mathbb{E}_P[ Y_i(1) ]
         \leq  \mathbb{E}_P[ B(Y_1,D_1,D_2) ] .
   \label{BoundMainExample}
\end{align}
While equation \eqref{BoundMainExample} appears to treat units asymmetrically, our final bounds symmetrize across observations through averaging:
$$
  \frac 1 2 \left[ B(Y_1,D_1,D_2) + B(Y_2,D_2,D_1) \right] .
$$
Beyond mere validity, we seek bound functions that cannot be improved upon, as formalized in the following definition.
\begin{definition}
    We say that the function $\widetilde B: \mathbb{R} \times \{0,1\}^2 \rightarrow \mathbb{R}$  \underline{dominates} 
    the function $B: \mathbb{R} \times \{0,1\}^2 \rightarrow \mathbb{R}$
      if for all  $P \in {\cal P}$ we have
\begin{align*}
  \mathbb{E}_P[ Y_i(1) ]
         \leq  \mathbb{E}_P[ \widetilde B(Y_1,D_1,D_2) ]
         \leq  \mathbb{E}_P[ B(Y_1,D_1,D_2) ] ,
\end{align*}
and the last inequality is strict for at least one $P \in {\cal P}$.
\end{definition}

\begin{theorem}
    \label{th:OurBounds}
    Let $B: \mathbb{R} \times \{0,1\}^2 \rightarrow \mathbb{R}$
    be such that \eqref{BoundMainExample} is satisfied
    for all distributions $P \in {\cal P}$ satisfying assumptions (i), (ii), (iii) above.
    Furthermore, assume that there exists no
    alternative such function that dominates $B$. Then, there exists $p_* \in (0,1]$ such that\footnote{The conclusions here should be interpreted as statements about equivalence classes of functions that are equal almost surely under all distributions $P \in {\cal P}$, that is, the bounds we present are unique up to modifications on sets of measure zero.}
     \begin{align*}
    B(Y_1,D_1,D_2)  &=  a_{\max} + \frac{2  p_*  -  D_2}{p_*^2} \;   D_1 \, (Y_1-a_{\max})  .
\end{align*} 
Conversely, any function $B(Y_1, D_1, D_2)$ of the form in the last display satisfies \eqref{BoundMainExample} and is not dominated, and thus we obtain a full characterization of all valid and non-dominated bounds of this type.
\end{theorem}
The proof is given in the appendix. This theorem characterizes all functions $B(Y_1,D_1,D_2)$ whose expectations provide valid and non-dominated upper bounds on $\mathbb{E}[ Y_i(1) ]$ under our assumptions, showing they can be parameterized by $p_* \in (0,1]$. For identification purposes, we could optimize over $p_*$:
\begin{align*}
  \mathbb{E}_P[ Y_i(1) ]
         &\leq  \min_{p_* \in (0,1]} 
         \mathbb{E}_P\left[ a_{\max} + \frac{2  p_*  -  D_2}{p_*^2} \;   D_1 \, (Y_1-a_{\max}) \right] ,
        \\
        &= \left\{  
        \begin{array}{l@{\;\;}l}
               \mathbb{E}_P[ Y_i(1) ] & \text{if $P(D_i=1)>0$,}
               \\
               a_{\max}   & \text{if $P(D_i=1)=0$.}
        \end{array} \right.  
\end{align*}
The last equality follows from choosing $p_* = P(D_i=1)$ in the minimum when positive. Through such ``intersection bounds'' we recover the well-known identification result for $\mathbb{E}_P[ Y_i(1) ]$ under unconfoundedness.

However, replicating this identification result is not our aim. As discussed previously, a key challenge in finite-sample inference is that the true propensity score $p = P(D_i=1)$ is unknown, varies with $X_i$, and may be close to zero --- leading to potentially large variances and non-uniformity in any corresponding estimator. Therefore, we focus on bounds that can be expressed as simple sample averages, as introduced in \eqref{BoundULplim}, which is the context for Theorem~\ref{th:OurBounds}.

An illuminating special case arises when $p_*=1$, yielding:
    \begin{align*}
        B(Y_1,D_1,D_2)  =  a_{\max} + (2    -  D_2) \,   D_1 \, (Y_1-a_{\max}).
    \end{align*}
When symmetrized across observations, this becomes:
\begin{align}
    \frac{1}{2}[B(Y_1,D_1,D_2) + B(Y_2,D_2,D_1)]
   &= \left\{ \begin{array}{ll}
       a_{\max}  & \text{if $(D_1,D_2)=(0,0)$,}
         \\
          Y_2 & \text{if $(D_1,D_2)=(0,1)$,}
          \\
          Y_1 & \text{if $(D_1,D_2)=(1,0)$,}
          \\
            \frac{1}{2}[Y_1 + Y_2]    & \text{if $(D_1,D_2)=(1,1)$.}
     \end{array} \right.
     \label{BasicSecondOrderBounds}
\end{align}
This special case has an intuitive interpretation: When neither unit is treated $(D_1,D_2)=(0,0)$, we can only use the worst-case bound $a_{\max}$. However, when at least one unit is treated, we can use the corresponding treatment outcome as an estimate for $\mathbb{E}_P[ Y_i(1) ]$. This requires unconfoundedness since we select between $Y_1$ and $Y_2$ based on $(D_1,D_2)$. Equation \eqref{BasicSecondOrderBounds} provides the simplest example of what we term a ``second-order bound'', where information is pooled across two observations using unconfoundedness.

These bounds share similarities with matching estimators, where outcomes with $D_1 \neq D_2$ for units sharing the same covariate value $X_1=X_2$ are matched to obtain counterfactual outcomes. The key distinction is that we do not require $D_1 \neq D_2$, necessitating worst-case bounds when $D_1=D_2=0$. However, this relaxation allows our bounds to remain valid without requiring overlap assumptions.

While the $p_*=1$ case yields straightforward bounds, Theorem~\ref{th:OurBounds} reveals a richer family of bound functions parameterized by $p_* \in (0,1]$. To better understand this family, we introduce the true propensity score $p:=P(D_i=1)$ and define the weight function $w^{(2)}: [0,1] \times (0,1] \rightarrow  (-\infty,1]$ as:
\begin{align*}
    w^{(2)}(p,p_*) := 1- \left( \frac{p_* - p}{ p_*}   \right)^{2}        . 
\end{align*} 
For $p \neq 0$ we then have\footnote{
The condition $p \neq 0$ is only required for our discussion here, not for Theorem~\ref{th:OurBounds}.
This is because the $1/p$ in \eqref{ExpectationOverD2} can  be canceled against the factor $p$ in
$ w^{(2)}(p,p_*)   = p \cdot \frac {2p_* - p} {p_*^2}  $
to avoid divison by zero when $p=0$.
}
 \begin{align}
    \mathbb{E}_P\left[ B(Y_1,D_1,D_2) \, \big| \, Y_1,D_1 \right] &=  a_{\max}   
    +  w^{(2)} \big(p ,p_*  \big)\;   \frac{D_1 \, (Y_1-a_{\max})} {p}  ,
    \label{ExpectationOverD2}
\end{align}  
and consequently
 \begin{align*}
    \mathbb{E}_P\left[ B(Y_1,D_1,D_2)  \right] &=  a_{\max}   
    +  w^{(2)} \big(p ,p_*  \big)\; \big\{ \mathbb{E}_P[Y(1)]-a_{\max} \big\}
    \\
    &=  \left[ 1 -  w^{(2)} \big(p ,p_*  \big) \right] \,a_{\max}   
    +  w^{(2)} \big(p ,p_*  \big) \, \mathbb{E}_P[Y(1)]   .
\end{align*} 
Thus, in expectation, our bounds form weighted averages between the worst-case bound $a_{\max}$ and the target parameter $\mathbb{E}_P[Y(1)]$, with weights determined by $w^{(2)} \big(p ,p_*  \big)$.

\begin{figure}[tb!] 
\begin{center}
\begin{tikzpicture}[scale=0.8]
\begin{axis}[
    axis lines = left,
    xlabel = $p$,
    ylabel = {$w^{(2)}(p,p_*)$},
    xmin=0, xmax=1.1,
    ymin=-0.3, ymax=1.1,
    legend pos=outer north east
]

\addplot [
    domain=0:1, 
    samples=100, 
    color=black,
]
{x};
\addlegendentry{Manski bounds}

\addplot [
    domain=0:1, 
    samples=100, 
    color=blue,
    style=ultra thick
    ]
{1-((x-0.25)/0.25)^2};
\addlegendentry{$p_*=0.25$}

\addplot [
    dashed,
    domain=0:1, 
    samples=100, 
    color=red,
    style=ultra thick
    ]
{1-((x-0.5)/0.5)^2};
\addlegendentry{$p_*=0.5$}

\addplot [
    dotted,
    domain=0:1, 
    samples=100, 
    color=dgreen,
    style=ultra thick
    ]
{1-((x-1)/1)^2};
\addlegendentry{$p_*=1$}

\end{axis}

\end{tikzpicture}

\caption{\label{fig:weights0}Weights $w^{(2)}(p,p_*)$  as a function of $p$, for different values of $p_*$}
\end{center}
\end{figure}
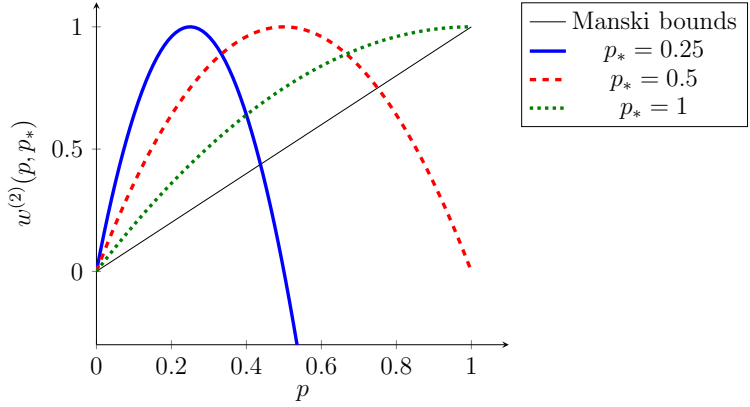 

Figure~\ref{fig:weights0} plots $w^{(2)}(p,p_*)$ as a function of $p$ for various values of $p_*$. When $p_*$ equals the true propensity score $p$, we have $w^{(2)} \big(p ,p_*  \big)=1$ and our upper bounds are sharp. When $p \neq p_*$, we have $w^{(2)} \big(p ,p_*  \big) < 1$, yielding valid but non-sharp bounds.

For comparison, the Manski bounds have $\mathbb{E}\left[ a_{\max} + D_i (Y_i -  a_{\max})   \right] = [1- p(x)] \, a_{\max} +  p(x) \; \mathbb{E}\left[ Y(1)  \right]$, corresponding to the weight function $w^{(1)}:p \mapsto p$, also shown in Figure~\ref{fig:weights0}. The figure reveals that the bounds in \eqref{BasicSecondOrderBounds}, corresponding to $p_*(x)=1$, are the only second-order bounds that uniformly dominate the Manski bounds across all possible values of the true propensity $p$.

Since we consider only undominated second-order bounds, none of the second-order bounds dominates any other second-order bound with a different $p_*$ value uniformly across data generating processes parameterized by $p$. If we have a reliable guess (or estimate) for the propensity score $p$, it should be used for $p_*$ to ensure reasonably tight bounds. The advantage of our bounds over alternatives (like inverse propensity score weighting) is their continued validity even when our guess (or estimate) for $p$ is incorrect.

It's worth noting that for $p_*<0.5$, the weights $w^{(2)} \big(p,p_* \big)$ become negative for large values of $p$, indicating that the expected bounds can perform worse than simply reporting $a_{\max}$, but nevertheless remaining valid.

The bounds introduced in this section illustrate the middle ground discussed in Section~\ref{sec:Setup} between minimal-assumption Manski bounds and full point identification through propensity score methods. By pooling information across pairs of observations, we achieve potential improvements over Manski bounds without requiring the extensive pooling needed for consistent propensity score estimation. We believe that this approach is particularly valuable in settings with high-dimensional covariates or limited local sample sizes, where reliable propensity score estimation may be infeasible but some degree of information pooling across observations remains possible. This two-observation case serves as a building block for the general bounds developed in subsequent sections.

\begin{remark}
    For the bounds discussed in this section and the result of Theorem~\ref{th:OurBounds}, it is important to observe that the expectation in \eqref{BoundMainExample} is taken jointly over $(Y_i, D_i)$, with $X_i$ implicitly conditioned on throughout. If, instead, we required the conditional bound
\begin{align}
  \mathbb{E}_P[ Y_i(1) ]
         \leq  \mathbb{E}_P\left[ \left. \frac 1 2 \Big( B(Y_1,D_1,D_2) + B(Y_2,D_2,D_1) \Big) \, \right| \, D_1,D_2 \right] 
   \label{BoundsUnconditional}
\end{align}    
 to hold, then the only undominated solution would be the one in Theorem~\ref{th:OurBounds} with $p_*=1$, which reduces to \eqref{BasicSecondOrderBounds} after symmetrization. 
In other words, the bounds in Theorem~\ref{th:OurBounds} for $p_* < 1$ do not satisfy the conditional statement in \eqref{BoundsUnconditional}, only the unconditional one in \eqref{BoundMainExample}. 
Indeed, when $D_1 + D_2 > 0$, the sample average of treated outcomes is the unique conditionally unbiased estimator of $\mathbb{E}_P[Y_i(1)]$, so any conditionally valid bound must coincide with it. When $D_1 = D_2 = 0$, only the trivial bound $a_{\max}$ is available. The $p_* = 1$ bound in \eqref{BasicSecondOrderBounds} is exactly this conditionally optimal bound. This illustrates why requiring conditional validity is too strong for our purposes: it eliminates the dependence on $p$ that is central to our argument.
\end{remark}

\begin{remark}\label{rem:discussion:p_star_1}
  Closely related to the previous remark, if $D_1 = 1$ and/or $D_2 = 1$, then the bounds in Theorem~\ref{th:OurBounds} with $p_* < 1$ become somewhat counterintuitive. In such cases, we observe an outcome under treatment, which provides an unbiased estimate of $\mathbb{E}_P[ Y_i(1) ]$ conditional on that realization of $(D_1, D_2)$. Yet the bounds for $p_* < 1$ are not equal to that
  unbiased estimate in that case. For instance, when $D_1 = D_2 = 1$, we obtain
  $$
  \frac{1}{2}[B(Y_1,1,1) + B(Y_2,1,1)]
  =  \left( \frac{1-p_*}{p_*} \right)^2
  a_{\max} + 
  \left[ 1 - \left( \frac{1-p_*}{p_*} \right)^2  \right]
  \, \frac 1 2 (Y_1+Y_2)   ,
  $$
 which is always weakly greater than the simple average $ \frac 1 2 (Y_1+Y_2)$. 
 In other words, if one insists on using the conditionally unbiased estimate
  for $\mathbb{E}_P[ Y_i(1) ]$ in cases where $D_1 = 1$ and/or $D_2 = 1$, then one has to use the bound with $p_*=1$.

  Nevertheless, as discussed above, when the true propensity score $p = P(D_i = 1)$ is strictly less than one, the choice $p_* = 1$ is generally suboptimal: only the choice $p_* = p$ yields bounds that are tight, i.e., $\mathbb{E}_P[ Y_i(1) ] = \mathbb{E}_P[ B(Y_1, D_1, D_2) ]$. In practice, however, $p$ is unknown, and selecting $p_* = p$ is infeasible without additional information. Instead, the goal should be to choose $p_*$ reasonably close to $p$ --- the bounds remain valid regardless, but they are tighter when $p_*$ is well chosen.
 \end{remark}

\begin{remark}
\label{remark:ChoiceP*}
In applications, a simple and effective default 
for choosing the reference propensity score is the global treatment share $p_*(x) = n^{-1}\sum_{i=1}^n D_i$, applied uniformly across all covariate values. This choice is strictly between zero and one when both treatment statuses are present, remains stable even when individual covariate cells are small, and performs well in our simulations and empirical examples in Sections~\ref{sec:MC} and~\ref{sec:EE}. 
More refined choices (such as shrinkage toward a global mean, smoothing across similar covariate values, or empirical Bayes procedures) may further tighten the bounds when appropriate structure is present in the data. A formal development of such data-driven strategies, however, is beyond the scope of the present paper.
\end{remark}

\section{Generalizations to higher order, ATE and ATT}
\label{sec:PopulationBounds}

 We now return to the general setting with covariates introduced in Section~\ref{sec:Setup}. The basic insights gained from analyzing the case of two observations sharing the same covariate value carry over naturally to this more general context, primarily requiring adjustments to notation.

To understand how the notation from Section~\ref{sec:MainIdea} extends to the current setting, note that the function $B(Y_1,D_1,D_2)$ introduced there becomes $B^{(2)}_{1,a}(Y_i,D_i,p(X_i),X_i)$ in our general notation here, where the superscript $(2)$ indicates that we are dealing with second-order bounds, the subscript 1 indicates that we are bounding $Y_i(1)$,
and the choice of constant $a \in \{a_{\min},a_{\max}\}$ 
in the second subscript
depends on whether we construct upper or lower bounds.
The parameter $p_*(x)$ continues to play the role of $p_*$ but can now vary with the covariate value $x$. The weight function $w^{(2)}(p,p_*)$ remains unchanged but is now applied separately for each covariate value.

\subsection{Short summary of second-order bounds}

The main idea of our second-order bounds was already discussed in Section~\ref{sec:MainIdea} for bounds on $\mathbb{E}[Y_i(1)]$. The generalization to $\mathbb{E}[Y_i(0)]$ follows naturally by symmetry, which jointly provides second-order bounds for the ATE. While the main text presents results for both ATE and ATT throughout, we defer a more detailed discussion of the second-order ATT bounds to Appendix~\ref{app:SecondOrderATT}. 

To implement these bounds, we require the researcher to specify a reference propensity score $p_* : {\cal X} \rightarrow (0,1)$, which can be either postulated based on prior knowledge or estimated from the data. While the resulting bounds remain valid regardless of the choice of $p_*(x)$, their sharpness depends critically on how close the true propensity score is to $p_*(x)$. 
For all theoretical results that follow, we assume that $p_*(x)$ is non-random, with $p_*(x)=1/2$ serving as a natural default choice in the absence of strong prior information
  (see also Remark~\ref{remark:ChoiceP*} for practical guidance on choosing $p_*(x)$).
Using this framework, we can now formally define our second-order bounds. For every $a \in \mathbb{R}$, we define
\begin{align}
    B^{(2)}_{0,a}(Y_i,D_i,p(X_i),X_i) &:=
   a + \frac{1 - 2 \, p_*(X_i)  +   p(X_i)    }  {\left[ 1- p_*(X_i) \right]^2} \; (1-D_i) \, (Y_i-a)        ,
\nonumber  \\[5pt]  
    B^{(2)}_{1,a}(Y_i,D_i,p(X_i),X_i) &:=
   a + \frac{ 2 \, p_*(X_i)  -  p(X_i)   }{  \left[ p_*(X_i) \right]^2     }  \; D_i \, (Y_i-a)   ,
\nonumber  \\[5pt]  
    C^{(2)}_{a}(Y_i,D_i,p(X_i),X_i)
      &:=    D_i\, (Y_i-a)     +  \frac{ \left[ p_*(X_i)  \right]^2   -  p(X_i)     }   {\left[ 1- p_*(X_i) \right]^2} \;  (1-D_i) \, (Y_i-a) ,
    \label{DefBCsecond}  
\end{align}
and
\begin{align}
    L^{(2)}(Y_i,D_i,p(X_i),X_i) &:= B^{(2)}_{1,a_{\min}}(Y_i,D_i,p(X_i), X_i)   -   B^{(2)}_{0,a_{\max}}(Y_i,D_i,p(X_i), X_i)  ,
   \nonumber \\
    U^{(2)}(Y_i,D_i,p(X_i), X_i) &:= B^{(2)}_{1,a_{\max}}(Y_i,D_i,p(X_i), X_i)   -   B^{(2)}_{0,a_{\min}}(Y_i,D_i,p(X_i), X_i)  .
    \label{DefLU2}
\end{align}    
Here, we introduce the second-order bound functions as dependent on the unknown propensity score \( p(X_i) \). Since \( p(X_i) \) is not observed, these bounds are initially infeasible in this form. However, because all second-order bounds are linear in \( p(X_i) \), they can be rendered feasible by substituting \( p(X_i) \) with the treatment status \( D_j \) of another unit \( j \neq i \) that shares the same covariate value $X_j = X_i$. 
The resulting bound function \( B^{(2)}_{1,a}(Y_i,D_i,D_j,X_i) \) then corresponds exactly to the bound derived in Theorem~\ref{th:OurBounds}. Expressing the bounds as functions of \( p(X_i) \) in this section is advantageous, as it eliminates the need to reference other units explicitly. The following proposition summarizes the key properties of these second-order bounds.

\begin{proposition}
    \label{prop:MainBounds}
    Let Assumption~\ref{ass:MAIN} hold. Let  $p_* : {\cal X} \rightarrow (0,1)$.\footnote{Strictly speaking, it is allowed that $p_*(x) = 1$ for $\mathbb{E} \, Y(1)$ and $p_*(x) = 0$ for  $\mathbb{E} \, Y(0)$ and ATT. For simplicity, we assume that $p_* : {\cal X} \rightarrow (0,1)$.}
    Let $d \in \{0,1\}$.
    Then,  
   \begin{align*}
      (a) &&  \mathbb{E}\left[  B^{(2)}_{d,a_{\min}}(Y_i,D_i,p(X_i),X_i) \, \Big| \, X_i  \right]  &\leq  \mathbb{E}\left[ Y_i(d)   \, \big| \, X_i  \right]   \leq \mathbb{E}\left[  B^{(2)}_{d,a_{\max}}(Y_i,D_i,p(X_i),X_i)   \, \Big| \, X_i  \right] ,  
    \\[5pt]
      (b) &&  \mathbb{E}\left[  L^{(2)} (Y_i,D_i,p(X_i),X_i) \, \Big| \, X_i  \right]  &\leq  \tau(X_i)   \leq \mathbb{E}\left[  U^{(2)} (Y_i,D_i,p(X_i),X_i)   \, \Big| \, X_i  \right] ,  
    \\[5pt]
    (c) &&    \mathbb{E}\left[ C^{(2)}_{a_{\max}}   \, \Big| \, X_i  \right]  &\leq   \pi(X_i)  \leq
          \mathbb{E}\left[ C^{(2)}_{a_{\min}}   \, \Big| \, X_i  \right]  .
      & \qquad \qquad\qquad\qquad   \qquad  \qquad  \qquad  
    \end{align*}
    If, in addition, $p(X_i) = p_*(X_i)$, then all the inequalities in this proposition become equalities.
\end{proposition}

The proof is provided in the appendix. 
Once we have bounds on  $ \tau(x) $ and $  \pi(x) $, then we can also construct bounds on ATE and ATT 
defined in  \eqref{ATestimands}.

\subsection{Higher-order bounds}

In Section~\ref{sec:MainIdea}, we derived second-order bounds by directly considering pairs of observations sharing the same covariate value. For higher-order bounds, it is more convenient to first formulate bounds at the ``population-level'' --- that is, expressing the bounds directly in terms of the unknown propensity score $p(x)$ rather than in terms of treatment indicators of other observations. This formulation can be thought of as having already taken conditional expectations over the treatment indicators of other observations, leaving us with expressions that depend on $p(X_i)$ directly.

This ``population-level'' formulation was already used in
the second-order bounds in \eqref{DefBCsecond} and \eqref{DefLU2} above, which are linear functions of $p(X_i)$.  For higher-order bounds, we generalize this idea by allowing $p(X_i)$ to enter as higher-order polynomials. The resulting bounds will be infeasible since they depend on the unknown propensity score $p(X_i)$. However, they have feasible sample analogs: when implementing the bounds in practice, each power $[p(X_i)]^r$ in our population-level expressions can be replaced by products of $r$ different treatment indicators $D_j$ from observations sharing the same covariate value. The advantage of deriving the bounds in that form is that it allows us to understand the structure of the bounds before considering their sample implementation that requires multiple observations.

Dropping the index $i$ throughout, and the arguments $(Y_i,D_i,p(X_i),X_i)$ from the bound functions,
our aim is to generalize the second-order bounds in 
\eqref{DefBCsecond} by considering,
for positive integers $q$,
\begin{align}
    B^{(q)}_{d,a}(\lambda) &=
   a +  \left\{  \sum_{r=0}^{q-1} \lambda_{r,d}(X) \,  [p(X)]^r \right\}  \, \mathbbm{1}\left\{D=d\right\} \, (Y-a)        ,   \quad  d \in \{0,1\} ,
\nonumber \\
    C^{(q)}_a( \lambda)
      &=    D\, (Y-a)     -  \left\{  \sum_{r=0}^{q-1}  \lambda_r(X) \,  [p(X)]^r \right\}   \,  (1-D) \, (Y-a) ,
      \label{AllOrderBoundsGeneral}
\end{align}
where the coefficients $\lambda_{r,d}(x), \lambda_r(x) \in \mathbb{R}$ still need to be determined for $q>2$.
Again, the motivation for \eqref{AllOrderBoundsGeneral} is that,  we can construct
unbiased estimates for $B^{(q)}_{d,a}(\lambda) $ and $C^{(q)}_a(\lambda)$ by replacing $ [p(X)]^r$ with
a product of treatment indicators  from $r$ different observations with the same  (or similar) regressor values.

Motivated by our finding for second-order bounds we again choose a reference propensity score $p_*(x)$ to find unique 
solutions for the coefficients $ \lambda_{r,d}(x)$ and $  \lambda_r(x)$.
Once we have chosen $p_*(x)$, then for the second-order bounds the coefficients are uniquely determined
by the properties of the bounds summarized in Proposition~\ref{prop:MainBounds} --- namely, the bounds should be valid for all population distributions satisfying 
 Assumption~\ref{ass:MAIN}, and the bounds should be binding if $p(x)=p_*(x)$. However, for $q>2$ those properties are not sufficient anymore 
 to uniquely determine the coefficients, because we now have additional degrees of freedom in the higher-order polynomial coefficients. 
To make use of this additional flexibility and to obtain unique coefficients again, we therefore demand the bounds to not only have good
properties when $p(x)=p_*(x)$, but also when $p(x) \in [p_*(x)-\epsilon, p_*(x) + \epsilon]$, for small $\epsilon>0$, that is, we want
to have good performance in a small neighborhood around the reference propensity score $p_*(x)$.

Specifically, we choose the optimal coefficients $\lambda_{r,d}(x)$ and $\lambda_r(x)$ such that the expected widths of the bounds
\begin{align*}
     & \mathbb{E}_{p(x)} \left[ B^{(q)}_{d,a_{\max}}(\lambda) 
      - B^{(q)}_{d,a_{\min}}(\lambda) \, \Big| \, X=x \right] ,
      \\
       & \mathbb{E}_{p(x)}  \left[ C^{(q)}_{a_{\min}}(\lambda) 
      - C^{(q)}_{a_{\max}}(\lambda)   \, \Big| \, X=x \right] ,
\end{align*}
are minimized not only at $p(x)=p_*(x)$, but also when considering the worst-case expected widths within an infinitesimal neighborhood
of the reference propensity score $p_*(x)$,
see part (iii) of Proposition~\ref{prop:HigherOrder} below for a formalisation of this.
 Here, $\mathbb{E}_{p(x)}$ refers to the expectation  over $(Y_i,X_i,D_i)$ with propensity score (i.e.\ distribution of $D_i|X_i=x$) specified by $p(x)$.

Once we have solved for the optimal coefficients accordingly, we obtain the following optimal $B^{(q)}_{d,a}(\lambda) $
and $ C^{(q)}_a(\lambda) $,  for  integers $q \geq 1$,\footnote{%
Formally, for $p(x)=1$ we have 
$B^{(q)}_{0,a} = a + \left\{  \frac {q - p_*(x) \, \mathbbm{1} \left\{ \text{$q$ is odd} \right\}  } {1-p_*(x)} \right\}   (1-D) (Y-a)$
and $ C^{(q)}_a  =  D\, (Y-a)  + \left[   \frac {q-1 + p_*(x) \, \mathbbm{1} \left\{ \text{$q$ is even} \right\}} {1-p_*(x)} \right] (1-D) (Y-a)$.
For $p(x)=0$ we have   $B^{(q)}_{1,a} =a + \left\{  \frac {q - [1-p_*(x)] \, \mathbbm{1} \left\{ \text{$q$ is odd} \right\}  } {p_*(x)} \right\}  D (Y-a)$.
From the formulas in \eqref{DefGeneralBC} we obtain those results for $p(x)=1$ and $p(x)=0$ as limits
 when $p(x) \rightarrow 1$ and $p(x) \rightarrow 0$. However, the details of those special cases do not actually matter, 
 because e.g.\ for $p(x)=1$ we also have $D=1$ with probability one, and therefore $B^{(q)}_{0,a} = a$
 and $C^{(q)}_a  =  D\, (Y-a) $.
}
\begin{align}
    B^{(q)}_{0,a}   &:=   a +  w^{(q)} \big(1-p(X),1-p_*(X) \big) \;  \frac{(1-D)(Y-a)}{1-p(X)}  ,
   \nonumber     \\ 
    B^{(q)}_{1,a}   &:=     a +  w^{(q)} \big(p(X),p_*(X) \big) \;  \frac{D(Y-a)}{p(X)} ,
 \nonumber   \\ 
    C^{(q)}_a    &:=   
      \displaystyle   D\, (Y-a) 
 -    \widetilde w^{(q)} \big(p(X),p_*(X) \big) \,  \frac{p(X) \, (1-D) \, (Y-a)} {1-p(X)} ,
      \label{DefGeneralBC}      
\end{align}     
where the weight functions   are given by
\begin{align}
   w^{(q)} \left(p,p_* \right)  &:= \left\{ 
    \begin{array}{l@{\quad}l}
 \displaystyle    1- \left( 1- p \right) \left( \frac{p_* - p }{ p_*}   \right)^{q-1}   
         & \text{if $q$ is odd,}
\\[20pt]
 \displaystyle   1- \left( \frac{p_*-p}{ p_*}   \right)^{q}  
         & \text{if $q$ is even,}
    \end{array} \right.
\nonumber  \\[10pt]
      \widetilde w^{(q)} \left(p,p_* \right)    &:=  \left\{ 
    \begin{array}{l@{\qquad \quad}l}
\displaystyle    
    1 - \left( \frac{p - p_*}  {1-p_*}  \right)^{q-1}      
         & \text{if $q$ is odd,}
   \\[20pt]
 \displaystyle 
        1 - \frac 1 p  \left( \frac{p - p_*}  {1-p_*}  \right)^q  
         & \text{if $q$ is even.}
    \end{array} \right.
   \label{DefGeneralWeights} 
\end{align}
For $q=1$ and $q=2$ the formulas in \eqref{DefGeneralBC}  just give the same functions $B^{(q)}_{d,a}$ and $ C^{(q)}_a $ that were already discussed above.
It may not be obvious from those general formulas, but $B^{(q)}_{d,a}$ and $ C^{(q)}_a $ are indeed polynomials of order $(q-1)$ in $p(X)$. For example,
for $q=3$ we find
\begin{align*}
      B^{(3)}_{0,a} &=  a + \left\{ 1 +   p(X) \, \frac{ 1 + p(X) - 2 \, p_*(X)  }{[1- p_*(X)]^2}    \right\}
                    (1-D) \, (Y-a)  ,
     \\[5pt]
      B^{(3)}_{1,a} &=  a + \left\{ 1 +   [1-p(X)] \, \frac{  2\, p_*(X) - p(X) }{[p_*(X)]^2}    \right\}
                    D \, (Y-a)  ,
    \\[5pt]
       C^{(3)}_a    &=            D\, (Y{-}a)    -   p(X) \, \frac{ 1 + p(X) - 2 \, p_*(X)  }{[1- p_*(X)]^2}  (1 - D) \, (Y - a)       ,
\end{align*}
which are all second order polynomials in $p(X)$.

We now want to formally state the optimality result for these bounds.
We define $L^{(q)}(Y_i,D_i,p(X_i),X_i) $ and $ U^{(q)}(Y_i,D_i,p(X_i), X_i) $  as in \eqref{DefLU2}, but with superscipt $(2)$ replaced by $(q)$,
and for  $\epsilon \geq 0$ and $p_*(x) \in (0,1)$, we let
${\cal B}_\epsilon(p_*(x)) := \Big\{ p(x) \in [0,1] \, \Big| \,   \left| p(x) - p_*(x) \right| \leq \epsilon \Big\}$
be the $\epsilon$-ball around $p_*(x)$.

\begin{proposition}
    \label{prop:HigherOrder}
    Let Assumption~\ref{ass:MAIN}  hold. Let  $p_* : {\cal X} \rightarrow (0,1)$. Then:
    \begin{itemize}
    \item[(i)]
    For integers $q \geq 1$  we have
  \begin{align*}
      (a) &&  \mathbb{E}\left[  B^{(q)}_{d,a_{\min}}(Y_i,D_i,p(X_i),X_i) \, \Big| \, X_i  \right]  &\leq  \mathbb{E}\left[ Y_i(d)   \, \big| \, X_i  \right]   \leq \mathbb{E}\left[  B^{(q)}_{d,a_{\max}}(Y_i,D_i,p(X_i),X_i)   \, \Big| \, X_i  \right] ,  
    \\[5pt]
      (b) &&  \mathbb{E}\left[  L^{(q)} (Y_i,D_i,p(X_i),X_i) \, \Big| \, X_i  \right]  &\leq  \tau(X_i)   \leq \mathbb{E}\left[  U^{(q)} (Y_i,D_i,p(X_i),X_i)   \, \Big| \, X_i  \right] ,  
    \\[5pt]
    (c) &&    \mathbb{E}\left[ C^{(q)}_{a_{\max}}   \, \Big| \, X_i  \right]  &\leq   \pi(X_i)  \leq
          \mathbb{E}\left[ C^{(q)}_{a_{\min}}   \, \Big| \, X_i  \right]  .
      & \qquad \qquad\qquad\qquad   \qquad  \qquad  \qquad  
    \end{align*}

   \item[(ii)] If $q>1$ and   $p(X_i) = p_*(X_i)$, then all the inequalities in part (i) of the proposition  become equalities.
    
   \item[(iii)]
    Let $\lambda_{r,d}(x) \in \mathbb{R}$ 
    and $\lambda_r(x)  \in \mathbb{R}$ be such that  $ B^{(q)}_{d,a}(\lambda) $ and $ C^{(q)}_a(\lambda)$ defined 
    in  \eqref{AllOrderBoundsGeneral} satisfy the inequalities in part (i) for all population distribution that
    satisfy Assumption~\ref{ass:MAIN}. Then there exists $\epsilon>0$ such that 
    for all $p(X_i) \in {\cal B}_\epsilon(p_*(X_i))$ 
    and $d \in \{0,1\}$ we have
\begin{align*}
        \mathbb{E}_{p(X_i)} \left[ B^{(q)}_{d,a_{\max}}
      - B^{(q)}_{d,a_{\min}}  \, \Big| \, X_i \right] 
  &  \leq
       \mathbb{E}_{p(X_i)} \left[ B^{(q)}_{d,a_{\max}}(\lambda) 
      - B^{(q)}_{d,a_{\min}}(\lambda)  \, \Big| \, X_i  \right] ,
 \\
         \mathbb{E}_{p(X_i)} \left[ C^{(q)}_{a_{\min}}
      - C^{(q)}_{a_{\max}}  \, \Big| \, X_i  \right] 
   &\leq
      \mathbb{E}_{p(X_i)}  \left[ C^{(q)}_{a_{\min}}(\lambda) 
      - C^{(q)}_{a_{\max}}(\lambda)   \, \Big| \, X_i \right] .
\end{align*}  
That is, within a small neighborhood of $p_*(X_i)$,
 the expected width of the bounds in part (i) 
 is smaller or equal to the expected width of any other set of valid $q$'th order bounds.
 \end{itemize}

\end{proposition}

\medskip
 
 The proof is given in the appendix.
To better understand the result of Proposition~\ref{prop:HigherOrder},
consider the lower bound on  $\mathbb{E}\left[ Y(1)\, \big| \, X\right]  $, which is given by
\begin{align*}
     \mathbb{E}\left[ B^{(q)}_{1,a_{\min}} \, \big| \, X  \right] 
     &= [1-w^{(q)}(p(X),p_*(X))] \, a_{\min} +  w^{(q)}(p(X),p_*(X)) \; \mathbb{E}\left[ Y(1)\, \big| \, X  \right] .
\end{align*}     
Thus, $  \mathbb{E}\left[ B^{(q)}_{1,a_{\min}} \, \big| \, X=x \right] $ is a weighted average
between $a_{\min}$   and $\mathbb{E}\left[ Y(1)\, \big| \, X=x \right]  $.
 The weights   always satisfy  $w^{(q)}(p,p_*) \leq 1$, which together with $a_{\min} \leq Y(1)$  guarantees that
 $  \mathbb{E}\left[ B^{(q)}_{1,a_{\min}} \, \big| \, X=x \right]  \leq \mathbb{E}\left[ Y(1)\, \big| \, X=x \right] $.

\begin{figure}[tb!] 
\begin{center}
\begin{tikzpicture}[scale=0.8]
\begin{axis}[
    axis lines = left,
    xlabel = $p$,
    ylabel = {$w^{(q)}(p,p_*)$},
    xmin=0, xmax=1.1,
    ymin=-0.3, ymax=1.1,
    legend pos=outer north east
]

\addplot [
    domain=0:1, 
    samples=100, 
    color=black,
]
{x};

\addplot [
    domain=0:1, 
    samples=100, 
    color=blue,
    style=ultra thick
    ]
{1-((x-0.4)/0.4)^2};

\addplot [
    dashed,
    domain=0:1, 
    samples=100, 
    color=red,
    style=ultra thick
    ]
{1-((1-x)*((x-0.4)/0.4)^2)};

\addplot [
    dotted,
    domain=0:1, 
    samples=100, 
    color=dgreen,
    style=ultra thick
    ]
{1-((x-0.4)/0.4)^4};

\addplot[dashed]
  coordinates{(0.4,-0.3) (0.4,1)};

\end{axis}

\end{tikzpicture}
\begin{tikzpicture}[scale=0.8]
\begin{axis}[
    axis lines = left,
    xlabel = $p$,
    ylabel = {$\widetilde w^{(q)}(p,p_*)$},
    xmin=0, xmax=1.1,
    ymin=-0.3, ymax=1.1,
    legend pos=outer north east
]

\addplot [
    domain=0:1, 
    samples=100, 
    color=black,
]
{0};
\addlegendentry{$q=1$}

\addplot [
    domain=0:1, 
    samples=100, 
    color=blue,
    style=ultra thick
    ]
{1-((x-0.4)/(1-0.4))^2/x};
\addlegendentry{$q=2$}

\addplot [
    dashed,
    domain=0:1, 
    samples=100, 
    color=red,
    style=ultra thick
    ]
{1-((x-0.4)/(1-0.4))^2};
\addlegendentry{$q=3$}

\addplot [
    dotted,
    domain=0:1, 
    samples=100, 
    color=dgreen,
    style=ultra thick
    ]
{1-((x-0.4)/(1-0.4))^4/x};
\addlegendentry{$q=4$}

\addplot[dashed]
  coordinates{(0.4,-0.3) (0.4,1)};

\end{axis}

\end{tikzpicture}

\caption{\label{fig:weights}Weights $w^{(q)}(p,p_*)$ and $\widetilde w^{(q)}(p,p_*)$ as a function of $p$, for $p_*=0.4$ and $q \in \{1,2,3,4\}$}
\end{center}
\end{figure}
 
Figure~\ref{fig:weights} shows  $w^{(q)}(p,p_*)$ as a function of $p$ for $p_*=0.4$ and different values of $q$. 
For $p=0$ we always have $w^{(q)}(p,p_*)=0$, because in that case we only have observations with $D=0$ for $X=x$,
implying that we cannot learn anything about $Y(1)$ from the data.
For   $p = p_*$ we have $w^{(q)}(p,p_*)=1$ for $q \geq 2$, that is,   the lower bound is sharp in that case.
For $p$ close to $p_*$ the weights are closer to one (implying that the bounds are sharper) the larger we choose $q$.
For the $k$'th derivative of $w^{(q)}(p,p_*)$ at $p=p_*$ we have
$$
     \frac{\partial^k w^{(q)}(p_*,p_*)} {\partial^k p} = 0 , 
    \qquad \text{for} \;  \left\{ 
    \begin{array}{l@{\quad}l}
    k \in \{1,\ldots,q-2\} 
         & \text{if $q$ is odd,}
\\ 
  k \in \{1,\ldots,q-1\} 
         & \text{if $q$ is even,}
    \end{array} \right.
$$
which explain why for $p$ close to $p_*$ the weights are closer to one the higher we choose $q$.
However, if $p$ is far away from $p_*$, then the weights $w^{(q)}(p,p_*)$ for $q \geq 2$ can be far away from one, and can even be smaller
than $w^{(1)}(p)$, that is, the bounds can be worse than Manski bounds if $p$ is far away from $p_*$.

The discussion for ATT bounds is analogous. In that case we have
\begin{align*}
   \mathbb{E}\left[ C^{(q)}_a  \big|  X  \right]  
  &=
   \Big[1-\widetilde w^{(q)} \big(p(X),p_*(X) \big) \Big]    \mathbb{E}\left[ C^{(1)}_a  \big|   X    \right]  
    +
   \widetilde w^{(q)} \big(p(X),p_*(X) \big) \,  \pi(X) ,
\end{align*} 
that is, conditional on $X$, the ATT bounds are a linear combination between their Manski bounds and the true ATT contribution for $X$.
Figure~\ref{fig:weights} also shows the weights $\widetilde w^{(q)}(p,p_*)$ as a function of $p$, for $p_*=0.4$ and various values of $q$.

\begin{remark}
    We have chosen to consider bounds that are optimal in a small neighborhood of a given reference propensity
    score $p_*(x)$. Alternative  bounds can be constructed based on other optimality 
    criteria. For example, subject to the bounds being valid for all population distribution that satisfy 
    Assumption~\ref{ass:MAIN},
    one could minimize the expected width of the bounds under a chosen prior on the propensity score.
    From a frequentist perspective, it is ultimately a matter of taste  what optimality criteria to use here.
    We find it convenient to  parameterize the bounds in terms of the reference propensity
    score $p_*(x)$, because it is easy to interpret and 
      leads to easy analytic formulas for the bounds.
\end{remark}

\begin{remark}
    Even the local optimality of our bounds needs to be interpreted carefully. This is because in part (iii) of Proposition~\ref{prop:HigherOrder}  we   only compare to other bounds of the form \eqref{AllOrderBoundsGeneral},
   and it is natural to ask about the existence of other bounds, say for   $\mathbb{E} \, Y(d)$, that are not of the
    form $\mathbb{E}\left[ B^{(q)}_{d,a}(\lambda) \right]$. Such bounds indeed exist, and the most obvious example is the following:
    Let $B^{(q)}_{d,a}(p_*) := B^{(q)}_{d,a}$ be as defined in \eqref{DefGeneralBC}, but with the dependence on $p_*$
    now made explicit. Let ${\cal P}_*$ be a set of functions $p_* : {\cal X} \rightarrow (0,1)$. Then we have
    \begin{align}
      \sup_{p_* \in {\cal P}_*}  \mathbb{E}\left[ B^{(q)}_{d,a_{\min}}(p_*) \right]  &\leq  \mathbb{E} \, Y(d)  \leq
          \inf_{p_* \in {\cal P}_*}   \mathbb{E}\left[ B^{(q)}_{d,a_{\max}}(p_*) \right] .
          \label{IntersectionBounds}
    \end{align}
    Thus, by forming intersections of the bounds discussed so far we can obtain new valid bounds, and  those intersection bounds
     are
    generally tighter (see \citealt{chernozhukov2013intersection}).
    We do not consider such intersection bounds any further in this paper, and leave the question of constructing truly ``optimal bounds''
    (in some sense) to future research.
\end{remark}

\begin{remark}
\citet{Armstrong:Kolesar:21} propose fixed-length confidence intervals that are optimal in finite samples under normal errors with known variance. Their method remains valid asymptotically under lack of overlap, assuming a Lipschitz bound on the conditional mean function. Their approach is complementary to ours: they condition on treatment assignments, whereas our bounds average over them and allow incorporation of prior information on the propensity score. In principle, smoothness assumptions like theirs could also be combined with our framework to further tighten bounds, but we leave this to future work.
\end{remark}
  
\section{Implementation of the Bounds}\label{sec:implementation}

In this section we construct sample analogs of the bounds in Proposition~\ref{prop:HigherOrder},
and use those sample bounds to obtain asymptotically valid confidence intervals on the average treatment effects.
The bounds constructed in this section are valid for both discrete and continuous covariates $X_i$.
However, if the covariates are continuously distributed, then every observed value $X_i$ is typically only observed once,
in which case the bounds here simply become  Manski worst-case bounds. 

The interesting case, for the purpose of this section, is therefore the case where 
 the set of possible covariate values ${\cal X}$ is discrete.
However, we consider an asymptotic setting where the number of 
 covariate values  grows to infinity jointly with the total sample size. 
This is the challenging case from the perspective of treatment effect estimation,
in particular when the average number of observations available for each observed $x \in {\cal X}$ remains small.

In Section~\ref{sec:Cluster} we explain how the sample bounds for discrete covariate values from this section
can be generalized to continuous covariate values via clustering, that is,
by approximating the continuous set  ${\cal X}$ with a finite set. In that way we obtain non-trivial bounds also
for the case of continuous covariates.

\subsection{Sample analogs of the bounds of Section~\ref{sec:PopulationBounds}}
 
We require some additional notation to formulate the sample bounds.
 Analogously to $X^{(n)} :=(X_1,\ldots,X_n)$, we also define 
 $D^{(n)} :=(D_1,\ldots,D_n)$, 
the  observed sample of binary treatments.
Remember that
${\cal X}_* = \{X_i \, : \, i =1,\ldots,n\} \subset {\cal X}$ is the set of actually observed covariate values  in the sample,
and $m=| {\cal X}_* |$ is its cardinality.  
As already mentioned above, in our asymptotic analysis we let $m \rightarrow \infty$
as $n \rightarrow \infty$. This implies that $ {\cal X}_*$  changes with the sample size (we can allow ${\cal X}$ to change with $n$ as well), but we do not make
that explicit in our notation.
For $x \in {\cal X} $ we define
$$
{\cal N}(x) := \left\{ i \in \{1,\ldots,n\} \, \Big| \, X_i = x  \right\} ,
$$
the set of observations $i$ for which the observed covariate value is equal to $x$.\footnote{%
${\cal N}(x)$ is empty for $x \notin {\cal X}_*$.}
Let $n(x) := \left| {\cal N}(x) \right|$ be number of observations with $X_i=x$, and let
\begin{align*}
    n_0(x)  &:=   \sum_{i \in {\cal N}(x)}  (1-D_i) 
    \;\; \text{ and }  \;\;
      n_1(x)  :=   \sum_{i \in {\cal N}(x)}   D_i  = n(x) -   n_0(x)   
\end{align*}
be the number of observations with $X_i=x$, and $D_i=0$ or $D_i=1$, respectively. 

 To construct our sample bounds, we furthermore require the researcher to choose a ``bandwidth parameter''  $Q \in \{1,2,3,\ldots,\infty\}$.
If $\max_{ x \in {\cal X}_*} n( x)$ remains bounded as $n \rightarrow \infty$, then we can choose $Q=\infty$,
which simplifies many of the expressions in this section, and the reader may think of this case as the baseline case
which makes the connection to Section~\ref{sec:PopulationBounds} most obvious.
 
For each covariate value $ x \in {\cal X}_*$ we need to choose the order $q( x) \in \{1,2,3,\ldots\}$ of the bounds in 
Proposition~\ref{prop:HigherOrder} that we want to implement. To implement bounds of a certain order $q( x)$ we require at least that many observations for that covariate value, that is, we need to choose $q( x) \leq n( x)$. Choosing the maximal value $q( x) = n( x)$
is optimal from the perspective of expected width of the bounds, but it is not advisable in general since it
can lead to upper and lower bound estimates with very large variance.
In our implementation of the bounds we therefore choose 
\begin{align}
q( x) :=  \min\{Q,n( x)\},
    \label{choiceQX}
\end{align}
 that is, we choose the maximum order that satisfies
both $q( x) \leq n({ x})$ and $q( x) \leq Q$.  In practice, we recommend choosing $Q$ as small
as $Q=3$ or $Q=4$, but the choice $Q=\infty$ gives some theoretical optimality properties for the expected width of the bounds 
(but usually at the cost of higher variance).

Having chosen the order $q( x)$
for each $ x \in {\cal X}_*$,
 we then construct sample weights $\widehat w_0( x)$, $\widehat w_1( x)$, $\widehat v( x) $,
which are functions of the chosen order $q( x)$, the chosen reference propensity score $p_*( x)$, and the values
$n( x)$, $n_0( x)$, $n_1( x)$ obtained from the sample, such that
 \begin{align}
    \mathbb{E}\left[ \widehat w_0( x) \, \Big| \, X^{(n)} \right] 
    &=  w^{(q( x))} \big(1-p( x),1-p_*( x) \big)  ,
\nonumber \\
    \mathbb{E}\left[ \widehat w_1( x) \, \Big| \, X^{(n)}  \right] 
    &=  w^{(q( x))} \big(p( x),p_*( x) \big)  ,
\nonumber \\
    \mathbb{E}\left[  \widehat v( x)  \, \Big| \, X^{(n)}   \right] 
    &=    p( x) \, \widetilde w^{(q( x))} \big(p( x),p_*( x) \big)     ,
   \label{SampleWeightsExpectations} 
\end{align}
where   
$q( x)$ is given in \eqref{choiceQX}, and the weight functions $w^{(q)}$ and $\widetilde w^{(q)}$ on the right-hand side
were defined in \eqref{DefGeneralWeights}.
Here and in the following, the dependence of those sample weights on $p_*( x)$ and $q( x)$ (and thereby on $Q$) is not made explicit,
and the dependence on the sample $X^{(n)} $ and $D^{(n)}$ (through $n( x)$, $n_0( x)$, $n_1( x)$) is only indicated by the ``hat''. 
Explicit formulas for $\widehat w_0( x)$, $\widehat w_1( x)$, $\widehat v( x) $ are provided in the next subsection.

The natural  sample analogs of  the bounds $ B^{(q)}_{d,a} $ and $C^{(q)}_a$ in the last section are then given by
\begin{align}
     \widehat B_i(0,a)  &:=  a +   \widehat w_0( X_i) \;  \frac{ n( X_i)  \, (1-D_i) \, (Y_i-a) } {  \max\{1,n_0( X_i)\}    }  ,
   \nonumber  \\ 
     \widehat B_i(1,a)  &:=  a +    \widehat w_1( X_i) \; \frac{ n( X_i)  \, D_i \, (Y_i-a) } { \max\{1, n_1( X_i)\}    }  ,
    \nonumber \\ 
      \widehat C_i(a)    &:=   D_i \, (Y_i - a) 
 -     \widehat v( X_i) \,  \frac{  n( X_i)  \,  (1-D_i) \, (Y_i-a) } { \max\{1, n_0( X_i)\}    }      ,
    \label{DefGeneralBCsample}
\end{align}
for $a \in \mathbb{R}$.
 Notice that the arguments $d$, $a$ were subscripts in the population analysis, but for the sample version in this section we 
prefer to use the unit $i$ as the subscript instead. Also, the dependence on the order $q$ is not made explicit anymore here,
but we always have the choice \eqref{choiceQX} in mind.

In view of \eqref{SampleWeightsExpectations}, the expressions in \eqref{DefGeneralBCsample} are direct translations of 
the formulas in display \eqref{DefGeneralBC}, where the weights were replaced by sample weights, and
 the remaining occurrences of the unknown $1-p(x)$ and $p(x)$  were replaced 
by their sample analogs $n_0( x)/n( x)$ and $n_1( x)/n( x)$, respectively.
In all three expressions of display \eqref{DefGeneralBCsample} the maximum function in the denominator
is only included to avoid a potentially zero denominator. However,
$ n_0( X_i)=0$ implies $1-D_i=0$, and $n_1( X_i)=0$ implies $D_i=0$, that is, 
in all cases where the maximum function is required to avoid a zero denominator, the corresponding
numerator is zero anyways. In particular, we could replace $\max\{1, \ldots \}$ by 
$\max\{c, \ldots \}$ for any  constant $0 < c \leq 1$ without changing the sample bounds in \eqref{DefGeneralBCsample} at all.

The sample analogs of the  expectations over $ B^{(q)}_{d,a} $
and  $ C^{(q)}_a  $
are then given by 
\begin{align}
    \overline B(d,a) &:= \frac 1 n \sum_{i=1}^n  \widehat B_{i}(d,a) ,
    &
    \overline C(a) &:=   \frac 1 n \sum_{i=1}^n  \widehat C_{i}(a) ,
    \label{SampleBoundsBC}
\end{align}
and the final upper and lower sample bounds on the ATE read
\begin{align}
    \overline L^{\rm (ATE)} &:=    \overline B(1,a_{\min}) -  \overline B(0,a_{\max})  ,
 &
    \overline U^{\rm (ATE)} &:=     \overline B(1,a_{\max}) - \overline B(0,a_{\min}) .
    \label{SampleBoundsATE}
\end{align}
Similarly, for the ATT, the lower and upper sample bounds on $\frac 1 n \sum_{i=1}^n \pi(X_i)$
are given by $  \overline C(a_{\max}) $ and $  \overline C(a_{\min}) $, respectively.
To estimate the lower- and upper bounds on the ATT itself we still need to plug-in the sample analog of
  the denominator $\frac 1 n \sum_{i=1}^n  p(X_i)$,
which gives
\begin{align}
      \overline L^{\rm (ATT)} &:=     \frac{ \overline  C(a_{\max})    } {\frac 1 n \sum_{i=1}^n D_i}  ,
      &
      \overline U^{\rm (ATT)} &:=     \frac{ \overline  C(a_{\min})    } {\frac 1 n \sum_{i=1}^n D_i}  .
    \label{SampleBoundsATT}
\end{align}
In Section~\ref{sec:asympt} we show that the sample bounds just constructed are unbiased and consistent estimates
(as $m \rightarrow \infty$) of the corresponding population bounds from the last section, and we will also use those sample bounds to construct asymptotically
valid confidence intervals for ATE and ATT.

\subsection{Construction of the sample weights \texorpdfstring{$\widehat w_0(x)$, $\widehat w_1(x)$, $\widehat v(x) $}{hat w0(x), hat w1(x), hat v(x)}}
 
A key ingredient of the sample bounds just introduced are the sample weights that satisfy \eqref{SampleWeightsExpectations},
and which we want to define in this section. For ease of exposition we start with the simplest case $q(x)=n(x)$, which can be even or odd,
and then generalize the formulas to the case $q(x) =  \min\{Q,n(x)\}$ afterwards.
 
\subsubsection{Case  $q(x)=n(x)$ and $n(x)$ even}\label{subsec:qEVEN}

Let  $q(x)=n(x)$, and assume that $n(x)$ is even.
We consider $\widehat w_1(x)$ first.
By setting
 \begin{align}
\widehat w_1(x)  & =    1-      \prod_{i \in  {\cal N}(x)} \frac{p_*(x)-D_i}{p_*(x)}   
    \label{BasicSampleWeight}
\end{align} 
and using that, under Assumption~\ref{ass:MAIN}, we have
 $ \mathbb{E}\left(D_i \, \big| \, X^{(n)}  \right) = p(X_i)$,
we find that
\begin{align*}
   \mathbb{E}\left[ \widehat w_1(x) \, \Big| \, X^{(n)}  \right]
   &=    1-   \prod_{i \in  {\cal N}(x)}  \frac{p_*(x)-p(X_i)}{ p_*(x)}   
    =   1- \left( \frac{p_*(x)-p(x)}{ p_*(x)}   \right)^{q(x)}   
   \\  &=  w^{(q(x))} \big(p(x),p_*(x) \big) ,
\end{align*}   
where we used that the set $ {\cal N}(x)$ has $n(x) = q(x)$ elements, and the definition of the 
population weights in \eqref{DefGeneralWeights}. Thus, $\widehat w_1(x)$ satisfies the desired result in 
\eqref{SampleWeightsExpectations}. Finally, we can rewrite equation \eqref{BasicSampleWeight} as
 \begin{align}
\widehat w_1(x)  & :=    1-    \left(  \frac{p_*(x)-1}{p_*(x)}   \right)^{n_1(x)} ,
    \label{BasicSampleWeightRewritten}
\end{align} 
which from now on will serve as our definition of $\widehat w_1(x)$ in the current case. By analogous arguments one obtains,
for the current case of $q(x)=n(x)$ and $n(x)$ even, that
 \begin{align*}
\widehat w_0(x)  & :=    1-    \left(  \frac{p_*(x)}{p_*(x)-1}   \right)^{n_0(x)} ,
\\
  \widehat v(x)    & :=  \frac{n_1(x)}{n(x)}
           -       \left(     \frac{p_*(x) }{p_*(x)-1}  \right)^{n_0(x)} ,
\end{align*} 
and one can easily verify that those expressions 
satisfy \eqref{SampleWeightsExpectations}.

\subsubsection{Case \texorpdfstring{$q(x)=n(x)$ and $n(x)$ odd}{hat q x=n(x) and n(x) odd}}\label{subsec:qODD}

For $q(x)=n(x)$  odd  we have $ w^{(q(x))}(p,p_*)= 1- \left( 1- p \right) \left( \frac{p_* - p }{ p_*}   \right)^{q(x)-1} $
according to \eqref{DefGeneralWeights},
and we then need to change \eqref{BasicSampleWeight} to
 \begin{align}
\widehat w_1(x) & =    1-   \frac 1 {n(x)}  \sum_{i \in {\cal N}(x)}  (1-D_i)  \prod_{j \in  {\cal N}(x) \setminus \{i\}} \frac{p_*(x)-D_j}{p_*(x)}   .
   \label{BasicSampleWeight2}
\end{align} 
Under  Assumption~\ref{ass:MAIN},
it is again easy to see that the approximate unbiasedness condition for $\widehat w_1(x)$ in
\eqref{SampleWeightsExpectations} is satisfied here.
In equation \eqref{BasicSampleWeight2},
the sum over $i$  only gives a contribution for the $n(x) - n_{1}(x)$ instances where $D_i=0$,
in which case there still are $n_1(x)$ units $j \in  {\cal N}(x) \setminus \{i\}$ with $D_j=1$.
We can therefore rewrite this equation as
 \begin{align}
\widehat w_1(x)  & :=    1-    \frac {n(x) - n_{1}(x)}  {n(x)}   \left(  \frac{p_*(x)-1}{p_*(x)}   \right)^{n_1(x)} ,
    \label{BasicSampleWeightRewritten2}
\end{align} 
which from now on is our definition of $\widehat w_1(x)$ for the case $q(x)=n(x)$  odd.
By analogous arguments one obtains,
for the current case, that
 \begin{align*}
\widehat w_0(x)  & :=    1-    \frac {n(x) - n_{0}(x)}  {n(x)}  \left(  \frac{p_*(x)}{p_*(x)-1}   \right)^{n_0(x)} ,
\\
  \widehat v(x)    & :=  \frac{n_1(x)}{n(x)}
           -    \frac {n(x) - n_{0}(x)}  {n(x)}     \left(     \frac{p_*(x) }{p_*(x)-1}  \right)^{n_0(x)} ,
\end{align*} 
and one can again verify that those expressions 
satisfy \eqref{SampleWeightsExpectations}.

\subsubsection{General case \texorpdfstring{$q(x) =  \min\{Q,n(x)\}$}{hat q x =  min{Q,n(x)}}}

For $Q=\infty$ we have $q(x) = n(x)$, in which case all the required formulas for the sample weights are already 
provided in Subsections~\ref{subsec:qEVEN} and \ref{subsec:qODD} above. The generalization to finite $Q$ discussed
in the following is not conceptually difficult, but it requires some combinatorial arguments.
Remember that we choose the order $q(x)$ of the bounds according to \eqref{choiceQX}. 
For even order $q(x) =q$, we generalize the formula for $\widehat w_1(x)$ in  \eqref{BasicSampleWeight} as follows:
 \begin{align}
\widehat w_1(x) & =    1-   { n(x) \choose q }^{-1} \sum_{{\cal S}_q} \prod_{i \in {\cal S}_q} \frac{p_*(x)-D_i}{p_*(x)}  ,
    \label{DefAlternativeSampleWeights}
\end{align}
where the sum is over all subsets ${\cal S}_q \subset {\cal N}(x)$ with $q$ elements.
For odd  order $q(x) = q$, we generalize the formula for $\widehat w_1(x)$ in \eqref{BasicSampleWeight2} to
 \begin{align}
\widehat w_1(x) & =    1-   \frac 1 {n(x)}  \sum_{i \in {\cal N}(x)}  (1-D_i) \,  { n(x) - 1 \choose q-1 }^{-1} \sum_{{\cal S}_{q-1,i}} \; \prod_{j \in {\cal S}_{q-1,i}} \frac{p_*(x)-D_j}{p_*(x)}   .
    \label{DefAlternativeSampleWeights2}
\end{align}
where the sum is over all subsets ${\cal S}_{q-1,i} \subset  {\cal N}(x) \setminus \{i\}$ with $q-1$ elements. 

Under  Assumption~\ref{ass:MAIN},
it is again straightforward to verify that those formulas for $\widehat w_1(x)$ guarantee that
$
  \mathbb{E}\left[ \widehat w_1(x) \, \Big| \, X^{(n)}  \right] 
    =  w^{(q(x))} \big(p(x),p_*(x) \big)  $.    
If  $q(x) < n(x)$, then alternative choices for the sample weight $\widehat w_1(x) $ exist that have the same conditional expectation
-- for example, instead of averaging over ${\cal S}_q$ and ${\cal S}_{q-1,i}$,
 one could  randomly choose one subset of $q(x)$ observations out of the set ${\cal N}(x)$ and implement the formulas
in Subsections~\ref{subsec:qEVEN} and \ref{subsec:qODD} using only that subset of observations. 
To avoid that ambiguity in the definition of the sample weights we have chosen the formulas in  \eqref{DefAlternativeSampleWeights} and \eqref{DefAlternativeSampleWeights2} such that the binary treatment values $D_i$ of all units $i \in {\cal N}(x)$ enter exchangeably into 
$\widehat w_1(x)$, that is, the sample weights remain unchanged if we swap the data of any two observations in the same cluster ${\cal N}(x)$.
 This requirement also guarantees that it is possible to rewrite $\widehat w_1(x)$ such that the $D_i$ only enter through 
their summary statistics   $n_1(x)  =   \sum_{i \in {\cal N}(x)}   D_i $ and $n(x)$. Namely, one can rewrite
\eqref{DefAlternativeSampleWeights} and \eqref{DefAlternativeSampleWeights2} as
\begin{align}
      \widehat w_1(x)  &:= 1-   \sum_{k=0}^{2 \, \lfloor q(x) / 2 \rfloor}     \omega_{k,n_1(x),n(x),Q}   \left(     \frac{p_*(x) - 1}{p_*(x)}  \right)^{k}  ,
   \label{SampleWeightsDef} 
\end{align}
where $ \lfloor q(x) / 2 \rfloor$ is the integer part of $q(x) / 2$, and the
combinatorial coefficients $  \omega_{k,n_1(x),n(x),Q}    \in [0,1]$  are implicitly determined from \eqref{DefAlternativeSampleWeights} and \eqref{DefAlternativeSampleWeights2}, and one can show that
\begin{align}
   \omega_{k,n_{1},n(x),Q} & =  
    \left\{ 
    \begin{array}{ll}
 \displaystyle  
    \frac {n(x)-n_{1}}  {n(x)}     { n(x)-1 \choose q-1 }^{-1}    { n_{1} \choose k }  { n(x)-1-n_{1} \choose q-1-k }
     & \text{if  $q$  is odd,}
\\[10pt]
 \displaystyle   
 { n(x) \choose q }^{-1}    { n_{1} \choose k }  { n(x)-n_{1} \choose q-k }
         &  \text{if   $q$  is even,}
    \end{array} \right.
    \label{ResultsCombinatorialOmega}
\end{align}
where  $n_{1}=n_{1}(x)$ and $q=q(x)$  also depend on $x$.
 Appendix~\ref{sec:SampleWeights} provides a derivation of this formula for $  \omega_{k,n_1(x),n(x),Q}$.
Implementing $ \widehat w_1(x)$ via \eqref{SampleWeightsDef} and \eqref{ResultsCombinatorialOmega} is much faster than via 
\eqref{DefAlternativeSampleWeights} and \eqref{DefAlternativeSampleWeights2}, and can be done quickly also for relatively large values of $n(x)$
and $n_1(x)$.

Analogously we have
\begin{align}
      \widehat w_0(x)  &:= 1-   \sum_{k=0}^{2 \, \lfloor \min\{Q,n(x)\} / 2 \rfloor}   \omega_{k,n_0(x),n(x),Q}   \left(     \frac{p_*(x) }{p_*(x)-1}  \right)^{k}  ,
\nonumber \\
         \widehat v(x)    &:=  \frac{n_1(x)}{n(x)}
           -      \sum_{k=0}^{2 \, \lfloor \min\{Q,n(x)\} / 2 \rfloor}     \omega_{k,n_0(x),n(x),Q}    \left(     \frac{p_*(x) }{p_*(x)-1}  \right)^{k} ,
   \label{SampleWeightsDef2} 
\end{align}
where the combinatorial coefficients $\omega_{k,n_0(x),n(x),Q} \in [0,1]$ are again those in \eqref{ResultsCombinatorialOmega}, only
the argument $n_1(x)$ was changed to $n_0(x)$.
The equations in \eqref{SampleWeightsDef} and \eqref{SampleWeightsDef2} provide general definitions of the sample weights
that satisfy \eqref{SampleWeightsExpectations}.

\subsubsection{Discussion of the sample weights}

We want to briefly discuss some properties of the sample weights, again mostly focusing on   $\widehat w_1(x) $ for concreteness.
If we choose $Q=\infty$, then the formula for $\widehat w_1(x) $ is given in \eqref{BasicSampleWeightRewritten}
for even $n(x)$, and in   \eqref{BasicSampleWeightRewritten2} for odd $n(x)$. 
For $p_*(x)<\frac 1 2$ we have $\left|  \frac{p_*(x)-1}{p_*(x)} \right|>1$, implying that the absolute value of  $\widehat w_1(x) $ grows
exponentially with $n_1(x)$.
Analogously, for $Q=\infty$ and $p_*(x) > \frac 1 2$ the absolute values of the weights $\widehat w_0(x) $  and $\widehat v(x)$ grow exponentially with $n_0(x)$.
Only for $p_*(x) = 1/2$ are all the sample weights bounded, independent of the realization of $n_0(x)$ and $n_1(x)$.

Thus, for $Q=\infty$ the weights can take very large negative or positive values, potentially resulting in sample bounds for ATE and ATT with very large  variance. This is the main reason why we introduce the bandwidth parameter $Q$, which in practice we recommend to set relative small,
say $Q =3$ or  $Q =4$. Once we have chosen a finite value of $Q$, then our sample weights in \eqref{SampleWeightsDef}
and \eqref{SampleWeightsDef2} are all bounded, independent of the realization of $n_0(x)$ and $n_1(x)$
--- notice that the combinatorial coefficients $\omega_{k,n_{0/1}(x),n(x),Q}$ are all bounded between zero and one.

An interesting alternative way to guarantee that the weights  $\widehat w_{0}(x) $ and $\widehat w_{1}(x) $ both remain bounded is to 
choose $Q= \infty$, but $p_*(x)=1/2$ for all $x \in {\cal X}_*$. That is not our leading recommendation, because in many applications 
one might prefer values of $p_*(x)$ different from $1/2$ to obtain better bounds. If the parameter of interest is ATT, then
we can choose $Q= \infty$ and $\widehat v(x)$ will remain bounded as long as $p_*(x) \leq \frac 1 2$ for all $x \in {\cal X}_*$.
This could indeed be an interesting option in applications on ATT estimation. Nevertheless, the variance of the bounds will usually be smaller 
when a finite value of $Q$ is chosen. Furthermore, as illustrated in the following concrete examples for $\widehat w_{1}(x) $,
only for finite $Q$ do the sample weights converge to the population weights as $n(x) \rightarrow \infty$.

\begin{figure}[tb!] 
\begin{center}
\begin{tikzpicture}[scale=0.58]
\begin{axis}[
    axis lines = left,
    xlabel = $ n_1(x) / n(x)$,
    ylabel = {$  \widehat w_1(x)$},
    xmin=0, xmax=1.1,
    ymin=-11, ymax=9,
    legend pos=outer north east
]

\addplot [
    domain=0:1, 
    samples=100, 
    color=black,
    style=ultra thick
    ]
{1-((x-0.4)/0.4)^6};

\addplot [ 
                only marks,
                color=blue,
                mark=*,
                mark options={solid},
                mark size=3pt
                ]
                coordinates{
(0, 0.)(1/6, 2.5)(1/3, -1.25)(1/2, 4.375)(2/3, -4.0625)(5/
  6, 8.59375)(1, -10.3906)
                };

\addplot [ 
                only marks,
                color=red,
                mark=*,
                mark options={solid},
                mark size=2pt
                ]
                coordinates{
(0, 0.)(1/12, 1.25)(1/6, 1.07955)(1/4, 0.909091)(1/3, 
  0.975379)(5/12, 1.07126)(1/2, 1.02129)(7/12, 0.893111)(2/3, 
  0.944602)(3/4, 1.30682)(5/6, 1.4027)(11/
  12, -0.898438)(1, -10.3906)
                };

\addplot[dashed]
  coordinates{(0.4,-11) (0.4,10)};

\end{axis}

\end{tikzpicture}
\begin{tikzpicture}[scale=0.58]
\begin{axis}[
    axis lines = left,
    xlabel = $ n_1(x) / n(x)$,
    ylabel = {$  \widehat w_1(x)$},
    xmin=0, xmax=1.1,
    ymin=0, ymax=2.1,
    legend pos=outer north east
]

\addplot [
    domain=0:1, 
    samples=100, 
    color=black,
    style=ultra thick
    ]
{1-((x-0.5)/0.5)^6};

\addplot [ 
                only marks,
                color=blue,
                mark=*,
                mark options={solid},
                mark size=3pt
                ]
                coordinates{
                 (0, 0.) (1/6, 2.) (1/3, 0.) (1/2, 2.) (2/3, 0.) (5/6, 2.) (1, 0.)
                };

\addplot [ 
                only marks,
                color=red,
                mark=*,
                mark options={solid},
                mark size=2pt
                ]
                coordinates{
                (0, 0.)(1/12, 1.)(1/6, 1.09091)(1/4, 1.)(1/3, 0.969697)(5/
  12, 1.)(1/2, 1.02165)(7/12, 1.)(2/3, 0.969697)(3/4, 1.)(5/
  6, 1.09091)(11/12, 1.)(1, 0.)
                };

\addplot[dashed]
  coordinates{(0.5,0) (0.5,2)};

\end{axis}

\end{tikzpicture}
\begin{tikzpicture}[scale=0.58]
\begin{axis}[
    axis lines = left,
    xlabel = $ n_1(x) / n(x)$,
    ylabel = {$  \widehat w_1(x)$},
    xmin=0, xmax=1.1,
    ymin=0, ymax=2.1,
    legend style={at={(0.6,1)},anchor=west}
]

\addplot [
    domain=0:1, 
    samples=100, 
    color=black,
    style=ultra thick
    ]
{1-((x-0.6)/0.6)^6};
\addlegendentry{$w^{(q)}(p,p_*)$, $q=6$}

\addplot [ 
                only marks,
                color=blue,
                mark=*,
                mark options={solid},
                mark size=3pt
                ]
                coordinates{
                 (0, 0.)(1/6, 1.66667)(1/3, 0.555556)(1/2, 1.2963)(2/3, 
  0.802469)(5/6, 1.13169)(1, 0.912209)
                };
\addlegendentry{$Q=6$, $n(x)=6$}

\addplot [ 
                only marks,
                color=red,
                mark=*,
                mark options={solid},
                mark size=2pt
                ]
                coordinates{
(0, 0.)(1/12, 0.833333)(1/6, 1.03535)(1/4, 1.02694)(1/3, 
  0.995137)(5/12, 0.990616)(1/2, 1.00187)(7/12, 1.00626)(2/3, 
  0.997838)(3/4, 0.992019)(5/6, 1.00698)(11/12, 1.02195)(1, 
  0.912209)
                };
\addlegendentry{$Q=6$, $n(x)=12$}

\addplot[dashed]
  coordinates{(0.6,0) (0.6,2)};

\end{axis}

\end{tikzpicture}

\caption{\label{fig:SampleWeights}Sample Weights $  \widehat w_1(x)$
plotted as a function of $n_1(x) / n(x)$ for $p_*(x)=0.4$ (left), $p_*(x)=0.5$ (middle) and $p_*(x)=0.6$ (right).
The corresponding population weights $w^{(q)}(p(x),p_*(x))$ are also plotted as a function of $p(x)$.
}
\end{center}
\end{figure}

Figure~\ref{fig:SampleWeights} plots the weights $  \widehat w_1(x)$ for $Q=6$,
$n(x) \in \{6,12\}$, and for three different values for the reference propensity score $p_*(x)$.
The plot shows that as $n(x)$ becomes large the weights  $  \widehat w_1(x)$ as a function of $\widehat p(x) = n_1(x)/n(x)$ converge
to the population weights $w^{(q)}(p(x),p_*(x))$ as a function of $p(x)$. This, in particular, implies that $  \widehat w_1(x)$ becomes a
smooth function of $n_1(x)$ for large values of $n(x)$. However, for small $n(x)=Q=6$ the weights $  \widehat w_1(x)$ heavily fluctuate
as a function of $n_1(x)$. Furthermore, for $p_*(x)<0.5$ the weights $  \widehat w_1(x)$  can take on very small and very large values (notice the
different scale of the plot  for $p_*(x) = 0.4$), but for $p_*(x)\geq 0.5$ the weights remain within the bounded interval $[0,2]$.

\subsection{Asymptotically valid confidence intervals}
\label{sec:asympt}

Remember that  $m= | {\cal X}_* |$ is the number of different covariate values in our sample.
Our treatment effect bounds are then based on weight functions that combine the observed treatment status 
$D_i$ for observations $i \in {\cal N}(x)$ of the same covariate value  $x \in {\cal X}_*$ in a non-linear way.
 However, if we condition on realization of the covariates $X^{(n)}$, then
across different covariate values the bounds are just averages of independent observations.
Given that the bounds have this structure, it is useful to think of $m$ as our effective sample size,
and of each  $x \in  {\cal X}_*$ as labelling one effective observation. It is therefore convenient to rewrite the sample bounds in \eqref{SampleBoundsBC}
not as cross-sectional averages over $i \in \{1,\ldots,n\}$, but as sample averages over $x \in  {\cal X}_*$. 
For that purpose, for $d \in \{0,1\}$ and $a \in \mathbb{R}$, we define\footnote{%
We are slightly abusing notation here, for example, $\widehat B_{x}(d,a)$ for $x=1$ (assuming $1 \in {\cal X}_*$) is not the same 
as $ \widehat B_i(d,a)$ for $i=1$. However, it will always be clear from the subscript letter which object is meant.
}
\begin{align}
      \widehat B_{x}(d,a) &:= \frac 1 {n(x)} \sum_{i \in {\cal N}(x)}  \widehat B_i(d,a) ,
      &
        \widehat C_{x}(a) &:= \frac 1 {n(x)} \sum_{i \in {\cal N}(x)}   \widehat C_i(a) ,
    \label{AggregateCluster}    
\end{align}
which allows us to rewrite the sample bounds in \eqref{SampleBoundsBC} as 
\begin{align*}
    \overline B(d,a) &:= \frac 1 n \sum_{x \in  {\cal X}_*}  n(x) \, \widehat B_{x}(d,a) ,
    &
    \overline C(a) &:=   \frac 1 n \sum_{x \in  {\cal X}_*} n(x) \,  \widehat C_{x}(a) .
\end{align*}
Using the definitions of $\widehat B_i(d,a) $ and $\widehat C_i(a) $ in \eqref{DefGeneralBCsample} we furthermore have
\begin{align}
     \widehat B_{x}(d,a)  &=  a +     \widehat w_d(x) \, \left[ \, \overline Y_{x}(d)-a\right]   ,
  \nonumber  \\
      \widehat C_{x}(a)    &=   \frac{n_1(x)}{n(x)} \, \left[ \overline Y_{x}(1) - a \right] 
 -      \widehat v(x) \, \left[ \overline Y_{x}(0)-a \right]       ,
 \label{SampleBoundsFiniteSup}
\end{align}
where
\begin{align*}
    \overline Y_{x}(d) := \left\{ 
    \begin{array}{ll}
     \displaystyle   \frac 1 {n_d(x)}  \sum_{i \in {\cal N}(x)}   \mathbbm{1}\{D_i=d\}  \, Y_i  
       &
       \text{if }  n_d(x)  >0 ,
       \\[15pt]
      \mathbb{E}\left[ Y_i(d)  \, \big| \, X_i=x \right] 
       &\text{if }  n_d(x)  =0.
     \end{array}  \right.
\end{align*}
Notice that for $n_d(x)=0$ we have $ \widehat w_d(x) =0$,
and for $n_0(x)=0$ we have $  \widehat v(x) = 0$.
Therefore, $\overline Y_{x}(d)$ only enters into the bounds in \eqref{SampleBoundsFiniteSup} when $ n_d(x)  >0$.
In that case,  $ \overline Y_{x}(d)$ is simply the average of the $n_d(x)$ observed outcomes $Y_i$ for which $X_i=x$ and $D_i=d$.
However,
for our theoretical discussion it is useful to also define  $ \overline Y_{x}(d)$ for the case $n_d(x)  =0$,
because with that definition we have that, under Assumption~\ref{ass:MAIN}, 
\begin{align}
    \mathbb{E}\left[  \overline Y_{x}(d)  \, \big| \, D^{(n)}, \, X^{(n)} \right]  =   \mathbb{E}\left[ Y_i(d)  \, \big| \,  X_i=x \right]
    \label{meanIndepY}
\end{align}    
Equation \eqref{meanIndepY} states that 
  $  \overline Y_{x}(d)$ is mean-independent of $D^{(n)}$ and $X^{(n)} $.
  The properties of $\widehat w_{0/1}(x)$ and $\widehat v(x) $
in display \eqref{SampleWeightsExpectations} 
together with \eqref{meanIndepY}
 guarantee that the expected values of $  \widehat B_{x}(d,a)$
and $  \widehat C_{x}(a)  $ are equal to the  expectations of 
the population bounds $B^{(q)}_{0,a} $  and $C^{(q)}_a $ in Section~\ref{sec:PopulationBounds}.

Next, we want to show consistency of those sample bounds and use them to construct confidence intervals.
For that purpose, it is convenient to define
\begin{align}
    \theta^{(0)} &:= \frac 1 n \sum_{i=1}^n \mathbb{E} \left[ Y_i(0) \, \big| \, X_i \right] ,
      &
    \theta^{(1)} &:=  \frac 1 n \sum_{i=1}^n \mathbb{E} \left[ Y_i(1) \, \big| \, X_i \right]   ,
    \nonumber  \\
    \theta^{\rm (ATE)} &:=   {\rm ATE}   ,
    &
    \theta^{\rm (ATT)} &:=  {\rm ATT}  ,
    \label{DefThetaParameter}
\end{align}
which are the four parameters of interest that we focus on in this paper after conditioning on the realization of all the covariates 
$X^{(n)}= (X_1,\ldots,X_n)$. 
For each of those parameters  we have already introduced upper and lower bound estimates
in     \eqref{SampleBoundsBC}, \eqref{SampleBoundsATE}, \eqref{SampleBoundsATT}. For $ \theta^{(0)}$
and $\theta^{(1)}$ we now denote those bounds by
\begin{align*}
    \overline L^{(d)} &:= \overline B(d,a_{\min})  , 
    &
    \overline U^{(d)} &:= \overline B(d,a_{\max})  , 
    &
    &\text{where $d \in \{0,1\}$} .
\end{align*}
Using the above definitions we have,
  for $r \in \{0,1,{\rm ATE}\}$,
\begin{align*}
   \overline L^{(r)} &=    \frac 1 m  \sum_{x \in {\cal X}_*}  L^{(r)}_{x} ,
&
   \overline U^{(r)} &=    \frac 1 m  \sum_{x \in {\cal X}_*}  U^{(r)}_{x} ,
\end{align*}
where
\begin{align}
    L^{(d)}_{x} &:=  \frac{m \, n(x)} n \,    \widehat  B_{x}(d,a_{\min}) ,
    &
      L^{\rm (ATE)}_{x} &:= \frac{m \, n(x)} n   \left[  \widehat  B_{x}(1,a_{\min}) - \widehat B_{x}(0,a_{\max}) \right]  ,
  \nonumber \\
    U^{(d)}_{x} &:=   \frac{m \, n(x)} n \,    \widehat  B_{x}(d,a_{\max}),
    &
   U^{\rm (ATE)}_{x} &:=  \frac{m \, n(x)} n  \left[    \widehat  B_{x}(1,a_{\max}) -  \widehat B_{x}(0,a_{\min}) \right] ,
   \label{DefLUATE}
\end{align}
for $d \in \{0,1\}$. 
Our results on the ``population bounds'' in the last section together with \eqref{SampleWeightsExpectations}, \eqref{SampleBoundsFiniteSup}
and \eqref{meanIndepY} guarantee that    
\begin{align}
    \underbrace{  \mathbb{E} \left[ \overline L^{(r)} \, \Big| \, X^{(n)} \right] }_{\displaystyle =: \theta^{(r)}_L}  \, &\leq \, \; \; \theta^{(r)}  \; \;  \, \leq  \,      \underbrace{  \mathbb{E} \left[ \overline U^{(r)} \, \Big| \, X^{(n)} \right] }_{\displaystyle =: \theta^{(r)}_U}  ,
     \qquad  \qquad \text{for $r \in \{0,1,{\rm ATE}\}$},
\nonumber \\[20pt]
      \underbrace{   \frac{  \mathbb{E} \left[  \overline  C(a_{\max}) \, \Big| \, X^{(n)} \right]  } { \frac 1 n \sum_{i=1}^n  p(X_i)}  }_{\displaystyle =: \theta^{(\rm ATT)}_L}
       \, &\leq \,  \theta^{(\rm ATT)}   \, \leq  \,  
    \underbrace{    \frac{ \mathbb{E} \left[ \overline  C(a_{\min})  \, \Big| \, X^{(n)} \right] } { \frac 1 n \sum_{i=1}^n  p(X_i)}  }_{\displaystyle =: \theta^{(\rm ATT)}_U}.
     \label{ValidBoundsConditionalX} 
\end{align}
When comparing the last line with the definition of the actual sample bounds $ \overline L^{\rm (ATT)} $ and $\overline U^{\rm (ATT)}$ in \eqref{SampleBoundsATT}
we notice that we need to account for the 
randomness of the denominator term $\frac 1 n \sum_{i=1}^n D_i$ as well when constructing confidence intervals, and we therefore write those bounds as
(see appendix \ref{app:Asymptotic} for details)
\begin{align}
   \overline L^{\rm (ATT)} &=     \frac{  \mathbb{E} \left[ \, \overline  C(a_{\max}) \, \big| \, X^{(n)} \right]  } { \frac 1 n \sum_{i=1}^n  p(X_i)}   + 
    \frac 1 m  \sum_{x \in {\cal X}_*}  L^{\rm (ATT)}_{x}  + o_P(m^{-1/2}) ,
 \nonumber \\
   \overline U^{\rm (ATT)} &=      \frac{  \mathbb{E} \left[ \, \overline  C(a_{\min}) \, \big| \, X^{(n)} \right]  } { \frac 1 n \sum_{i=1}^n  p(X_i)}  
    +  \frac 1 m  \sum_{x \in {\cal X}_*}  U^{\rm (ATT)}_{x}    + o_P(m^{-1/2})  ,
     \label{ATTinfluenceFunction1}
\end{align}
where
\begin{align}
   L^{\rm (ATT)}_{x}  &:=    \frac{ m \, n(x) \,  \widehat C_{x}(a_{\max})    } {  \sum_{i=1}^n D_i}  
         - \frac{ m \, n \, n_1(x)  \, \overline  C(a_{\max})    } {\left( \sum_{i=1}^n D_i\right)^2}     ,
 \nonumber \\
   U^{\rm (ATT)}_{x}  &:=    \frac{ m \, n(x) \,  \widehat C_{x}(a_{\min})    } {  \sum_{i=1}^n D_i}  
         - \frac{ m \, n \, n_1(x)  \, \overline  C(a_{\min})    } {\left( \sum_{i=1}^n D_i\right)^2}     .
     \label{ATTinfluenceFunction2}
\end{align}
Under  Assumption~\ref{ass:MAIN}(iii) we have that  $(D_i,Y_i(0),Y_i(1))$ is independent across $i$, conditional on $X^{(n)}= (X_1,\ldots,X_n)$.
This, in particular, guarantees that $ L^{(r)}_{x} $ and $U^{(r)}_{x}$, for $r \in \{0,1,{\rm ATE},\allowbreak {\rm ATT}\}$, are independent across $x \in {\cal X}_*$,
conditional $X^{(n)}$. This independence is crucially used for the asymptotic convergence results stated in the following theorem. 
For that reason, all the stochastic statements in the theorem are conditional on $X^{(n)}$.  
 Notice also that we have in mind a triangular array in our {asymptotic theory},\todo[size=tiny]{change from `asymptotic' to `asymptotic theory'} 
 where the support of the regressors may change as the sample size increases.

\begin{theorem}
     \label{th:Asympt}
     Let $r \in \{0,1,{\rm ATE},{\rm ATT}\}$.
      Let  Assumption~\ref{ass:MAIN} hold, and assume that as $m \rightarrow \infty$ we have that $Q$ is fixed,
      $p_*(x)$ is bounded away from zero and one, uniformly over $x \in {\cal X}_*$.
 Also assume 
      $ \frac 1 {m}   \sum_{x \in {\cal X}_*} \left( \frac{m \, n(x)} n \right)^4  = O_P(1)$,\footnote{
      Here, we allow for $X^{(n)}$ to be random, but when
      conditioning on the realization of covariates, we could also simply write $O(1)$ here. 
      }
    and  
    $ \left[\frac 1 m  \sum_{x \in {\cal X}_*} {\rm Var}  \left(   M^{(r)}_{x}  \, \Big| \, X^{(n)} \right) \right]^{-1} \, 
      = o_P( m^{1/3} )$,
     for ${M_{x}^{(r)}} \in \{   L^{(r)}_{x},    U^{(r)}_{x} \}  $.\todo[size=tiny]{Superscript $(r)$ is added to $M_x$.}
      Then, conditional on the realization of all the covariates $X^{(n)}$,
 the sample bounds are asymptotically normally distributed:
\begin{align*}
   \frac{  \overline L^{(r)}  -  \theta^{(r)}_L   } {\left\{ {\rm Var}  \left[ \displaystyle \frac 1 m  \sum_{x \in {\cal X}_*}  L^{(r)}_{x}\, \Bigg| \, X^{(n)} \right]  \right\}^{1/2}}   
     \, &\Rightarrow \, {\cal N}\left(0,1 \right) ,
  &
  \frac{  \overline U^{(r)}  -  \theta^{(r)}_U   } {\left\{ {\rm Var}  \left[ \displaystyle  \frac 1 m  \sum_{x \in {\cal X}_*}  U^{(r)}_{x} \, \Bigg| \, X^{(n)} \right]  \right\}^{1/2}}  
     \, &\Rightarrow \, {\cal N}\left(0,1 \right) .
\end{align*}
Furthermore, for ${M_{x}^{(r)}} \in \{   L^{(r)}_{x},    U^{(r)}_{x} \}  $ we have
\begin{align*}
{\rm Var}  \left[ \displaystyle \frac 1 m  \sum_{x \in {\cal X}_*}  M^{(r)}_{x}\, \Bigg| \, X^{(n)} \right] 
   &\leq 
   \frac { {\rm SVar}  \left(   M^{(r)}_{x}  \right)} m  \left[ 1 + o_P(1) \right] ,
\end{align*} 
where 
\begin{align*}
  {\rm SVar} \left( M^{(r)}_{x}   \right)
   &:=
    \frac 1 {m}  \sum_{x \in {\cal X}_*}   \left( M^{(r)}_{x} \right)^2 -  \left( \frac 1 m  \sum_{x \in {\cal X}_*}   M_{x}^{(r)} \right)^2 .
\end{align*}

\end{theorem}

Here, the assumptions that $Q$ is fixed and that $p_*(x)$ is bounded away from zero and one guarantee that
our sample weights $\widehat w_d(x) $ and $\widehat v(x)$, and therefore also $\widehat B_{x}(d,a)$ and $ \widehat C_{x}(a)$
defined in \eqref{SampleBoundsFiniteSup}, are uniformly bounded.
However, the  averages  $\frac 1 m  \sum_{x \in {\cal X}_*}  L^{(r)}_{x}$
and  $\frac 1 m  \sum_{x \in {\cal X}_*}  U^{(r)}_{x}$ that give our bounds are over the $ L^{(d)}_{x} $
and $U^{(d)}_{x}$  defined in \eqref{DefLUATE}, and those feature the additional factors $\frac{m \, n(x)} n \in [0,\infty)$.
Thus, covariate values that appear often in the sample  get more weight than covariate values that appear less often. Notice that $\frac{n} m = \frac 1 m  \sum_{x \in {\cal X}_*} n(x)$ is the average number of observations for a given covariate value,
that is, the factor $\frac{m \, n(x)} n  $ simply rescales the $n(x)$ such that they average to one:
$ \frac 1 m  \sum_{x \in {\cal X}_*} \frac{m \, n(x)} n =1$. The assumption   $ \frac 1 {m}   \sum_{x \in {\cal X}_*} \left( \frac{m \, n(x)} n \right)^4  = O_P(1)$
requires that the fourth moment of $\frac{m \, n(x)} n$ remains bounded
asymptotically, that is, it demands that the $n(x)$ are not distributed too heterogeneously across covariates.
For example, if $X_i$ is uniformly distributed over ${\cal X}_*$, then each $n(x)$ has a Binomial distribution with parameters $n$
and $1/m$, and it is easy to verify that the assumption is satisfied. More generally, the assumption holds as long as the probabilities $P(X_i =x)$ 
are not too heterogeneous across $x$.

Notice that for ${M_{x}^{(r)}} \in \{   L^{(r)}_{x},    U^{(r)}_{x} \}  $ 
we have $$ {\rm Var}  \left(   \frac 1 {\sqrt{m}}  \sum_{x \in {\cal X}_*}  M^{(r)}_{x}   \, \Big| \, X^{(n)} \right) 
= \frac 1 m  \sum_{x \in {\cal X}_*} {\rm Var}  \left(   M^{(r)}_{x}  \, \Big| \, X^{(n)} \right), $$
that is, our assumption 
$ \left[\frac 1 m  \sum_{x \in {\cal X}_*} {\rm Var}  \left(   M^{(r)}_{x}  \, \Big| \, X^{(n)} \right) \right]^{-1} \, 
      = o_P( m^{1/3} )$
 simply demands that the variance of       $\frac 1 {\sqrt{m}}  \sum_{x \in {\cal X}_*}  M^{(r)}_{x} $ is not too small.
Here, $ \frac 1 {\sqrt{m}}$ is a natural rescaling, because the $M^{(r)}_{x}$ have zero mean and are independent 
across $x$, conditional on $X^{(n)} $. However, the  $M^{(r)}_{x} $ may contribute heterogeneously to the variance because
of the factors  $\frac{m \, n(x)} n$ in their definition, and also because of the weights $\widehat w_d(x) $ and $\widehat v(x)$. The assumption therefore allows 
for the possibility that $ \frac 1 m  \sum_{x \in {\cal X}_*} {\rm Var}  \left(   M^{(r)}_{x}  \, \Big| \, X^{(n)} \right)$ converges to zero
as $n,m \rightarrow \infty$, but not too fast.
 
Using \eqref{ValidBoundsConditionalX} and  Theorem~\ref{th:Asympt}
we obtain the following asymptotically valid confidence interval for  $\theta^{(r)} $  of confidence level $(1-\alpha) \in (0,1)$,\footnote{%
Here, we use the convention $[a,b]=\emptyset$ if $a>b$.
}
\begin{align}
          {\rm CI}^{(r)}_{\rm basic}
          := \left[  \overline L^{(r)} -  \frac{  \widehat \sigma^{(r)}_L  } {\sqrt{m}} \; \Phi^{-1}\left(1-\frac \alpha 2\right)  ,
          \;  \overline U^{(r)} +  \frac{  \widehat \sigma^{(r)}_U  } {\sqrt{m}} \; \Phi^{-1}\left(1-\frac \alpha 2\right) 
          \right] ,
          \label{CIbasic} 
\end{align}
 where $\widehat \sigma^{(r)}_L := \sqrt{  {\rm SVar}  \left(  L^{(r)}_{x}   \right) }$,
$\widehat \sigma^{(r)}_U := \sqrt{  {\rm SVar}  \left(  U^{(r)}_{x}   \right) }$.
The following corollary states that $ {\rm CI}^{(r)}_{\rm basic}$
contains $\theta^{(r)} $ with probability at least $1-\alpha$  in large samples.

\begin{corollary}
\label{cor:CoverageBasic}
Let $\alpha \in (0,1)$.
Under the assumptions of Theorem~\ref{th:Asympt} we have
\begin{align*}
         \lim_{n \rightarrow \infty}
        {\rm Pr}\left(  \theta^{(r)}
        \in   {\rm CI}^{(r)}_{\rm basic} \, \Big| \, X^{(n)} \right) \,  \geq \, 1-\alpha ,
\end{align*}
for $r \in \{0,1,{\rm ATE},{\rm ATT}\}$. 
\end{corollary}

Thus, 
those confidence intervals $ {\rm CI}^{(r)}_{\rm basic}$ are asymptotically valid, but they may be conservative for three reasons: (i) the true  $\theta^{(r)}$ may be an interior point of the expected
bounds, implying 100\% coverage in large samples;
(ii) we are using an upper bound estimate for the variance of the upper and lower bounds when constructing the confidence interval,
and (iii) we are using Bonferroni inequalities when dividing the statistical problem into one-sided confidence interval constructions for the
upper and lower bounds --- notice the $\alpha/2$ in both the upper and lower bounds in \eqref{CIbasic}.\footnote{%
One could improve on those $\alpha/2$ critical values by adapting the methods in \cite{imbens2004confidence}
and \cite{stoye2009more} to our case. However, we want to keep the confidence interval construction simple here, and there is also
the more important issue that $ {\rm CI}^{(r)}_{\rm basic}$ can be empty in our case, which we address using \cite{stoye2020}.
}

Here, the issues (i) and (iii) are very typical for bound estimation, and (ii) is impossible to fully overcome in our setting,
unless $n_d(x)$ are sufficiently large for all $d$ and $x$. For example, if $n_d(x)=1$, then only a single outcome $Y_i$ is observed for which
we have $D_i=d$
and $X_i=x$, implying that unbiased estimation of the variance of that outcome is impossible, but since $Y_i$ enters into 
$ \overline L^{(r)}   $ and $   \overline U^{(r)} $ we can in general not expect to 
estimate the variances of these bounds consistently.\footnote{%
Another problem is that the true propensity scores $p(x)$ are unknown, rendering the distribution
of the sample weights  $ \widehat w_d(x)$ also unknown.
}

We therefore believe that one needs to be content with conservative confidence intervals in our setting, and that
our construction so far has the advantage of being relatively simple and robust. However,
a potentially more severe problem in practice is that the confidence interval $ {\rm CI}_{\rm basic}$ may be empty, that is, the lower
bound may be larger than the upper bound, because  nothing in our construction guarantees that $\overline L^{(r)} $
cannot be larger than $ \overline U^{(r)}  $ in finite samples. While our theory guarantees that this problem cannot occur asymptotically,
it is still   undesirable to have a potentially empty confidence interval in  applications.

We therefore use the method in   \cite{stoye2020} to obtain a valid confidence interval that is never empty.
The general version of that method requires knowing the correlation $\rho$ between $ \overline L^{(r)}   $ and $   \overline U^{(r)} $,
which we cannot estimate consistently in our setting 
(for the same reasons for which we can only obtain upper bounds on the variances of $ \overline L^{(r)}   $ and $   \overline U^{(r)} $).
We therefore apply  \cite{stoye2020}'s method with $\rho=1$, which corresponds to the worst case:
Let
\begin{align*}
     \widehat \theta^{(r)}_* &:= \frac{\widehat \sigma^{(r)}_U \,  \overline L^{(r)}  + \widehat \sigma^{(r)}_L \,  \overline U^{(r)} } {\widehat \sigma^{(r)}_L + \widehat \sigma^{(r)}_U} ,
     &
     \widehat \sigma^{(r)}_* &:= \frac{2 \, \widehat \sigma^{(r)}_L \, \widehat \sigma^{(r)}_U} {\widehat \sigma^{(r)}_L + \widehat \sigma^{(r)}_U} ,
\end{align*}
and
\begin{align*}
          {\rm CI}^{(r)}_*
          := \left[ \widehat \theta^{(r)}_* -  \frac{  \widehat \sigma^{(r)}_* } {\sqrt{m}} \; \Phi^{-1}\left(1-\frac \alpha 2\right)  ,
          \;  \widehat \theta^{(r)}_* +  \frac{  \widehat \sigma^{(r)}_* } {\sqrt{m}} \; \Phi^{-1}\left(1-\frac \alpha 2\right) 
          \right] ,
\end{align*}
and define the final confidence interval to be reported for $\theta$ as the union of ${\rm CI}_{\rm basic}$ and $ {\rm CI}_*$, that is,
$$
    {\rm CI}^{(r)}_{\theta} := {\rm CI}^{(r)}_{\rm basic} \, \cup \, {\rm CI}^{(r)}_* .
$$
Then, by construction, ${\rm CI}_{\theta}$ is never empty, because ${\rm CI}_*$ is never empty,
and Corollary~\ref{cor:CoverageBasic} implies that
\begin{align*}
         \lim_{n \rightarrow \infty}
        {\rm Pr}\left(  \theta^{(r)}
        \in    {\rm CI}^{(r)}_{\theta} \right) \,  \geq \, 1-\alpha .
\end{align*}
We refer to \cite{stoye2020} for a further justification of this specific confidence interval construction. 
We have thus shown how to construct valid non-empty confidence intervals for all of those objects of interest.

Notice also that
   for the constructions of confidence intervals here we have assumed that $p_*(x)$ is non-random. 
    If $p_*(x)$ is estimated, then the randomness of $p_*(x)$ should be accounted for when constructing
    those confidence intervals, either via an application of the delta method, or via a bootstrap procedure. 

\section{Clustering the covariate observations}
\label{sec:Cluster}

Unconfoundedness only places restrictions on the observed data if there are at least some repeated covariate values. When each covariate vector is unique, we essentially revert to first-order (Manski) bounds, which remain valid without additional assumptions but do not leverage unconfoundedness to tighten those bounds.

A common way to exploit unconfoundedness when covariates are nearly unique is to coarsen them by binning. For example, one might group ages into years rather than days. This process discards some information but remains transparent. Alternatively, one can adopt automated methods such as clustering or nearest-neighbor matching. For clustering, we partition units into groups of similar \(X_i\) values and label each unit by its cluster identity \(\overline{X}_i\). The main steps of our approach remain unchanged, except that we substitute \(\overline{X}_i\) for \(X_i\). 

Although clustering is straightforward in practice, it introduces dependence among the labeled \(\overline{X}_i\) because the clustering procedure relies on the entire sample. A fully rigorous treatment would require additional smoothness assumptions or a formal model for the underlying clustering structure. We leave these issues for future work, noting that established methods (e.g., sample-splitting or matching) can mitigate some of the complications. In summary, binning or clustering offers a practical way to address rare or unique covariates when applying our bounds, but more theoretical investigation is warranted.

The specific clustering procedure we employ in our simulations and empirical application proceeds as follows.
First, we studentize each observed covariate, then use the Euclidean distance \(\| X_i - X_j \|\) to measure the closeness of observations \(i\) and \(j\). 
With this distance measure, we apply hierarchical, agglomerative clustering with complete linkage 
to the observed covariate sample \(\bigl(X_1,\ldots,X_n\bigr)\).
We refer to, e.g., \citet{kaufman2009finding} and \citet{everitt2011cluster} for an introduction to hierarchical clustering methods, and to \citet{mullner2013fastcluster} and \citet{R:cluster} for software implementations.
Hierarchical, agglomerative clustering begins with singleton clusters and iteratively merges pairs of clusters until all observations lie in a single cluster. A user-selected number of clusters, \(m\), can be obtained by ``cutting'' the resulting tree. Different forms of hierarchical clustering differ in how they measure inter-cluster distance. Complete linkage uses the maximum distance between any two points, one in each cluster, which tends to produce relatively compact clusters \citep[][Chapter~4]{everitt2011cluster}.

The only tuning parameter in this clustering procedure is the number of clusters \(m\in \{1,2,\dots,n\}\). This plays the same role as the number of unique covariate values in our earlier analysis. In practice, we recommend choosing
\begin{align}
\label{choice-cluster}
m \;=\;\bigl\lceil \tfrac{n}{L} \bigr\rceil,
\end{align}
where \(L\) is a constant (e.g., \(L=10\)) and \(\lceil \cdot \rceil\) denotes the ceiling function. This ad hoc rule aims for around \(L\) observations per cluster on average, and letting \(L\) remain fixed ensures \(m\to \infty\) as \(n\to \infty\), consistent with the large-\(m\) asymptotic theory in Section~\ref{sec:asympt}.

Given this partition \(\{1,\dots,n\} = \mathcal{N}_1 \cup \cdots \cup \mathcal{N}_m\), we label each cluster by its average covariate value. Concretely, for each \(g\in \{1,\dots,m\}\) and \(i\in \mathcal{N}_g\),
\[
\overline{X}_i \;:=\; \frac{1}{\lvert \mathcal{N}_g\rvert} \sum_{j \in \mathcal{N}_g} X_j,
\]
and let \(\overline{\mathcal{X}}=\{\overline{X}_i : i=1,\dots,n\}\) be the set of these cluster averages. By construction, \(\lvert \overline{\mathcal{X}} \rvert=m\) and each \(\overline{X}_i\) uniquely identifies the cluster that observation \(i\) belongs to. For \(\overline{x}\in \overline{\mathcal{X}}\), define the corresponding cluster as
\[
\mathcal{N}(\overline{x}) \;:=\; \bigl\{ i \in \{1,\dots,n\}\,\big\vert\, \overline{X}_i = \overline{x} \bigr\},
\]
and let \(n(\overline{x})=\lvert \mathcal{N}(\overline{x})\rvert\) be its number of observations. Notice that if no observation is ``close'' to \(i\) in terms of covariates, then \(i\) may end up in a singleton cluster, i.e.\ \(n(\overline{x})=1\).

Once the partition is obtained and labeled, the construction of our sample bounds proceeds exactly as in Section~\ref{sec:implementation}, except that we replace each \(X_i\) by \(\overline{X}_i\), each set \(\mathcal{X}_*\) by \(\overline{\mathcal{X}}\), and so on. 

\begin{remark}
    Our clustering algorithm aligns conceptually with data-generating processes characterized by large discrete covariate spaces, where the occurrence of each particular covariate value is rare, leading naturally to observation counts within clusters that follow an approximate Poisson distribution.
\end{remark}

\section{Monte Carlo Experiments}\label{sec:MC}

In this section, we report results from Monte Carlo experiments.
The scalar covariate $X_i$ is randomly generated from $\text{Unif}[-3,3]$. 
The binary treatment variable $D_i$ is then obtained from the following two models:
\begin{align}\label{true-PS-MC}
\begin{split}
\text{(DGP A)} &\;\;\;\; \mathbb{E} \left[ D_i | X_i = x\right] =  p(x) =  0.5, \\
\text{(DGP B)} &\;\;\;\; \mathbb{E} \left[ D_i | X_i = x\right] =  p(x) \\
&\qquad 
=     0.75 \times \mathbbm{1}\{ x \geq 2 \} + 0.5 \times \mathbbm{1}\{ |x| < 2 \} 
             +  1  \times \mathbbm{1}\{ x \leq -2 \}.
\end{split}
\end{align}
To generate the outcome variable, define
\begin{align*}
Y_{di}^\ast &= d + 1 - p(X_i) + V_{di},
\end{align*}
where $V_{di} \sim N(0,1)$, $d \in \{0,1\}$, and $(V_{1i}, V_{0i})$ are independent of $(D_i, X_i)$.
Finally, the observed outcome variable is generated by 
$$
Y_i = D_i \mathbbm{1}\{ Y_{1i}^\ast > 0 \} + (1-D_i) \mathbbm{1}\{ Y_{0i}^\ast > 0 \}.
$$ 
To study the effect of misspecification and the lack of overlap, we take the reference propensity score $p_*(x)=0.5$.
That is, under DGP A, the model is correctly specified and the overlap condition is satisfied;
whereas, under DGP B, the model is misspecified and the overlap condition is not satisfied. 
When $X_i \leq - 2$, $p(X_i) = 1$ in DGP B.
By simulation design, $a_{\min} = 0$ and $a_{\max} = 1$. 
In the Monte Carlo experiments, we focus on the ATT.

Recall from \eqref{ATestimands} that in our definition, the true ATT is given by 
\begin{equation}\label{eq:ATTdef:MC}
{\rm ATT}
= \frac{\frac{1}{n}\sum_{i=1}^n \tau(X_i) p(X_i)}{\frac{1}{n}\sum_{i=1}^n p(X_i)},
\end{equation}
where $p(x)=\mathbb{E}(D_i\mid X_i=x)$ and 
$\tau(x)=\mathbb{E}[Y_i(1)-Y_i(0)\mid X_i=x]$.
To obtain the closed-form expression for $\tau(x)$ in the Monte Carlo design, 
let $\Phi(\cdot)$ denote the standard normal CDF. For any $x$ and $d \in \{0,1\}$,
\begin{align*}
\mathbb{E}\!\left[Y_i(d)\mid X_i=x\right]
&= \mathbb{P}\!\left(Y_{di}^*>0\mid X_i=x\right)
= \mathbb{P}\!\left(V_{di}>-(d+1-p(x))\right)
\nonumber\\
&= \Phi\!\left(d+1-p(x)\right).
\label{eq:EYd}
\end{align*}
Hence,
\begin{equation}\label{eq:tauclosed}
\tau(x)
= \mathbb{E}\!\left[Y_i(1)-Y_i(0)\mid X_i=x\right]
= \Phi\!\left(2-p(x)\right)-\Phi\!\left(1-p(x)\right).
\end{equation}
Then, the finite-$n$ true ATT in our setting is obtained by 
combining \eqref{true-PS-MC} and \eqref{eq:tauclosed} into \eqref{eq:ATTdef:MC}.

Define $\widehat{p} = n^{-1} \sum_{i=1}^n D_i$.  
We consider the following point estimators:
\begin{align*}
\widehat{\rm ATT}_{\mathrm{Oracle}} &= 
\left(  n \widehat{p} \, \right)^{-1}
\sum_{i=1}^n 
D_i \left[ \mathbbm{1}\{ Y_{1i}^\ast > 0 \} - \mathbbm{1}\{ Y_{0i}^\ast > 0 \} \right], \\
\widehat{\rm ATT}_{\mathrm{RPS}} &= 
\left(  n \widehat{p} \, \right)^{-1}
\sum_{i=1}^n \left\{ D_i  - \frac{p_*(X_i)}{1-p_*(X_i)} (1-D_i)  \right\} Y_i.
\end{align*}
Here, $\widehat{\rm ATT}_{\mathrm{Oracle}}$ is an infeasible oracle estimator of ATT, whereas 
$\widehat{\rm ATT}_{\mathrm{RPS}}$ is an estimator using the known (parametric) propensity score 
$p_*(\cdot) = 0.5$ (for both DGPs). 
We also consider the nearest neighbor estimator of ATT:
\begin{align*}
\widehat{\rm ATT}_{\text{NN}} &= 
\left(  n \widehat{p} \, \right)^{-1}
\sum_{i=1}^n 
D_i \left[ Y_i - \widehat Y_{0 i} \right], 
\end{align*}
where $\widehat Y_{0 i}$ is the nearest neighbor estimator of $\mathbb{E}[Y \mid X=X_i, D=0]$.
For the bounds, $[\mathrm{LB}(Q),\mathrm{UB}(Q)]$ denotes the $Q$th-order bounds constructed with the reference propensity score $p_*(x)=0.5$, while $[\mathrm{LBc}(Q),\mathrm{UBc}(Q)]$ denotes the alternative bounds constructed with $p_*(x)=0$.\footnote{Recall Remark~\ref{rem:discussion:p_star_1} regarding the use of $p_*=1$ for estimating $\mathbb{E}_P[ Y_i(1) ]$. Since our object of interest is the ATT, the relevant quantity is instead $\mathbb{E}_P[ Y_i(0) \mid D_i=1 ]$. In this case, one must consider $p_*=0$ for $\mathbb{E}_P[ Y_i(0) \mid D_i=1 ]$ if one insists on using estimators that are conditionally unbiased. See also the general weighting schemes in \eqref{DefGeneralBC} and \eqref{DefGeneralWeights}.}
We report results for $Q=1,2,3$ throughout.
The number $m$ of clusters is chosen according to \eqref{choice-cluster} with $L=10$.
The sample size is $n=1{,}000$, and each design is based on $1{,}000$ Monte Carlo replications.

\begin{table}[htbp]
\caption{Monte Carlo Results}\label{tab-mc}
\begin{center}
\begin{tabular}{lrrr|rrr}
  \hline\hline
 & \multicolumn{3}{c}{DGP A} & \multicolumn{3}{c}{DGP B} \\
 & Mean & Median & St.Dev. & Mean & Median & St.Dev. \\
  \hline
Oracle & -0.000 & 0.000 & 0.023 & -0.000 & -0.001 & 0.024 \\
RPS & -0.001 & -0.001 & 0.051 & 0.211 & 0.212 & 0.035 \\
NN & 0.003 & 0.002 & 0.033 & -0.057 & -0.118 & 0.128 \\
LB(1) & -0.308 & -0.307 & 0.012 & -0.350 & -0.349 & 0.013 \\
UB(1) & 0.691 & 0.691 & 0.011 & 0.620 & 0.621 & 0.013 \\
LB(2) & 0.008 & 0.008 & 0.030 & -0.043 & -0.043 & 0.027 \\
UB(2) & -0.001 & 0.000 & 0.034 & 0.169 & 0.168 & 0.030 \\
LB(3) & 0.009 & 0.009 & 0.029 & -0.007 & -0.006 & 0.028 \\
UB(3) & 0.000 & 0.001 & 0.030 & 0.158 & 0.159 & 0.029 \\
LBc(1) & -0.308 & -0.307 & 0.012 & -0.350 & -0.349 & 0.013 \\
UBc(1) & 0.691 & 0.691 & 0.011 & 0.620 & 0.621 & 0.013 \\
LBc(2) & -0.151 & -0.151 & 0.017 & -0.218 & -0.218 & 0.017 \\
UBc(2) & 0.345 & 0.345 & 0.020 & 0.405 & 0.405 & 0.018 \\
LBc(3) & -0.072 & -0.072 & 0.022 & -0.132 & -0.132 & 0.021 \\
UBc(3) & 0.172 & 0.172 & 0.025 & 0.290 & 0.289 & 0.022 \\
  \hline
\end{tabular}
\end{center}
Notes: Reported statistics are computed after subtracting the true ATT, i.e., they summarize $\widehat{\mathrm{ATT}}-\mathrm{ATT}$ across Monte Carlo replications. Oracle refers to the infeasible estimator using both potential outcomes (latent binary outcomes). RPS is the estimator using the reference propensity score $p_*(x)=0.5$ as if it were true. NN is the nearest neighbor estimator. $[\mathrm{LB}(Q),\mathrm{UB}(Q)]$ are the $Q$-order bounds with $p_*(x)=0.5$, and $[\mathrm{LBc}(Q),\mathrm{UBc}(Q)]$ are the $Q$-order conservative bounds with $p_*(x)=0$. The sample size is $n=1{,}000$ with $1{,}000$ Monte Carlo replications.
\end{table}

Table~\ref{tab-mc} summarizes the Monte Carlo results.\footnote{In Online Appendix~\ref{sec:MC:additional}, we report additional Monte Carlo experiments that further investigate the finite-sample performance of the proposed inference methods.}
All reported statistics are computed after subtracting the true ATT from each estimator or bound.
That is, for each method, the mean, median, and standard deviation summarize the distribution of the centered quantity
$\widehat{\rm ATT}-{\rm ATT}$ across Monte Carlo replications.
Accordingly, values close to zero indicate good finite-sample performance.

In DGP~A, where the overlap condition holds and the reference propensity score is correctly specified, the oracle, RPS, NN, and higher-order bound estimators with $Q=2,3$ all have means and medians close to zero.
In contrast, the first-order Manski bounds $\mathrm{LB}(1)$ and $\mathrm{UB}(1)$ are wide and centered far from zero, reflecting the fact that they do not exploit the unconfoundedness assumption.\footnote{$\mathrm{LBc}(1)$ and $\mathrm{UBc}(1)$ coincide with $\mathrm{LB}(1)$ and $\mathrm{UB}(1)$ because the reference propensity score $p_*(x)$ does not play any role when $Q=1$.}
The alternative bounds $\mathrm{LBc}(Q)$ and $\mathrm{UBc}(Q)$ with $Q=2,3$ are tighter than the Manski bounds but remain relatively wide because the reference propensity score $p_*(x)=0$ is far from the true propensity.

In DGP~B, the overlap condition fails and the ATT is not point identified.
The NN estimator performs poorly, exhibiting substantial dispersion and a median far from zero.
The RPS estimator, which relies on a misspecified propensity score, is also unreliable: its mean lies outside the range implied by the higher-order bound estimators with $Q=2,3$.
In contrast, the proposed bounds with $Q=2,3$ remain informative and correctly reflect partial identification.
The lower bounds become tighter as $Q$ increases, while remaining below the oracle mean.

Overall, the results from DGPs A and B illustrate that the proposed bound approach does not require the overlap condition and can substantially improve upon parametric estimators when the propensity score is misspecified.
While the conservative choice $p_*(x)=0$ provides a useful worst-case benchmark, it may not be as competitive as $p_*(x)=0.5$, which is the case in the current Monte Carlo designs.
These findings suggest that higher-order bound estimators constructed with a reasonable interior reference propensity offer a useful compromise between point identification under strong ignorability and worst-case Manski-type bounds.

\section{Empirical Examples}\label{sec:EE}

\subsection{Effects of Right Heart Catheterization}\label{sec:RHC}

In this section, we apply our methods to \cite{Connors1996}'s study of the efficacy of 
right heart catheterization (RHC), which is a diagnostic procedure for directly measuring cardiac function in critically ill patients.
This dataset has been subsequently used in the context of limited overlap by 
\cite{crump2009dealing},
\cite{Rothe:2017},
\cite{Li-et-al:2018},
and
\cite{Ma_Sasaki_Wang_2024}
among others.
 The dataset is publicly available on the Vanderbilt Biostatistics website at 
\url{https://hbiostat.org/data/}.

In this example, the dependent variable is 1 if a patient survived after 30 days of admission,
and 0 if a patient died within 30 days.
The binary treatment variable is 1 if RHC was applied within 24 hours of admission, and 0 otherwise.
The sample size was $n = 5735$, and 2184 patients were treated with RHC. 
There are a large number of covariates: \cite{hirano2001estimation} constructed 72 variables from the dataset and the same number of covariates were considered in \cite{crump2009dealing} and \cite{Li-et-al:2018}, and \cite{Ma_Sasaki_Wang_2024},
and 50 covariates were used in \cite{Rothe:2017}.
In our exercise, we constructed the same 72 covariates. 
 For the purpose of illustrating our methodology, we assume that the unconfoundedness assumption holds in this example.\footnote{%
\citet{bhattacharya2008treatment,BSV-2012}  raise the concern that 
catheterized and noncatheterized patients may differ on unobserved dimensions
and propose different bounds using a day of admission 
as an instrument for RHC.}

In this section, we focus on ATT. We first estimate ATT by the 
normalized inverse probability weighted estimator\footnote{See, e.g., equation (3) and discussions in 
\citet{busso2014new} for details of the normalized inverse probability weighted ATT estimator.}:
\begin{align*}
\widehat{\text{ATT}}_{\text{PS}} 
:= \frac{\sum_{i=1}^n D_i Y_i}{\sum_{i=1}^n D_i}
- \frac{\sum_{i=1}^n (1-D_i) W_i Y_i}{\sum_{i=1}^n (1-D_i) W_i},
\end{align*}
where $W_i := \widehat{p}(X_i)/[1-\widehat{p}(X_i)]$
and $\widehat{p}(X_i)$ is the estimated propensity score for observation $i$ based on
a logit model with all 72 covariates being added linearly as in the aforementioned papers.
The estimator $\widehat{\text{ATT}}_{\text{PS}}$ requires that the assumed propensity score model be correctly specified and the overlap condition is satisfied.
The resulting estimate is
$\widehat{\text{ATT}}_{\text{PS}}  = -0.0639$.\footnote{The unnormalized ATT estimate is $-0.0837$ using the same propensity scores.}

We now turn to our methods.
We take the reference propensity score to be $
\widehat{p}_{\text{RPS}}(X_i) = n^{-1} \sum_{i=1}^n D_i$
for each observation $i$. That is, we assign the sample proportion of the treated to the 
reference propensity scores uniformly for all observations. Of course, this is likely to be misspecified; however, it has the advantage that $\widehat{p}_{\text{RPS}}(X_i)$ is never close to 0 or 1. The resulting inverse reference-propensity-score weighted ATT estimator is\footnote{When the sample proportion is used as the propensity score estimator, there is no difference between
unnormalized and normalized versions of ATT estimates. In fact, it is simply the mean difference between treatment and control groups. }
\begin{align*}
\widehat{\text{ATT}}_{\text{RPS}} 
:= \frac{\sum_{i=1}^n D_i Y_i}{\sum_{i=1}^n D_i}
- \frac{\sum_{i=1}^n (1-D_i) Y_i}{\sum_{i=1}^n (1-D_i)}
= -0.0507.
\end{align*} 
None of the covariate values in the observed sample are identical among patients (that is, $n(X_i)=1$ for all observations here). 
We therefore implement the clustering method described in Section~\ref{sec:Cluster}.
As recommended in Section~\ref{sec:Cluster}, we choose 
the number $m$ of clusters by \eqref{choice-cluster}: 
$ m = \left\lceil \frac{n}{L} \right\rceil$ with $L = 5, 10, 20$.
In addition, we consider $Q = 1,\ldots,4$. 

\begin{table}[htb!]
\caption{ATT Bounds: Right Heart Catheterization Study}\label{tab-rhc}
\begin{center}
\begin{tabular}{rrrrrr}
  \hline
  \hline
L & Q & LB & UB & CI-LB & CI-UB \\
  \hline
\multicolumn{6}{l}{Panel A. $p_*(x)=\bar D$} \\ 
5 & 1 &  -0.638 & 0.282 & -0.700 & 0.330 \\ 
 & 2 &  -0.131 & -0.000 & -0.174 & 0.033 \\ 
 & 3 &  -0.034 & -0.048 & -0.076 & -0.007 \\ 
 & 4 &  -0.006 & -0.073 & -0.079 & -0.006 \\ 
\hline
10 & 1 &  -0.664 & 0.307 & -0.766 & 0.376 \\ 
 & 2 &  -0.169 & 0.004 & -0.216 & 0.039 \\ 
 & 3 &  -0.077 & -0.039 & -0.117 & -0.006 \\ 
 & 4 &  -0.049 & -0.057 & -0.090 & -0.016 \\ 
\hline
20 & 1 &  -0.675 & 0.316 & -0.843 & 0.430 \\ 
 & 2 &  -0.178 & -0.005 & -0.238 & 0.034 \\ 
 & 3 &  -0.099 & -0.046 & -0.149 & -0.007 \\ 
 & 4 &  -0.065 & -0.060 & -0.112 & -0.017 \\ 
\hline
\multicolumn{6}{l}{Panel B. $p_*(x)=0$} \\ 
5 & 1 &  -0.638 & 0.282 & -0.700 & 0.330 \\ 
 & 2 &  -0.300 & 0.088 & -0.343 & 0.121 \\ 
 & 3 &  -0.173 & 0.012 & -0.208 & 0.042 \\ 
 & 4 &  -0.119 & -0.025 & -0.151 & 0.005 \\ 
\hline
10 & 1 &  -0.664 & 0.307 & -0.766 & 0.376 \\ 
 & 2 &  -0.326 & 0.104 & -0.390 & 0.143 \\ 
 & 3 &  -0.188 & 0.020 & -0.235 & 0.052 \\ 
 & 4 &  -0.126 & -0.020 & -0.166 & 0.011 \\ 
\hline
20 & 1 &  -0.675 & 0.316 & -0.843 & 0.430 \\ 
 & 2 &  -0.336 & 0.106 & -0.438 & 0.156 \\ 
 & 3 &  -0.198 & 0.017 & -0.270 & 0.048 \\ 
 & 4 &  -0.133 & -0.024 & -0.191 & 0.008 \\ 
\hline
    \end{tabular}
\end{center}
Notes: LB and UB correspond to the lower and upper bound estimates, where
CI-LB and CI-UB represent the lower and upper 95\% confidence interval estimates.  
Estimates are shown for selected values of $L = 5, 10, 20$ and $Q=1,\ldots,4$.
\end{table}

Table~\ref{tab-rhc} reports the estimated ATT bounds for selected values of $L$ and $Q$ using two reference propensity scores.
We first discuss Panel~A, which corresponds to the reference propensity score $p_*(x)=\bar D$.
When $Q=1$, our estimated bounds correspond to Manski bounds, which include zero and are wide, with interval lengths close to one for all values of $L$.
Our bounds with $Q=1$ are different across $L$ 
because we apply hierarchical clustering before obtaining Manski bounds.
With $Q=2$, the bounds shrink so that the estimated upper bound is zero for all cases of $L$;
with $Q = 3$, they shrink even further so that the upper end point of the 95\% confidence interval excludes zero.
Among three different values of $L$, the case of $L=5$ gives the tightest confidence interval but in this case, the lower bound is larger than the upper bound, indicating that the estimates might be biased. In view of that, we take the bound estimates with $L=10$ as our preferred estimates [$-0.077,  -0.039$] with the 95\% confidence interval 
$[-0.117,-0.006]$.
When $Q=4$,  the lower bound estimates exceed the upper bound estimates with $L = 5, 10$. However, the estimates with $L = 20$ give an almost identical confidence interval to our preferred estimates. It seems that the pairs of $(L, Q) = (10, 3)$ and 
$(L, Q) = (20, 4)$ provide reasonable estimates. 

Panel~B reports the corresponding results obtained using the conservative reference propensity score $p_*(x)=0$.
As expected, the resulting bounds are wider for all values of $L$ and $Q$, reflecting the additional conservatism of this choice.
 
The study of \cite{Connors1996}
offered a conclusion that RHC could cause an increase in patient mortality.
Based on our preferred estimates, we can exclude large beneficial effects with confidence. This conclusion is based solely on the unconfoundedness condition, but not on 
the overlap condition, nor on the correct specification of the logit model. 
Overall, our estimates seem to be consistent with the qualitative findings in \cite{Connors1996}
 under the maintained assumption that the unconfoundedness assumption holds.

\subsection{Impacts of a Temporary Employment Program}\label{sec:EE:NSW}

In this section, we apply our methods to the well-known 
\citet{LaLonde-AER} dataset, available on Rajeev Dehejia's web page 
at \url{http://users.nber.org/~rdehejia/nswdata2.html}.
The LaLonde dataset comes from the National Supported Work Demonstration (NSW), a randomized controlled temporary employment program. The binary treatment variable indicates whether an individual is assigned to the treatment or control group. The original outcome variable (\texttt{RE78}) is post-experimental earnings in 1978; in our application, we define the outcome as whether an individual was employed in 1978, i.e., whether earnings in 1978 were positive (\texttt{RE78} $> 0$). 
Because NSW is a randomized controlled trial (RCT), we first estimate the average treatment effect by computing simple mean differences, yielding a 95\% confidence interval of $[0.013, 0.143]$. This interval is relatively wide but excludes zero.

\begin{table}[htb!]
\caption{ATE Bounds: Dehejia--Wahba Subsample and PSID2 Controls}\label{tab-nsw-psid2}
\begin{center}
\begin{tabular}{rrrrrr}
  \hline
  \hline
L & Q & LB & UB & CI-LB & CI-UB \\
  \hline
\multicolumn{6}{l}{Panel A. Dehejia--Wahba subsample} \\ 
5 & 1 &  -0.425 & 0.458 & -0.628 & 0.659 \\ 
 & 2 &  0.053 & 0.040 & -0.073 & 0.173 \\ 
 & 3 &  0.109 & 0.008 & -0.055 & 0.180 \\ 
 & 4 &  0.093 & 0.021 & -0.088 & 0.204 \\ 
\hline
10 & 1 &  -0.458 & 0.501 & -0.788 & 0.804 \\ 
 & 2 &  0.079 & 0.092 & -0.017 & 0.197 \\ 
 & 3 &  0.100 & 0.069 & -0.014 & 0.184 \\ 
 & 4 &  0.127 & 0.030 & -0.038 & 0.191 \\ 
\hline
20 & 1 &  -0.476 & 0.515 & -0.893 & 0.885 \\ 
 & 2 &  0.084 & 0.087 & -0.010 & 0.192 \\ 
 & 3 &  0.098 & 0.077 & -0.007 & 0.183 \\ 
 & 4 &  0.066 & 0.067 & -0.052 & 0.177 \\ 
\hline
\multicolumn{6}{l}{Panel B. NSW treated and PSID2 controls} \\ 
5 & 1 &  -0.473 & 0.395 & -0.595 & 0.626 \\ 
 & 2 &  -0.504 & 0.111 & -0.794 & 0.365 \\ 
 & 3 &  -0.181 & -0.143 & -0.330 & 0.052 \\ 
 & 4 &  -0.359 & -0.210 & -0.645 & 0.036 \\ 
\hline
10 & 1 &  -0.479 & 0.463 & -0.668 & 0.770 \\ 
 & 2 &  -0.452 & 0.309 & -0.876 & 0.671 \\ 
 & 3 &  -0.178 & 0.077 & -0.346 & 0.310 \\ 
 & 4 &  -0.378 & 0.048 & -0.785 & 0.374 \\ 
\hline
20 & 1 &  -0.482 & 0.495 & -0.706 & 0.823 \\ 
 & 2 &  -0.359 & 0.386 & -0.776 & 0.728 \\ 
 & 3 &  -0.141 & 0.195 & -0.338 & 0.417 \\ 
 & 4 &  -0.241 & 0.213 & -0.651 & 0.523 \\ 
\hline
    \end{tabular}
\end{center}
Notes: In each panel, the reference propensity score is set to the sample treatment share, $p_*(x)=\bar D$. LB and UB denote the estimated lower and upper bounds on the ATE, and CI-LB and CI-UB denote the endpoints of the 95\% confidence interval. Estimates are reported for $L\in\{5,10,20\}$ and $Q\in\{1,2,3,4\}$.
\end{table}

\subsubsection{Case Study Using the Dehejia-Wahba Subsample} 

\cite{DehejiaWahba-JASA} and \cite{DehejiaWahba-RESTAT} extract a subset of LaLonde's NSW experimental data that includes information on RE74 (earnings in 1974). 
If we assume that the Dehejia-Wahba sample preserves the initial randomization, we can impose that the reference propensity score is independent of covariates. However, this may not be the case, so our approach provides a robust method to check whether the Dehejia-Wahba sample can be viewed as a random sample from an RCT.

We define the reference propensity score to be the sample proportion of treatment in the Dehejia-Wahba sample. 
The covariates are:
age in years,
years of education, 
indicators for black, hispanic, married, and no degree,
and earnings in 1974 as well as in 1975. 
If we treat this reference propensity score as a consistent estimator under preservation of randomization, the average treatment effect is again obtained by simple mean differences, producing a 95\% confidence interval of $[0.026, 0.196]$, which is wide but excludes zero.

Next, we obtain our bounds on the average treatment effect (ATE). 
As before, it is necessary to choose $Q$ and $L$. Based on previous numerical results, we set $Q=3$ and $L=10$.
Our bounds yield a 95\% confidence interval of $[-0.014,0.184]$, which is similar to the interval $[0.02,0.20]$ obtained under the assumption that the Dehejia-Wahba sample is a random sample from NSW. This result suggests two points: first, there is no evidence that the random sampling assumption is violated in the Dehejia-Wahba sample; and second, our inference method does not substantially widen the confidence interval to achieve robustness, although the null effect is now included. Furthermore, our bounds remain similar if we change $Q$ to 2 or 4 or $L$ to 5 or 20, indicating that our findings are robust to the choice of tuning parameters. 
See details in Panel A of Table \ref{tab-nsw-psid2}.

\subsubsection{Case Study Using the Population Survey of Income Dynamics Control Group} 

The Dehejia-Wahba sample can be viewed as a scenario where the propensity score is known and satisfies the overlap condition. We now turn to a different scenario where it is likely that the propensity score is unknown and may not satisfy the overlap condition. 
Specifically, we use one of the non-experimental comparison groups constructed by LaLonde from the Population Survey of Income Dynamics, the PSID2 controls. As in the previous subsection, we estimate the reference propensity score using the sample proportion and then obtain our bound estimates. 
The resulting confidence interval is $[-0.346, 0.310]$ with $Q=3$ and $L=10$, which is much larger than the interval $[-0.01,0.19]$ obtained with the Dehejia-Wahba sample. Note that the sample proportion is unlikely to be correctly specified in the NSW-treated/PSID2-control sample. Therefore, our inference method appears to produce a wider confidence interval to remain robust against possible misspecification of the propensity scores and/or a lack of overlap.  
As a benchmark, we also compute the Manski bounds by setting $Q=1$.
The resulting confidence interval for the Manski bounds is $[-0.668, 0.770]$ with $Q=1$ and $L=10$, which is even larger. Recall that the Manski bounds do not impose the unconfoundedness assumption and do not rely on any pooling information (so the specification of the reference propensity score does not matter). See Panel B of Table \ref{tab-nsw-psid2} for other values of $Q$ and $L$.
We conclude this section by noting that our empirical findings are broadly consistent with those in \cite{Ma_Sasaki_Wang_2024}. 
In particular, our results indicate that limited overlap is a salient concern in the NSW-treated/PSID-control sample, as evidenced by substantially wider bounds, while comparable concerns do not appear to arise in settings such as \cite{Connors1996}’s study, as discussed in the previous section.

\appendix
\renewcommand{\theequation}{\thesection.\arabic{equation}}
\setcounter{equation}{0}
\renewcommand{\thetheorem}{\thesection.\arabic{theorem}}
\setcounter{theorem}{0}

\renewcommand{\baselinestretch}{1.22}

\section*{Appendix}

\section{Second-order ATT bounds}
\label{app:SecondOrderATT}

In the main text, we primarily focused on ATE bounds. Here, we extend the second-order approach to ATT and analyze how these bounds compare to the first-order ATT bounds.

\medskip
To refine the first-order ATT bounds in \eqref{FirstBoundsATT}, we introduce the second-order adjustment:
\begin{align}
    C^{(2)}_{a}
      &=    D\, (Y-a)     + \left[  \lambda_0(X) +   \lambda_1(X)  \,  p(X) \right]  \,  (1-D) \, (Y-a) ,
      \label{SecondOrderBoundGeneralC}
\end{align}
where the coefficients \( \lambda_0(x), \lambda_1(x) \in \mathbb{R} \) must be chosen such that
\begin{align}
     \mathbb{E}\left[ C^{(2)}(a_{\max})    \, \big| \, X=x    \right]     &\, \leq \,  \mathbb{E}\left[ D\, (Y(1) -  Y(0)) \, \big| \, X=x\right] \, \leq \,
 \mathbb{E}\left[ C^{(2)}(a_{\min})    \, \big| \, X=x    \right]  .
           \label{SecondOrderConditionC}
\end{align}
Following the same logic as for second-order ATE bounds, we determine the optimal coefficients under Assumptions (i) and (ii):
\begin{align}
     \lambda_0(x) &= \left[   \frac{ p_*(x)       }   { 1- p_*(x) }  \right]^2 ,
    &
     \lambda_1(x) &=  -  \frac{ 1     }   {\left[ 1- p_*(x) \right]^2}  ,
    \label{SoultionLambdaTilde}
\end{align}
where \( p_*(x) \in [0,1) \) can be chosen arbitrarily. Plugging these coefficients back into \eqref{SecondOrderBoundGeneralC} gives
\begin{align*}
 C^{(2)}_{a}    &= D\, (Y-a) 
 -    \widetilde w^{(2)} \big(p(x),p_*(x) \big) \,  \frac{p(X) \, (1-D) \, (Y-a)} {1-p(X)} ,
\end{align*} 
where the weight function \( \widetilde w^{(2)}: (0,1] \times [0,1)  \rightarrow  (-\infty,1] \) is given by
\begin{align*}
    \widetilde  w^{(2)}(p,p_*) :=  1 - \frac 1 p \, \left( \frac{p-p_*} {1-p_*}  \right)^2   . 
\end{align*} 
Under Assumption~\ref{ass:MAIN}, we calculate
\begin{align*}
 \mathbb{E}\left[ C^{(2)}_{a}  \Big|  X{=}x   \right]  
  & =  p(x) 
  \Big\{    \mathbb{E}\left[  Y(1)-a  \big|  X=x   \right] 
 -    \widetilde w^{(2)} \big(p(x),p_*(x) \big) \,  \mathbb{E}\left[  Y(0)-a  \big|  X=x   \right]  \hspace{-0.13cm} \Big\}
  \\
   & = 
   \Big[1-\widetilde w^{(2)} \big(p(x),p_*(x) \big) \Big]    \mathbb{E}\left[ C^{(1)}_a  \big|   X=x   \right]  
    +
   \widetilde w^{(2)} \big(p(x),p_*(x) \big) \,  \pi(x) .
\end{align*} 
Thus, conditional on \( X=x \), the second-order ATT bounds are weighted averages between 
the first-order bounds and \( \pi(x) \).
The weight \( \widetilde w^{(2)}(p(x),p_*(x)) \) equals one when \( p(x) = p_*(x) \),
implying that if \( p(x) = p_*(x) \), the second-order bound holds with equality.

\begin{figure}[tb!] 
\begin{center}
\begin{tikzpicture}[scale=0.8]
\begin{axis}[
    axis lines = left,
    xlabel = $p$,
    ylabel = {$\widetilde w^{(2)}(p,p_*)$},
    xmin=0, xmax=1.1,
    ymin=-0.3, ymax=1.1,
    legend pos=outer north east
]

\addplot [
    domain=0:1, 
    samples=100, 
    color=black,
]
{0};
\addlegendentry{Manski bounds}

\addplot [
    dotted,
    domain=0:1, 
    samples=100, 
    color=dgreen,
    style=ultra thick
    ]
{1-((x-0)/(1-0))^2/x};
\addlegendentry{$p_*=0$}

\addplot [
     dashed,
    domain=0:1, 
    samples=100, 
    color=red,
    style=ultra thick
    ]
{1-((x-0.5)/(1-0.5))^2/x};
\addlegendentry{$p_*=0.5$}

\addplot [
    domain=0.5:1, 
    samples=100, 
    color=blue,
    style=ultra thick
    ]
{1-((x-0.75)/(1-0.75))^2/x};
\addlegendentry{$p_*=0.75$}

\end{axis}

\end{tikzpicture}

\caption{\label{fig:weights0ATT}Weights $\widetilde w^{(2)}(p,p_*)$  as a function of $p$, for different values of $p_*$}
\end{center}
\end{figure}  

Figure~\ref{fig:weights0ATT} plots \( \widetilde w^{(2)}(p(x),p_*(x)) \) as a function of \( p(x) \) for different values of \( p_*(x) \).
The Manski bounds correspond to setting \( \widetilde w^{(2)} = 0 \). Only for \( p_*(x)=0 \) do the second-order bounds uniformly improve upon the Manski bounds. However, whenever \( p(x) \) is close to the chosen \( p_*(x) \), the second-order bounds improve over the Manski bounds.

\section{Proofs for Section~\ref{sec:MainIdea}}

\begin{proof}[\bf Proof of Theorem~\ref{th:OurBounds}]
\# First, we show that for any valid bound function $B$, we must have $B(y,0,d_2) = a_{\max}$ for all $y \in [a_{\min},a_{\max}]$ and $d_2 \in \{0,1\}$. Suppose there exists some $y_0 \in [a_{\min},a_{\max}]$ and $d_2 \in \{0,1\}$ such that $B(y_0,0,d_2) \neq a_{\max}$.

If $B(y_0,0,d_2) < a_{\max}$, we can construct a distribution $P \in {\cal P}$ where $Y_1(0) = y_0$, $Y_i(1) = a_{\max}$ for all $i$, and $P(D_1=0) = 1$, $P(D_2=d_2) = 1$. Under this distribution, we have
\[
\mathbb{E}_P[Y_i(1)] = a_{\max} > B(y_0,0,d_2) = \mathbb{E}_P[B(Y_1,D_1,D_2)],
\]
which contradicts the bound condition.
If instead $B(y_0,0,d_2) > a_{\max}$, we can construct an alternative function $\tilde{B}$ defined as
\[
\tilde{B}(y,d_1,d_2) =
\begin{cases}
B(y,d_1,d_2), & \text{if } d_1 = 1, \\
a_{\max}, & \text{if } d_1 = 0.
\end{cases}
\]
This function still satisfies \eqref{BoundMainExample} for all $P \in {\cal P}$ (since $\mathbb{E}_P[Y_i(1)] \leq a_{\max}$), but strictly dominates $B$, contradicting the assumption that $B$ is undominated. Therefore, we must have $B(y,0,d_2) = a_{\max}$.
\medskip

\# Next, we show that for any fixed $d_2 \in \{0,1\}$, the function $y \mapsto B(y,1,d_2)$ must be of the form $a_{\max} + \lambda_{d_2}(y-a_{\max})$ for some coefficient $\lambda_{d_2} \geq 0$.

We first establish that $B(a_{\max},1,d_2) = a_{\max}$. If $B(a_{\max},1,d_2) < a_{\max}$, we could construct a distribution with $Y_i(1) = a_{\max}$ for all $i$, $P(D_1=1) = 1$, and $P(D_2=d_2) = 1$. This would give $\mathbb{E}_P[Y_i(1)] = a_{\max} > B(a_{\max},1,d_2) = \mathbb{E}_P[B(Y_1,D_1,D_2)]$, contradicting the bound condition. If $B(a_{\max},1,d_2) > a_{\max}$, lowering this value to $a_{\max}$ would maintain validity and strictly dominate $B$.

Now we prove linearity. Suppose that $B(y,1,d_2)$ is not linear on $[a_{\min},a_{\max}]$. Then there exist $y_1,y_2 \in [a_{\min},a_{\max}]$ and $\alpha \in (0,1)$ such that
\[
B(\alpha y_1 + (1-\alpha)y_2, 1, d_2) \neq \alpha B(y_1,1,d_2) + (1-\alpha)B(y_2,1,d_2).
\]
If $B(\alpha y_1 + (1-\alpha)y_2, 1, d_2) < \alpha B(y_1,1,d_2) + (1-\alpha)B(y_2,1,d_2)$, we can construct a distribution $P \in {\cal P}$ with $P(D_1=1) = 1$, $P(D_2=d_2) = 1$, and $P(Y_1 = \alpha y_1 + (1-\alpha)y_2 | D_1=1) = 1$. We can also construct a distribution $P' \in {\cal P}$ with the same treatment assignments but with $P'(Y_1 = y_1 | D_1=1) = \alpha$ and $P'(Y_1 = y_2 | D_1=1) = 1-\alpha$. By unconfoundedness, $\mathbb{E}_P[Y_i(1)] = \mathbb{E}_{P'}[Y_i(1)]$, but $\mathbb{E}_P[B(Y_1,D_1,D_2)] < \mathbb{E}_{P'}[B(Y_1,D_1,D_2)]$, contradicting the optimality of $B$.

If $B(\alpha y_1 + (1-\alpha)y_2, 1, d_2) > \alpha B(y_1,1,d_2) + (1-\alpha)B(y_2,1,d_2)$, we can define a dominating function $\tilde{B}$ identical to $B$ except at the point $(\alpha y_1 + (1-\alpha)y_2, 1, d_2)$, where $\tilde{B}$ takes the value $\alpha B(y_1,1,d_2) + (1-\alpha)B(y_2,1,d_2)$. This function satisfies the bound condition but strictly dominates $B$.

Thus, $B(y,1,d_2)$ must be linear in $y$, and with $B(a_{\max},1,d_2) = a_{\max}$, we have
\[
B(y,1,d_2) = a_{\max} + \lambda_{d_2} (y - a_{\max}),
\]
for some $\lambda_{d_2}$. Moreover, $\lambda_{d_2} \geq 0$ must hold, as otherwise we could construct a distribution with $Y_i(1) = a_{\min}$ where the bound condition would be violated.
\medskip

\# Next, to determine constraints on the coefficients $\lambda_0$ and $\lambda_1$, consider a distribution with $P(D_i=1) = p$ for $i \in \{1,2\}$. By the independence assumption (iii), we have $P(D_1=1,D_2=1) = p^2$, $P(D_1=1,D_2=0) = p(1-p)$, $P(D_1=0,D_2=1) = (1-p)p$, and $P(D_1=0,D_2=0) = (1-p)^2$. Computing expectations under this distribution, we obtain
\begin{align*}
\mathbb{E}_P[B(Y_1,D_1,D_2)] &= p^2[a_{\max} + \lambda_1\mathbb{E}_P(Y_1-a_{\max}|D_1=1,D_2=1)] \\
&+ p(1-p)[a_{\max} + \lambda_0\mathbb{E}_P(Y_1-a_{\max}|D_1=1,D_2=0)] \\
&+ (1-p)p[a_{\max}] + (1-p)^2[a_{\max}]
\end{align*}
By unconfoundedness, $\mathbb{E}_P[Y_1|D_1=1,D_2=d_2] = \mathbb{E}_P[Y_1(1)]$ for any $d_2$, so
\[
\mathbb{E}_P[B(Y_1,D_1,D_2)] = a_{\max} + \left[p^2\lambda_1 + p(1-p)\lambda_0\right] (\mathbb{E}_P[Y_1(1)] - a_{\max}).
\]
For the bound condition to hold, $\mathbb{E}_P[Y_1(1)]$ must be less than or equal to this expectation. Since $\mathbb{E}_P[Y_1(1)] - a_{\max} \leq 0$ by assumption (ii), this implies
\[
p^2\lambda_1 + p(1-p)\lambda_0 \leq 1, \quad \text{for all } p \in (0,1).
\]
Using this, we now determine the optimal coefficients. Since $B$ is undominated, this inequality must hold with equality for some $p_* \in (0,1]$, otherwise we could increase $\lambda_0$ or $\lambda_1$ slightly to obtain a strictly dominating function. So we have
\[
(p_*)^2\lambda_1 + p_*(1-p_*)\lambda_0 = 1.
\]
For the bound to be valid for all $p \in (0,1)$, the function $f(p) = p^2\lambda_1 + p(1-p)\lambda_0$ must be maximized at $p_*$. Taking the derivative and setting it to zero, we obtain
\[
2p_*\lambda_1 + \lambda_0 - 2p_*\lambda_0 = 0.
\]
Solving this system gives
\[
\lambda_0 = \frac{2}{p_*}, \quad \lambda_1 = \frac{2p_*-1}{(p_*)^2}.
\]
These are the only values that satisfy all constraints. If $\lambda_0 > \frac{2}{p_*}$ or $\lambda_1 > \frac{2p_*-1}{(p_*)^2}$, the bound condition would be violated for some $p$ near $p_*$. If $\lambda_0 < \frac{2}{p_*}$ or $\lambda_1 < \frac{2p_*-1}{(p_*)^2}$, we could construct a dominating function by increasing these coefficients.
\medskip

\# Combining the above, we conclude that any function $B$ satisfying the theorem's conditions must take the form:
\[
B(y,d_1,d_2) = a_{\max} + \frac{2p_* - d_2}{(p_*)^2} d_1 (y - a_{\max}),
\]
for some $p_* \in (0,1]$. This proves both existence and uniqueness of the bound function up to the choice of $p_*$ (and modifications on sets of measure zero).

\#
Finally, to establish the reverse direction: any function of the form
$$
B(y,d_1,d_2) = a_{\max} + \frac{2p_* - d_2}{p_*^2} \, d_1 \, (y - a_{\max}),
$$ 
for some \( p_* \in (0,1] \), satisfies the bound condition \eqref{BoundMainExample} by construction. Moreover, it is not dominated, since any increase in the coefficients would violate validity for some values of \( p \), while any decrease would lead to a strictly weaker (i.e., dominated) bound.
\end{proof}

\section{Proofs for Section~\ref{sec:PopulationBounds}}

\subsection{Proofs of the main text results in Section~\ref{sec:PopulationBounds}}

\begin{proof}[\bf Proof of Proposition~\ref{prop:MainBounds}]
    This proposition is the special case $q=2$ of part (i) and (ii) of 
    Proposition~\ref{prop:HigherOrder}. We therefore refer to the proof of Proposition~\ref{prop:HigherOrder} below.
\end{proof}

Before presenting the proof of  Proposition~\ref{prop:HigherOrder} it is useful to provide two intermediate lemmas.
Those  lemmas explain the properties of the weight functions 
 $ w^{(q)} \left(p,p_* \right)$ and $    \widetilde w^{(q)} \left(p,p_* \right) $ that were defined in the main text,
 and are crucial for the proof of part (iii) of  Proposition~\ref{prop:HigherOrder}.

\begin{lemma}
   \label{lemma:DefW}
   Let $q \in \{1,2,\ldots\}$.
   For $\lambda = (\lambda_0,\ldots,\lambda_{q-1}) \in \mathbb{R}^q$
   and $p \in [0,1]$  we define $ v(p,\lambda)  :=   \sum_{r=0}^{q-1} \lambda_r \,  p^{r+1}$,
   and for $p \in (0,1]$ we define   
     $  \widetilde v(p,\lambda)  :=  \sum_{r=0}^{q-1}  \lambda_r \,  p^{r-1} \, (1-p)$. Let $p_* \in (0,1)$.
   Then, the functions  $ w^{(q)} \left(p,p_* \right)$ and $    \widetilde w^{(q)} \left(p,p_* \right) $
   defined in \eqref{DefGeneralWeights} are the unique solutions to the following optimization problems.
     
  \begin{itemize}
       \item[(i)] The solution to the optimization problem
        \begin{align*}
              \overline \lambda  = \argmin_{\lambda \in  \mathbb{R}^q} \, \left| \frac{\partial^{q-1}  v(p_*,\lambda)} {\partial^{q-1} p}  \right|
               \qquad    \text{subject to} & \qquad
                   v(p_*,\lambda) = 1 ,
                \\
              \text{and} &  \qquad
                   \frac{\partial^k  v(p_*,\lambda)} {\partial^k p} = 0, \;\;  \text{for $k \in \{1,\ldots,q-2\}$,}
                \\
              \text{and} &  \qquad
                        v(p,\lambda)  \leq 1,        \; \; \text{for $p \in [0,1]$,}
        \end{align*}  
        satisfies
        $$
             v(p,  \overline \lambda )  = w^{(q)}(p,p_*) .
        $$
        
       \item[(ii)] The solution to the optimization problem
        \begin{align*}
              \widetilde \lambda  = \argmin_{\lambda \in  \mathbb{R}^q} \, \left| \frac{\partial^{q-1}    \widetilde v(p_*,\lambda)} {\partial^{q-1} p}  \right|
               \qquad    \text{subject to} & \qquad
                     \widetilde v(p_*,\lambda) = 1 ,
                \\
              \text{and} &  \qquad
                   \frac{\partial^k    \widetilde v(p_*,\lambda)} {\partial^k p} = 0, \;\;  \text{for $k \in \{1,\ldots,q-2\}$,}
                \\
              \text{and} &  \qquad
                          \widetilde v(p,\lambda)  \leq 1,        \; \; \text{for $p \in (0,1]$,}
        \end{align*}  
        satisfies
        $$
               \widetilde v(p,    \widetilde  \lambda )  =   \widetilde w^{(q)}(p,p_*) .
        $$        
  \end{itemize}   
 
\end{lemma}
 
The proof of Lemma~\ref{lemma:DefW} is provided in Appendix~\ref{app:ProofIntermediate}.
For the statement of the next lemma, remember that
for $p_* \in (0,1)$ and $\epsilon>0$ we defined
${\cal B}_\epsilon(p_*)$ to be the $\epsilon$-ball around~$p_*$.

\begin{lemma}
   \label{lemma:PropertiesW}
   Let $q \in \{1,2,\ldots\}$ and $p_* \in (0,1)$.
   For $\lambda = (\lambda_0,\ldots,\lambda_{q-1}) \in \mathbb{R}^q$ let 
   $ v(p,\lambda)$ and $\widetilde v(p,\lambda) $ be as defined in Lemma~\ref{lemma:DefW}.
   \begin{itemize}
       \item[(i)]
   Let $\lambda   \in \mathbb{R}^q$
   be such that  for all $p \in [0,1]$ we have  $v(p,\lambda)  \leq 1$.
    Then, there exists $\epsilon>0$ such that for all $p \in {\cal B}_\epsilon(p_*)$
    we have
$$
    v(p,\lambda)  \leq  w^{(q)}(p,p_*) .
$$

     \item[(ii)]
    Let $\lambda   \in \mathbb{R}^q$
   be such that  for all $p \in (0,1]$ we have   $\widetilde v(p,\lambda)  \leq 1$.
    Then, there exists $\epsilon>0$ such that for all $p \in {\cal B}_\epsilon(p_*)$
    we have
$$
    \widetilde v(p,\lambda)  \leq  \widetilde w^{(q)}(p,p_*) .
$$
\end{itemize}
\end{lemma}

The proof of Lemma~\ref{lemma:PropertiesW} is provided in Appendix~\ref{app:ProofIntermediate}.

\begin{proof}[\bf Proof of Proposition~\ref{prop:HigherOrder}]
\underline{\# Part (i):}
  Under Assumption~\ref{ass:MAIN}(i) we find for the bounds defined in \eqref{DefGeneralBC} that
\begin{align*}
    \mathbb{E}\left[  B^{(q)}_{0,a} -a  \, \big| \, X=x  \right]  
      &=    w^{(q)} \big(1-p(x),1-p_*(x) \big)    \mathbb{E}\left[  Y(0)-a   \, \big| \, X=x  \right]   ,
   \nonumber     \\ 
      \mathbb{E}\left[  B^{(q)}_{1,a} -a  \, \big| \, X=x  \right]    &=      w^{(q)} \big(p(x),p_*(x) \big) \;  
        \mathbb{E}\left[  Y(1)-a   \, \big| \, X=x  \right] ,
 \nonumber   \\ 
   \mathbb{E}\left[    C^{(q)}_a     \, \big| \, X=x  \right]  
   &=      p(x) \, \Bigg\{  \mathbb{E}\left[  Y(1)-a   \, \big| \, X=x  \right] 
   \\ & \qquad \qquad\qquad\qquad
 -    \widetilde w^{(q)} \big(p(x),p_*(x) \big) \,  \mathbb{E}\left[  Y(0)-a   \, \big| \, X=x  \right]  \Bigg\} .
\end{align*}
From the definition of the weight functions in \eqref{DefGeneralWeights} we have
\begin{align*}
        w^{(q)} \big(1-p(x),1-p_*(x) \big)  \leq 1 , \quad
        w^{(q)} \big(p(x),p_*(x) \big) \leq 1, \quad
        \text{ and } \quad
         \widetilde w^{(q)} \big(p(x),p_*(x) \big) & \leq 1 .
\end{align*}
Assumption~\ref{ass:MAIN}(ii)  guarantees that, for $d \in \{0,1\}$,
\begin{align*}
      \mathbb{E}\left[  Y(d)-a_{\min}   \, \big| \, X=x  \right]  &\geq 0 ,
      &
      \mathbb{E}\left[  Y(d)-a_{\max}   \, \big| \, X=x  \right]  &\leq 0 .
\end{align*}
Combining the results in the last three displays we find that
\begin{align*}
     \mathbb{E}\left[  B^{(q)}(d,a_{\min}) - a_{\min}  \, \big| \, X=x  \right]  
         &\leq     \mathbb{E}\left[  Y(d)  - a_{\min}  \, \big| \, X=x  \right]  ,
    \\
       \mathbb{E}\left[  B^{(q)}(d,a_{\max}) - a_{\max}  \, \big| \, X=x  \right]      
          &\geq     \mathbb{E}\left[  Y(d)  - a_{\max}  \, \big| \, X=x  \right]  ,
\end{align*}
and therefore
\begin{align}
     \mathbb{E}\left[  B^{(q)}(d,a_{\min})  \, \big| \, X  \right]
         &\leq     \mathbb{E}\left[  Y(d)   \, \big| \, X \right]
         \leq    \mathbb{E}\left[  B^{(q)}(d,a_{\max})  \, \big| \, X  \right]   .
         \label{ConditionalResultB}
\end{align}
Taking the expectation over $X$ gives the results of part (i)(a) of the proposition,
and part~(i)(b) immediately follows from that.

Similarly, we find
\begin{align*}
         \mathbb{E}\left[    C^{(q)}(a_{\min} )     \, \big| \, X=x  \right]  
   &\geq     p(x) \, \left\{  \mathbb{E}\left[  Y(1)-a_{\min}    \, \big| \, X=x  \right] 
 -      \mathbb{E}\left[  Y(0)-a_{\min}    \, \big| \, X=x  \right]  \right\}    
\\
   &=       p(x) \,    \mathbb{E}\left[  Y(1)  -   Y(0)   \, \big| \, X=x  \right]
    ,
\\
         \mathbb{E}\left[    C^{(q)}(a_{\max} )     \, \big| \, X=x  \right]  
   &\leq     p(x) \, \left\{  \mathbb{E}\left[  Y(1)-a_{\max}    \, \big| \, X=x  \right] 
 -      \mathbb{E}\left[  Y(0)-a_{\max}    \, \big| \, X=x  \right]  \right\}    
\\
   &=       p(x) \,    \mathbb{E}\left[  Y(1)  -   Y(0)   \, \big| \, X=x  \right] ,
\end{align*}
and therefore
\begin{align}
   \mathbb{E}\left[    C^{(q)}(a_{\max} )     \, \big| \, X=x  \right]   
   \leq
   \pi(x)   
   \leq 
   \mathbb{E}\left[    C^{(q)}(a_{\min} )     \, \big| \, X=x  \right]   ,
   \label{ConditionalResultC}
\end{align}
where $ \pi(x)$ is defined in display \eqref{ATestimands} of the main text.
Taking the expectation over $X$ gives the results of part (i)(c) of the proposition.

\medskip
\underline{\# Part (ii):}
      From the definition of the weight functions in \eqref{DefGeneralWeights} we find that for $p(x) = p_*(x)$  we have
\begin{align*}
        w^{(q)} \big(1-p(x),1-p_*(x) \big) = 1 ,\quad
        w^{(q)} \big(p(x),p_*(x) \big) = 1, \quad
        \text{ and } \quad
         \widetilde w^{(q)} \big(p(x),p_*(x) \big) &= 1 .
\end{align*}
By the same arguments as in part (i) of the proof we therefore find that
\eqref{ConditionalResultB} and \eqref{ConditionalResultC} hold with equality, and
all the inequalities in part (i) of the proposition  then also hold with equality.

\medskip
\underline{\# Part (iii):}
Define
\begin{align*}
   v^{(q)}(p,x) &:=   \sum_{r=0}^{q-1} \lambda_r(1,x) \,  p^{r+1} ,
   \\
   \widetilde v^{(q)}(p,x) &:=  - \sum_{r=0}^{q-1}  \lambda_r(x) \,  p^{r-1} \, (1-p) .
\end{align*}
The bounds in \eqref{AllOrderBoundsGeneral} can then be written as
\begin{align*}
    B^{(q)}_{1,a}(\lambda) &=
   a +    v^{(q)}(p(X),X)  \, \frac{D \, (Y-a)} {p(X)}        ,    
\nonumber \\
    C^{(q)}_a(\lambda)
      &=    D\, (Y-a)     -     \widetilde v^{(q)}(p(X),X) \,  \frac{p(X) \,  (1-D) \, (Y-a) }  {1-p(X)} .
\end{align*}     
Thus, $ v^{(q)}(p,x) $ and $   \widetilde v^{(q)}(p,x) $ take exactly the roles of 
 $  w^{(q)} \big(p,p_*(x) \big)$ and  $ \widetilde w^{(q)} \big(p,p_*(x) \big)$
in \eqref{DefGeneralBC}. By the same arguments as in the proof of Theorem~\ref{th:OurBounds}
and in part (i) of the proof of the current proposition we therefore find that these bounds are valid
(in the sense of satisfying the inequalities in part (i) of this proposition) for all  DGP's that satisfy
Assumption~\ref{ass:MAIN}(i) and (ii) if and only if we have
for all $x \in {\cal X}$ and $p \in [0,1]$ (or $p \in (0,1]$ for $\widetilde v$) that
\begin{align*}
      v^{(q)}(p,x)  &\leq 1 ,
      &
      \widetilde v^{(q)}(p,x)  &\leq 1 .
\end{align*}
Thus, $ v^{(q)}(p,x) $ and $ \widetilde v^{(q)}(p,x)$ satisfy all conditions on $v(p,\lambda)$ and $\widetilde v(p,\lambda)$
in Lemma~\ref{lemma:PropertiesW}.
Therefore, there exists $\epsilon>0$ such that for all $p(x) \in {\cal B}_\epsilon(p_*(x))$
    we have
\begin{align}
    w^{(q)}(p,p_*) - v^{(q)}(p,x)  &\geq 0,
    &
    & \text{and}
    &
    \widetilde w^{(q)}(p,p_*)   -  \widetilde v^{(q)}(p,x)   &\geq 0.
    \label{WeightsInequality}
\end{align}
Using this together with
\begin{align*}
        &\mathbb{E}_{p(x)} \left[  B^{(q)}_{1,a} -  B^{(q)}_{1,a}(\lambda)   \, \big| \, X=x  \right] 
       \\ & \qquad =    
        \left[ w^{(q)} \big(p(x),p_*(x) \big)  - v^{(q)}(p,x)  \right] \;  
        \mathbb{E}_{p(x)} \left[  Y(1)-a   \, \big| \, X=x  \right] ,
\end{align*}
and $   \mathbb{E}_{p(x)} \left[  Y(1)-a_{\min}   \, \big| \, X=x  \right] >0$,
and $  \mathbb{E}_{p(x)}\left[  Y(1)-a_{\max}   \, \big| \, X=x  \right] <0$ we obtain that
\begin{align*}
         \mathbb{E}_{p(x)}\left[  B^{(q)}_{1,a_{\min}} -  B^{(q)}_{1,a_{\min}}(\lambda)   \, \big| \, X=x  \right]     \geq 0 ,
         \\
         \mathbb{E}_{p(x)}\left[  B^{(q)}_{1,a_{\max}} -  B^{(q)}_{1,a_{\max}}(\lambda)   \, \big| \, X=x  \right]     \leq 0 ,
\end{align*}
where $p(x) \in {\cal B}_\epsilon(p_*(x))$ throughout, so that \eqref{WeightsInequality} holds. 
From this we find that
\begin{align*}
   &    \mathbb{E}_{p(x)} \left[ B^{(q)}_{d,a_{\max}} 
      - B^{(q)}_{d,a_{\min}}  \, \Big| \, X=x \right] 
  \\ & \qquad \qquad \qquad  \leq
       \mathbb{E}_{p(x)} \left[ B^{(q)}_{d,a_{\max}}(\lambda) 
      - B^{(q)}_{d,a_{\min}}(\lambda)  \, \Big| \, X=x  \right] 
\end{align*}
holds for $d=1$. The same result for $d=0$ follows by applying the transformation $Y \leftrightarrow 1-Y$ and 
$p(x) \leftrightarrow 1-p(x)$.

Similarly, we have
\begin{align*}
    \mathbb{E}_{p(x)}  \left[    C^{(q)}_a  - C^{(q)}_a(\lambda)    \, \big| \, X=x  \right]  
   &=  -    p(x)  
     \left[  \widetilde w^{(q)} \big(p(x),p_*(x) \big)  -  \widetilde v^{(q)}(p,x)   \right]   \mathbb{E}_{p(x)}  \left[  Y(0)-a   \, \big| \, X=x  \right]   ,
\end{align*}
and therefore, for $p(x) \in {\cal B}_\epsilon(p_*(x))$, we find that
\begin{align*}
       \mathbb{E}\left[    C^{(q)}_{a_{\min}}  - C^{(q)}_{a_{\min}}(\lambda)    \, \big| \, X=x  \right]   & \leq 0 ,
       &
       \mathbb{E}\left[    C^{(q)}_{a_{\max}}  - C^{(q)}_{a_{\max}}(\lambda)    \, \big| \, X=x  \right]   & \geq 0 ,
\end{align*}
which implies that
  \begin{align*}
   &      \mathbb{E}_{p(x)} \left[ C^{(q)}_{a_{\min}} 
      - C^{(q)}_{a_{\max}}  \, \Big| \, X=x  \right] 
\\   & \qquad \qquad \qquad  \leq
      \mathbb{E}_{p(x)}  \left[ C^{(q)}_{a_{\min}}(\lambda) 
      - C^{(q)}_{a_{\max}}(\lambda)   \, \Big| \, X=x \right] .
\end{align*}  
This concludes the proof of the proposition.     
\end{proof}

\subsection{Proofs of intermediate lemmas}
\label{app:ProofIntermediate}

\begin{proof}[\bf Proof of Lemma~\ref{lemma:DefW}]
\underline{\# Part (i) for $q$ even:}
Since  $ w^{(q)} \left(p,p_* \right)  =   1- \left( \frac{p_*-p}{ p_*}   \right)^{q}  $ is a $q$th-order polynomial
in $p$ and satisfies $ w^{(q)} \left(0,p_* \right)=0$ we can find coefficients $  \overline \lambda$
such that $ v(p,  \overline \lambda ) = w^{(q)} \left(p,p_* \right)  $. Furthermore, from the definition of $ w^{(q)} \left(p,p_* \right)  $
it is straightforward to verify that
\begin{align*}
 w^{(q)} \left(p_*,p_* \right) &= 1 ,
 \\
      w^{(q)} \left(p,p_* \right) &\leq 1 ,
      &
      & \text{for  $p \in [0,1]$,}
  \\
       \frac{\partial^k     w^{(q)} \left(p_*,p_* \right) } {\partial^k p} &= 0, 
       & 
       &   \text{for $k \in \{1,\ldots,q-1\}$}.
\end{align*}
This shows that $\overline \lambda$ with $ v(p,  \overline \lambda ) = w^{(q)} \left(p,p_* \right)  $
satisfies the optimization problem in part (i) of the lemma with objective function
$ \left| \frac{\partial^{q-1}  v(p_*,\lambda)} {\partial^{q-1} p}  \right|$ equal to zero at the optimum.
Since the objective function is non-negative this indeed must be a minimizer.
The solution is unique, because
$v(p_*,\lambda)=1$ and
 $   \frac{\partial^k   v(p_*,\lambda)} {\partial^k p} = 0$,
for $k \in \{1,\ldots,q-1\}$,
is a system of $q$ linear equations in $q$ unknowns $\lambda$ that has a unique solution.

\medskip
\underline{\# Part (i) for $q$ odd:}
The optimization problem has $q-1$ linear equality constraints:
\begin{align*}
v(p_*,\lambda) &= 1 ,
  \\
       \frac{\partial^k     v(p_*,\lambda)} {\partial^k p} &= 0, 
       & 
       &   \text{for $k \in \{1,\ldots,q-2\}$}.
\end{align*}
Any solution $\lambda = \lambda(\kappa)$ to this system of equations satisfies
\begin{align*}
     v(p, \lambda)  =   1- (1- \kappa \, p) \left( \frac{p_*-p}{ p_*}   \right)^{q-1}  ,
\end{align*}
where $\kappa \in \mathbb{R}$ is one remaining degree of freedom that is not determined from those equality constraints.
For this solution we have
\begin{align*}
     v(1, \lambda) &=  1- (1- \kappa ) \left( \frac{p_*-1}{ p_*}   \right)^{q-1}  ,
\end{align*}     
and the constraint $ v(1,\lambda)  \leq 1$ therefore requires that $\kappa \leq 1$. It is easy to check that for
$\kappa \leq 1$ we also have $ v(p,\lambda)  \leq 1$ for all other $p \in [0,1]$.
We furthermore find
\begin{align*}
     \left| \frac{\partial^{q-1}  v(p_*,\lambda)} {\partial^{q-1} p}  \right|
     &=   (q-1)! \; \frac{|1- \kappa p_*|} {p_*^{q-1}} .
\end{align*}
Minimizing this over $\kappa \leq 1$ gives the optimal value at the boundary point $\overline \kappa = 1$. We have therefore
shown that the unique solution to the minimization problem is given by
$$ v(p, \overline \lambda)   =    1- \left( 1- p \right) \left( \frac{p_* - p }{ p_*}   \right)^{q-1}   = w^{(q)} \left(p,p_* \right). $$

\medskip
\underline{\# Part (ii) for $q$ even:}
Since  $p \, \widetilde w^{(q)} \left(p,p_* \right)  =     p -    \left( \frac{p - p_*}  {1-p_*}  \right)^q $ is a $q$th-order polynomial
in $p$ and satisfies $\widetilde w^{(q)} \left(1,p_* \right)=0$ we can find coefficients $  \widetilde \lambda$
such that $\widetilde v(p,  \widetilde \lambda ) = \widetilde w^{(q)} \left(p,p_* \right)  $. Furthermore, from the definition of $\widetilde w^{(q)} \left(p,p_* \right)  $
it is straightforward to verify that
\begin{align*}
 \widetilde  w^{(q)} \left(p_*,p_* \right) &= 1 ,
 \\
      \widetilde  w^{(q)} \left(p,p_* \right) &\leq 1 ,
      &
      & \text{for  $p \in (0,1]$,}
  \\
       \frac{\partial^k     \widetilde  w^{(q)} \left(p_*,p_* \right) } {\partial^k p} &= 0, 
       & 
       &   \text{for $k \in \{1,\ldots,q-1\}$}.
\end{align*}
This shows that $\widetilde \lambda$ with $\widetilde v(p,  \widetilde \lambda ) = \widetilde w^{(q)} \left(p,p_* \right)  $
satisfies the optimization problem in part (ii) of the lemma with objective function
$ \left| \frac{\partial^{q-1}  \widetilde v(p_*,\lambda)} {\partial^{q-1} p}  \right|$ equal to zero at the optimum.
Since the objective function is non-negative this indeed must be a minimizer.
The solution is unique, because
$\widetilde v(p_*,\lambda)=1$ and
 $   \frac{\partial^k   \widetilde v(p_*,\lambda)} {\partial^k p} = 0$,
for $k \in \{1,\ldots,q-1\}$,
is a system of $q$ linear equations in $q$ unknowns $\lambda$ that has a unique solution.

\medskip
\underline{\# Part (ii) for $q$  odd:}
The optimization problem has $q-1$ linear equality constraints:
\begin{align*}
\widetilde v(p_*,\lambda) &= 1 ,
  \\
       \frac{\partial^k     \widetilde  v(p_*,\lambda)} {\partial^k p} &= 0, 
       & 
       &   \text{for $k \in \{1,\ldots,q-2\}$}.
\end{align*}
Any solution $\lambda = \lambda(\kappa)$ to this system of equations satisfies
\begin{align*}
     \widetilde   v(p, \lambda)  =   1 -  \left( \kappa + \frac{1-\kappa}  p \right) \left( \frac{p - p_*}  {1-p_*}  \right)^{q-1}   ,
\end{align*}
where $\kappa \in \mathbb{R}$ is one remaining degree of freedom that is not determined from those equality constraints.
For this solution we have
\begin{align*}
    \lim_{p \rightarrow 0} \,  \widetilde   v(p, \lambda)  =  
     \left\{  \begin{array}{ll}
         \infty & \text{if $\kappa>1$} ,
         \\
          1 -   \left( \frac{ p_*}  {1-p_*}  \right)^{q-1}  &  \text{if $\kappa=1$}
     \\
          - \infty & \text{if $\kappa<1$.}
     \end{array} \right.
\end{align*}
and the constraint $ \widetilde  v(p,\lambda)  \leq 1$ for all  $p \in (0,1]$
therefore requires that $\kappa \leq 1$. It is easy to check that for
$\kappa \leq 1$ this inequality is indeed satisfied for all $p \in (0,1]$.
We furthermore find
\begin{align*}
     \left| \frac{\partial^{q-1}  \widetilde v(p_*,\lambda)} {\partial^{q-1} p}  \right|
     &=   \frac{(q-1)! } {(1-p_*)^{q-1}}  \left| \kappa + \frac{1-\kappa}  {p_*}\right| .
\end{align*}
Minimizing this over $\kappa \leq 1$ gives the optimal value at the boundary point $\widetilde \kappa = 1$. We have therefore
shown that the unique solution to the minimization problem is given by
$$ \widetilde v(p, \widetilde \lambda)   =    1 - \left( \frac{p - p_*}  {1-p_*}  \right)^{q-1}    =  \widetilde  w^{(q)} \left(p,p_* \right). $$
\end{proof}

\begin{proof}[\bf Proof of Lemma~\ref{lemma:PropertiesW}]
  \underline{\# Part (i):}
  We define  the non-negative integer $K$ and the positive number $C$ as follows:  
  If  $v(p_*,\lambda) \neq 1$, then we set $K=0$ and $C = 1 - v(p_*,\lambda)$.
  Otherwise, let $K$ be the smallest integer such that 
  $$
      \frac{\partial^{K}     v(p_*,\lambda)} {\partial^{K} p} \neq 0,
   $$
   and set $$C = -  \frac{\partial^{K}     v(p_*,\lambda)} {\partial^{K} p}.$$   
   It must be the case that $K$ is even and that $C>0$, because otherwise the assumption   $v(p,\lambda)  \leq 1$, for all $p \in [0,1]$,
   would be violated.
   A  Taylor expansion of  $v(p,\lambda)$ around $p=p_*$ gives
   \begin{align}
        v(p,\lambda) = 1 - C \, (p-p_*)^K  + O\left( |p-p_*|^{K+1} \right).
        \label{ExpansionV}
   \end{align}  
  Next, let $q_*=q$ if $q$ is even, and let $q_*=q-1$ if $q$ is odd.  We have $w^{(q)}(p_*,p_*)=1$, and
  $$
        \frac{\partial^k  w^{(q)}(p_*,p_*)} {\partial^k \, p} = 0 ,
        \qquad
        \text{for all $k \in \{1,\ldots, q_*-1 \}$}.
  $$   
   Therefore, a  Taylor expansion of  $w^{(q)}(p,p_*)$ around $p=p_*$ gives
   \begin{align}
       w^{(q)}(p,p_*) = 1  + O\left( |p-p_*|^{q_*} \right) .
       \label{ExpansionW}
   \end{align}
   If $K<q_*$, then \eqref{ExpansionV} and \eqref{ExpansionW} imply that
   \begin{align*}
        v(p,\lambda)  &=  w^{(q)}(p,p_*) - C \, (p-p_*)^K  + O\left( |p-p_*|^{K+1} \right) .
   \end{align*}
   Since $C>0$ and $K$ is even, there must then exist $\epsilon>0$ such that   for all $p \in {\cal B}_\epsilon(p_*)$
   we have $ v(p,\lambda)  \leq  w^{(q)}(p,p_*)$.
 
 \medskip
 
    If $K=q_*$ and $q$ is even, then  $ v(p,\lambda)$ satisfies 
    $v(p_*,\lambda)=1$ and
 $   \frac{\partial^k   v(p_*,\lambda)} {\partial^k p} = 0$,
for all $k \in \{1,\ldots,q-1\}$.
   This is exactly the system of  $q$ linear equations in $q$ unknowns $\lambda$ whose solution
   is $\overline \lambda$. In that case, we therefore have $v(p,\lambda) =  w^{(q)}(p,p_*)$,
   and the statement of the lemma
   holds for any $\epsilon>0$.
   
 \medskip   
   
    If $K=q_*$ and $q$ is odd, then $ v(p,\lambda) $ satisfies all the constraints in the optimization problem in
   part (i) of Lemma~\ref{lemma:DefW}. 
   If  $ v(p,\lambda) $ is the solution to this optimization problem, then we again have 
   $v(p,\lambda) =  w^{(q)}(p,p_*)$,
   and the statement of the lemma
   holds for any $\epsilon>0$. Otherwise, $ v(p,\lambda) $ is not the solution to this  optimization problem,
   which implies that 
   $$
        C =  - \frac{\partial^{K}  v(p_*,\lambda)} {\partial^K p}  
        >  - \frac{\partial^{K}  w^{(q)}(p_*,p_*)} {\partial^K p} =: c  >0 .
   $$
   In that case, analogous to \eqref{ExpansionV} we have
   \begin{align*}
       w^{(q)}(p,p_*) = 1 - c \, (p-p_*)^K  + O\left( |p-p_*|^{K+1} \right) ,
   \end{align*}
   and therefore
   \begin{align*}
        v(p,\lambda)  &=  w^{(q)}(p,p_*) - (C-c) \, (p-p_*)^K  + O\left( |p-p_*|^{K+1} \right) .
   \end{align*}   
    Since $C-c>0$ and $K$ is even, there must again exist $\epsilon>0$ such that   for all $p \in {\cal B}_\epsilon(p_*)$
   we have $ v(p,\lambda)  \leq  w^{(q)}(p,p_*)$.
    We have therefore shown that the desired result holds in all possible cases.     
   
   \medskip
   
  \underline{\# Part (ii):}
  The proof of $   \widetilde v(p,\lambda)  \leq  \widetilde w^{(q)}(p,p_*)$ is analogous,
  using that $\widetilde w^{(q)}(p,p_*)$ is the solution to the 
   optimization problem in
   part (ii) of Lemma~\ref{lemma:DefW}.
\end{proof}

\section{Derivation of the sample weights in Section~\ref{sec:implementation}}
\label{sec:SampleWeights}

Here, we want to discuss where the formulas in \eqref{SampleWeightsDef} and \eqref{SampleWeightsDef2}
for $\widehat w_{0/1}(x)$ and $\widehat v(x) $ come from, and why the  $\omega$  coefficients need to be chosen
according to \eqref{ResultsCombinatorialOmega}.

Consider first the case where $q(x)=\min\{Q,n(x)\} $ is even,
in which case $\widehat w_1(x)$ is given by \eqref{DefAlternativeSampleWeights}.
As explained in the main text, this formula for $\widehat w_1(x)$ guarantees that 
the conditional expectation of  $\widehat w_1(x)$ is given by \eqref{SampleWeightsExpectations}, but 
for the purpose of practical implementation we want to express $\widehat w_1(x)$  not in terms of individual 
observations $D_i$, but in terms of the summary statistics $n(x)$ and $n_1(x)$.
For simplicity, we only write $q$ instead of $q(x)$ in the following.
We can rewrite the expression for $\widehat w_1(x)$ in \eqref{DefAlternativeSampleWeights} as 
 \begin{align*}
\widehat w_1(x) & =    1-   { n(x) \choose q }^{-1} \sum_{{\cal S}_q} \left( \frac{p_*(x)-1}{p_*(x)}   \right)^{n_1({\cal S}_q)}
\\
   &= 1-   { n(x) \choose q }^{-1}  \sum_{k=0}^{q}  \underbrace{ \left(  \sum_{{\cal S}_q} \mathbbm{1}\{n_1({\cal S}_q) = k \}  \right)
    }_{\displaystyle =: \alpha_{k,n_1(x),n(x),q} } \left(     \frac{p_*(x) - 1}{p_*(x)}  \right)^{k} ,
\end{align*}
where $n_1({\cal S}_q)$ is the number of observations $i \in {\cal S}_q$ with $D_i=1$, and
 $ \alpha_{k,n_1(x),n(x),q} \in \{1,2,\ldots\}$ is the number of subsets ${\cal S}_q$ for which we have $n_1({\cal S}_q) = k$.
By standard combinatorial arguments one finds that\footnote{%
We can generate all subsets  ${\cal S}_q \subset {\cal N}(x)$ with $q$ elements and $n_1({\cal S}_q) = k$ by first choosing
$k$ of the $n_1(x)$ units in $ {\cal N}(x)$ with $D_i=1$, which gives the factors  ${ n_1(x) \choose k } $,
and secondly choosing  $q-k$  of the $ n(x)-n_1(x) $ units in $ {\cal N}(x)$ with $D_i=0$, which gives the factor $ { n(x)-n_1(x) \choose q-k } $ in \eqref{CombinatorialArgument}.
Here, we use the standard convention for the binomial coefficient that
 ${a \choose b}=0$ for all integers $b > a \geq 0$, but  ${0 \choose 0}=1$.
}
\begin{align}
      \alpha_{k,n_1(x),n(x),q}  =  { n_1(x) \choose k }  { n(x)-n_1(x) \choose q-k } .
      \label{CombinatorialArgument}
\end{align}
We therefore obtain the definition of $\widehat w_1(x)$ in \eqref{SampleWeightsDef} by setting 
\begin{align*}
      \omega_{k,n_1(x),n(x),Q}  =   { n(x) \choose q }^{-1}     \alpha_{k,n_1(x),n(x),q} ,
\end{align*}
for $q=\min\{Q,n(x)\} $ even, and combining the last two displays gives the formulas for $\omega$
in \eqref{ResultsCombinatorialOmega} for that case.
Since   $  \alpha_{k,n_1(x),n(x),q}  \leq  { n(x) \choose q } $ it follows that $  \omega_{k,n_1(x),n(x),Q} \in [0,1]$.
The combinatorial argument for the case that $q(x)=\min\{Q,n(x)\} $  odd is analogous,
as are  the derivations for $  \widehat w_0(x)$ and $  \widehat v(x)  $, which give the same result for $  \omega_{k,n_{0/1}(x),n(x),Q} $.

\section{Proofs for Section~\ref{sec:asympt}}
\label{app:Asymptotic}

Display \eqref{DefThetaParameter} in the main text defined the parameters of interest $\theta^{(r)}$, which are labeled by the index
$r \in \{0,1,{\rm ATE},{\rm ATT}\}$. Our lower and upper bound estimates for $r \in \{0,1,{\rm ATE}\}$
can be written as simple sample averages over $x \in {\cal X}_*$,
\begin{align*}
   \overline L^{(r)} &=    \frac 1 m  \sum_{x \in {\cal X}_*}  L^{(r)}_{x} ,
&
   \overline U^{(r)} &=    \frac 1 m  \sum_{x \in {\cal X}_*}  U^{(r)}_{x} ,
\end{align*}
with $ L^{(r)}_{x} $ and $U^{(r)}_{x} $ defined in the main text. By contrast, 
the lower and upper bound estimates $ \overline L^{\rm (ATT)}$ and $\overline U^{\rm (ATT)}$
defined in \eqref{SampleBoundsATT} take the form of a ratio of sample averages, with numerator and denominator
given by
\begin{align*}
       \overline C(a) &=   \frac 1 n \sum_{i=1}^n  \widehat C_{i}(a) = \frac 1 m  \sum_{x \in {\cal X}_*}  \frac{m \, n(x) \, \widehat C_x(a)}  n ,
\\
     \frac 1 n \sum_{i=1}^n D_i &= \frac 1 m  \sum_{x \in {\cal X}_*}  \frac{m \, n_1(x)}  n .
\end{align*}
Since we assume $ \frac 1 n \sum_{i=1}^n  p(X_i) > 0$, 
and our assumptions also guarantee $ \frac 1 n \sum_{i=1}^n [D_i -  \frac 1 n \sum_{i=1}^n  p(X_i)] = O_P(1/\sqrt{n})$,
we can apply the delta method to find
\begin{align*}
     \frac 1{  \frac 1 n \sum_{i=1}^n D_i} &= 
     \frac  1 { \frac 1 n \sum_{i=1}^n  p(X_i) } 
     +   \frac{ \frac 1 n \sum_{i=1}^n  p(X_i)  -     \frac 1 n \sum_{i=1}^n D_i    }  {\left[ \frac 1 n \sum_{i=1}^n  p(X_i)\right]^2 } 
     + O_P(1/n) ,
\end{align*}
and therefore
\begin{align}
  &  \frac{ \overline C(a) }
   { \frac 1 n \sum_{i=1}^n D_i }
\nonumber  \\
  &=    \frac{ \mathbb{E} \left[ \, \overline  C(a) \, \big| \, X^{(n)} \right]     } { \frac 1 n \sum_{i=1}^n  p(X_i)}
  + \frac{ \frac 1 m  \sum_{x \in {\cal X}_*}  \frac{m \, n(x) \, \widehat C_x(a)} n - \mathbb{E} \left[ \, \overline  C(a) \, \big| \, X^{(n)} \right]   } { \frac 1 n \sum_{i=1}^n  p(X_i)}
 \nonumber \\ & \qquad  \qquad \qquad \qquad  \qquad \qquad  \qquad
     +   \frac{\mathbb{E} \left[ \, \overline  C(a) \, \big| \, X^{(n)} \right]  [ \frac 1 n \sum_{i=1}^n  p(X_i)  -     \frac 1 n \sum_{i=1}^n D_i]    }  {\left[ \frac 1 n \sum_{i=1}^n  p(X_i)\right]^2 }
     + O_P(1/n)
\nonumber  \\
  &=       \frac{ \mathbb{E} \left[ \, \overline  C(a) \, \big| \, X^{(n)} \right]     } { \frac 1 n \sum_{i=1}^n  p(X_i)}
  + \frac{ \frac 1 m  \sum_{x \in {\cal X}_*}  \frac{m \, n(x) \, \widehat C_x(a)} n   } { \frac 1 n \sum_{i=1}^n  p(X_i)}
 \nonumber \\ & \qquad  \qquad \qquad \qquad  \qquad \qquad  \qquad
     -   \frac{ \left[\mathbb{E} \left[ \, \overline  C(a) \, \big| \, X^{(n)} \right]  \right] \left[  \frac 1 m  \sum_{x \in {\cal X}_*}  \frac{m \, n_1(x)}  n  \right]    }  {\left[ \frac 1 n \sum_{i=1}^n  p(X_i)\right]^2 }
     + O_P(1/n)
 \nonumber \\
  &=       \frac{ \mathbb{E} \left[ \, \overline  C(a) \, \big| \, X^{(n)} \right]     } { \frac 1 n \sum_{i=1}^n  p(X_i)}
 +  \frac 1 m  \sum_{x \in {\cal X}_*}
  \left[   \frac{m \, n(x) \, \widehat C_x(a) } {n \,  \frac 1 n \sum_{i=1}^n  p(X_i)}
     -   \frac{m \, n_1(x) \,  \mathbb{E} \left[ \, \overline  C(a) \, \big| \, X^{(n)} \right]       }  {n \, \left[ \frac 1 n \sum_{i=1}^n  p(X_i)\right]^2 }  \right]
     + O_P(1/n) .
        \label{ATTinfluenceFunction0}
\end{align}
This shows that the influence function of the ratio $   \frac{ \overline C(a) }
   { \frac 1 n \sum_{i=1}^n D_i }$
  is given by  $$\frac{m \, n(x) \, \widehat C_x(a) } {n \,  \frac 1 n \sum_{i=1}^n  p(X_i)}
     -   \frac{m \, n_1(x) \, \mathbb{E} \left[ \, \overline  C(a) \, \big| \, X^{(n)} \right]     }  {n \, \left[ \frac 1 n \sum_{i=1}^n  p(X_i)\right]^2 }.$$
When this influence function is used to calculate the asymptotic variance of the ratio, the terms
$ \frac 1 n \sum_{i=1}^n  p(X_i)$ and $\mathbb{E} \left[ \, \overline  C(a) \, \big| \, X^{(n)} \right]  $ must again be replaced by their consistent
estimates $ \frac 1 n \sum_{i=1}^n D_i $ and $ \overline  C(a ) $. After that replacement we obtain
 \begin{align*}
   & \frac{ \overline C(a) }
   { \frac 1 n \sum_{i=1}^n D_i }
   \\
      &=       \frac{ \mathbb{E} \left[ \, \overline  C(a) \, \big| \, X^{(n)} \right]     } { \frac 1 n \sum_{i=1}^n  p(X_i)}
 +  \frac 1 m  \sum_{x \in {\cal X}_*}
  \left[   \frac{ m \, n(x) \,  \widehat C_{x}(a)    } {  \sum_{i=1}^n D_i}
         - \frac{ m \, n \, n_1(x)  \, \overline  C(a)    } {\left( \sum_{i=1}^n D_i\right)^2}    \right]
    + o_P(m^{-1/2}) ,
\end{align*}
which is exactly the expression for $\overline L^{\rm (ATT)} $
and $ \overline U^{\rm (ATT)} $ given in \eqref{ATTinfluenceFunction1}
and \eqref{ATTinfluenceFunction2} of the main text.
We have thus derived the expressions for $ L^{\rm (ATT)}_{x} $
and $U^{\rm (ATT)}_{x} $ given in the main text. We are now ready to prove Theorem~\ref{th:Asympt}.

\begin{proof}[\bf Proof of Theorem~\ref{th:Asympt}]

   \underline{\# Consider $r \in \{0,1,{\rm ATE}\}$}.  
    In that case we have     
\begin{align*}
   \overline L^{(r)} &=    \frac 1 m  \sum_{x \in {\cal X}_*}  L^{(r)}_{x} .
\end{align*}
Conditional on $X^{(n)}$, the  $L^{(r)}_{x}$ are independent across $x \in {\cal X}_*$, that is, 
$ \overline L^{(r)}$ is an average over $m$ independent terms $L^{(r)}_{x}$.
Let
$$F_m(\xi) := P\left\{  \left.   \frac{  \overline L^{(r)}  -   \mathbb{E}\left[ \overline L^{(r)} \Big| X^{(n)} \right]   } {  {\rm Var}  \left[ \overline L^{(r)} \, \Big| \, X^{(n)} \right]^{1/2}}   \leq \xi \, \right| \, X^{(n)} \right\},$$
which is the cdf of $ \overline L^{(r)} $ after centering to have zero mean and normalizing to have variance one.
According
to the Berry-Esseen theorem, the difference between $F_m(\xi) $ and the normal cdf $\Phi(\xi)$ is  bounded by
\begin{align}
    \sup_{\xi \in \mathbb{R} } \left| F_m(\xi) - \Phi(\xi) \right|
    \leq  \frac{ C \, \sum_{x \in {\cal X}_*}    \mathbb{E}\left[ \left. \left| L^{(r)}_{x} \right|^3 \right| X^{(n)} \right]  } {\left\{ \sum_{x \in {\cal X}_*}  {\rm Var}  \left[ L^{(r)}_{x}  \, \Big| \, X^{(n)} \right]  \right\}^{3/2}} ,
    \label{BerryEsseenTh}
\end{align}
where $C$ is a universal constant.
The assumptions that $Q$ is fixed and that
      $p_*(x)$ is bounded away from zero and one 
guarantee that       $  \widehat B_{x}(d,a) $ is uniformly bounded,
that is, there exists a constant $b>0$, independent of the sample size,
such that
\begin{align*}
    \max_{x \in {\cal X}_*, d \in \{0,1\}}
    \left\{
    \widehat  B_{x}(d,a_{\min}) ,   \widehat  B_{x}(d,a_{\max}) ,
    \widehat  B_{x}(1,a_{\min}) - \widehat B_{x}(0,a_{\max})  ,
    \widehat  B_{x}(1,a_{\max}) -  \widehat B_{x}(0,a_{\min}) 
    \right\} \leq b .
\end{align*}
Nevertheless, the $L^{(r)}_{x}$,  $r \in \{0,1,{\rm ATE}\}$, defined in 
\eqref{DefLUATE} may still not be bounded, because of the factors $ \frac{m \, n(x)} n$. 
Taking into account those factors we find that
\begin{align*}
   \frac 1 m \sum_{x \in {\cal X}_*}    \mathbb{E}\left[ \left. \left| L^{(r)}_{x} \right|^3 \right| X^{(n)} \right]
      \leq   \frac{b^3} m  \sum_{x \in {\cal X}_*}  \left( \frac{m \, n(x)} n \right)^3   = O_P(1),
\end{align*}
where we used that  $ \frac 1 {m}   \sum_{x \in {\cal X}_*} \left( \frac{m \, n(x)} n \right)^4  = O_P(1)$
 implies  $ \frac 1 {m}   \sum_{x \in {\cal X}_*} \left( \frac{m \, n(x)} n \right)^3  = O_P(1)$, by an application of Jensen's inequality.
 Together with  $ \left[\frac 1 m  \sum_{x \in {\cal X}_*} {\rm Var}  \left(   L^{(r)}_{x}  \, \Big| \, X^{(n)} \right) \right]^{-1} \, 
      = o_P( m^{1/3} )$,
this guarantees that the right-hand side of  \eqref{BerryEsseenTh} converges to zero, and we thus have
\begin{align*}
     \frac{  \overline L^{(r)}  -   \mathbb{E}\left[ \overline L^{(r)} \Big| X^{(n)} \right]   } {  {\rm Var}  \left[ \overline L^{(r)} \, \Big| \, X^{(n)} \right]^{1/2}} 
     \Rightarrow {\cal N}(0,1).
\end{align*}
For $r \in \{0,1,{\rm ATE}\}$ we have $ \theta^{(r)}_L  =  \mathbb{E}\left[ \overline L^{(r)} \Big| X^{(n)} \right] $
and $ \overline L^{(r)}  =  \displaystyle \frac 1 m  \sum_{x \in {\cal X}_*}  L^{(r)}_{x} $. The last display can therefore be rewritten as
$$
      \frac{  \overline L^{(r)}  -  \theta^{(r)}_L   } {\left\{ {\rm Var}  \left[ \displaystyle \frac 1 m  \sum_{x \in {\cal X}_*}  L^{(r)}_{x}\, \Bigg| \, X^{(n)} \right]  \right\}^{1/2}}   
     \,  \Rightarrow \, {\cal N}\left(0,1 \right) ,
$$
as stated in the theorem. 
Furthermore, we compute
\begin{align}
      m \,  {\rm Var}\left( \left. \overline L^{(r)} \right| X^{(n)}  \right) &= \frac 1 {m} \sum_{x \in {\cal X}_*}   {\rm Var}\left( \left. L^{(r)}_{x}  \right| X^{(n)}  \right)  
      \nonumber \\
       &=   \frac 1 {m} \sum_{x \in {\cal X}_*}  \mathbb{E} \left[ \left. \left( L^{(r)}_{x}   \right)^2  \right| X^{(n)}  \right]
        -  \frac 1 {m} \sum_{x \in {\cal X}_*} \left[ \mathbb{E} \left( \left. L^{(r)}_{x}  \right) \right| X^{(n)}  \right]^2
     \nonumber  \\
       &\leq   \frac 1 {m} \sum_{x \in {\cal X}_*}  \mathbb{E} \left[ \left. \left( L^{(r)}_{x}   \right)^2 \right| X^{(n)}  \right]
        -  \left[  \frac 1 m   \sum_{x \in {\cal X}_*}   \mathbb{E} \left(  \left. L^{(r)}_{x} \right| X^{(n)}  \right) \right]^2 ,
     \label{VarBoundJensen}   
\end{align} 
where in the last step we used that $  \frac 1 {m} \sum_{x  } \left( a_{x}  \right)^2 \geq \left(  \frac 1 {m} \sum_{x  } a_{x}  \right)^2 $, which holds for any $a_{x} \in \mathbb{R}$ according to  Jensen's inequality.
Using again that $L^{(r)}_{x} $ is independent across $x$ we have
\begin{align*}
   \mathbb{E} \left( \left.   \left\{   \frac 1 {m} \sum_{x \in {\cal X}_*} 
     \left( L^{(r)}_{x}   \right)^2 -   \mathbb{E} \left[ \left. \left( L^{(r)}_{x}   \right)^2 \right| X^{(n)}  \right] \right\}^2
      \right| X^{(n)}  \right)
      &\leq \frac 1 {m^2}   \sum_{x \in {\cal X}_*} 
    \mathbb{E} \left[  \left. \left( L^{(r)}_{x}   \right)^4     \right| X^{(n)}  \right]
    \\
    &=   \frac {b^4} {m^2}   \sum_{x \in {\cal X}_*} \left( \frac{m \, n(x)} n \right)^4 
    \\
     &=O_P(m^{-1}).
\end{align*}
We therefore have
\begin{align*}
     \frac 1 {m} \sum_{x \in {\cal X}_*}  \mathbb{E} \left[ \left. \left( L^{(r)}_{x}   \right)^2 \right| X^{(n)}  \right]
  &=    \frac 1 {m} \sum_{x \in {\cal X}_*}   \left( L^{(r)}_{x}   \right)^2  + O_P(1/\sqrt{m}) ,
\end{align*}
and analogously we conclude that
\begin{align*}
  \frac 1 m   \sum_{x \in {\cal X}_*}   \mathbb{E} \left( \left. L^{(r)}_{x} \right| X^{(n)}   \right) 
  &= \frac 1 m   \sum_{x \in {\cal X}_*}  L^{(r)}_{x}  + O_P(1/\sqrt{m}) .
\end{align*}
Combining the results of the last two displays with \eqref{VarBoundJensen} gives
\begin{align*}
       {\rm Var} \left(  \overline L^{(r)}  \right)
   &\leq 
   \frac { {\rm SVar}  \left(  L^{(r)}_{x}  \right)  } m  +  O_P(m^{-3/2}) .
\end{align*} 
Our assumption  $ \left[\frac 1 m  \sum_{x \in {\cal X}_*} {\rm Var}  \left(   L^{(r)}_{x}  \, \Big| \, X^{(n)} \right) \right]^{-1} \, 
      = o_P( m^{1/3} )$
guarantees that $ {\rm Var} \left(  \overline L^{(r)}  \right)$ does not converge to zero faster than $m^{-4/3} \gg m^{-3/2}$.
Therefore, on the right-hand side of the last display the term $   \frac { {\rm SVar}  \left(  L^{(r)}_{x}  \right)  } m$ must be asymptotically larger 
than the term $O_P(m^{-3/2}) $, and we thus have
\begin{align*}
       {\rm Var} \left(  \overline L^{(r)}  \right)
   &\leq 
   \frac { {\rm SVar}  \left(  L^{(r)}_{x}  \right)  } m   \left[ 1 + o_P(1) \right] .
\end{align*} 
We have thus shown the statement of the theorem for $\overline L^{(r)}$ and $r \in \{0,1,{\rm ATE}\}$. 
The proof for $\overline U^{(r)}$ and $r \in \{0,1,{\rm ATE}\}$ is  analogous.

\bigskip
     
   \underline{\# Next, consider $r = {\rm ATT}$}.  
  We can rewrite \eqref{ATTinfluenceFunction0} for $a=a_{\max}$ as
\begin{align*}
   \overline L^{\rm (ATT)} &=        \frac{ \mathbb{E} \left[ \, \overline  C(a_{\max}) \, \big| \, X^{(n)} \right]     } { \frac 1 n \sum_{i=1}^n  p(X_i)}   + \frac 1 m  \sum_{x \in {\cal X}_*}  \widetilde L^{\rm (ATT)}_{x} + O_P(1/n) ,
\end{align*}
where
\begin{align*}
   \widetilde L^{\rm (ATT)}_{x}  &:=       \frac{m \, n(x) \, \widehat C_x(a_{\max}) } {n \,  \frac 1 n \sum_{i=1}^n  p(X_i)}
     -   \frac{m \, n_1(x) \,  \mathbb{E} \left[ \, \overline  C(a_{\max}) \, \big| \, X^{(n)} \right]       }  {n \, \left[ \frac 1 n \sum_{i=1}^n  p(X_i)\right]^2 }  ,
\end{align*}  
which satisfies $ \mathbb{E} \left[ \,  \widetilde L^{\rm (ATT)}_{x} \, \big| \, X^{(n)} \right]  = 0$.
As for $r \in \{0,1,{\rm ATE}\}$ above, we then apply the  Berry-Esseen theorem to find that
$$
      \frac{ \displaystyle  \frac 1 m  \sum_{x \in {\cal X}_*}  \widetilde L^{\rm (ATT)}_{x}    } {\left\{ {\rm Var}  \left[ \displaystyle \frac 1 m  \sum_{x \in {\cal X}_*}  \widetilde L^{\rm (ATT)}_{x} \, \Bigg| \, X^{(n)} \right]  \right\}^{1/2}}   
     \,  \Rightarrow \, {\cal N}\left(0,1 \right) ,
$$
which then implies 
$$
      \frac{  \overline L^{\rm (ATT)} -     \theta^{\rm (ATT)}_L  } {\left\{ {\rm Var}  \left[ \displaystyle \frac 1 m  \sum_{x \in {\cal X}_*}  \widetilde L^{\rm (ATT)}_{x} \, \Bigg| \, X^{(n)} \right]  \right\}^{1/2}}   
     \,  \Rightarrow \, {\cal N}\left(0,1 \right) ,
$$
where $\theta^{\rm (ATT)}_L =   \mathbb{E} \left[ \left. \overline  C(a_{\max}) \, \right| \, X^{(n)} \right]    / \left(  \frac 1 n \sum_{i=1}^n  p(X_i) \right)   $,
as already defined in the main text.
As for the case $r \in \{0,1,{\rm ATE}\}$ above, one then also obtains that
\begin{align*}
      {\rm Var}  \left[ \displaystyle \frac 1 m  \sum_{x \in {\cal X}_*}  \widetilde L^{\rm (ATT)}_{x} \, \Bigg| \, X^{(n)} \right]
      &=  {\rm Var}  \left[ \displaystyle \frac 1 m  \sum_{x \in {\cal X}_*}  L^{\rm (ATT)}_{x} \, \Bigg| \, X^{(n)} \right]
      \\
      &\leq    \frac { {\rm SVar}  \left(  L^{\rm (ATT)}_{x}  \right)  } m   \left[ 1 + o_P(1) \right] .
\end{align*}
The proof for $\overline U^{(\rm ATT)}$ is analogous.
\end{proof} 

\begin{proof}[\bf Proof of Corollary~\ref{cor:CoverageBasic}]
According to \eqref{ValidBoundsConditionalX} we have
$\theta^{(r)}_L \leq \theta^{(r)} \leq \theta^{(r)}_U$, and according to Theorem~\ref{th:Asympt}
 our sample bounds $ \overline L^{(r)} $ and  $\overline U^{(r)} $ are asymptotically normally distributed estimates for $\theta^{(r)}_L$
 and $\theta^{(r)}_U$. We cannot estimate the  variances of $ \overline L^{(r)} $ and  $\overline U^{(r)} $ consistently, but 
 according to Theorem~\ref{th:Asympt} we have asymptotically valid upper bounds for those variances given by
 $ {\rm SVar}  \left(   L^{(r)}_{x}  \right) / m$ and $ {\rm SVar}  \left(   U^{(r)}_{x}  \right) / m$.  We can therefore construct
 valid one-sided confidence intervals for $\theta^{(r)}_L$ and $\theta^{(r)}_U$ of size $\alpha/2$, and combine them
 to obtain a valid confidence interval for  $\theta^{(r)} $  of confidence level $1-\alpha$, which gives ${\rm CI}^{(r)}_{\rm basic}$.
 \end{proof}

\section{Additional Monte Carlo Experiments: Inference}\label{sec:MC:additional}

In this section, we report additional Monte Carlo experiments that focus on 
finite sample performance of our proposed methods.
We consider both continuous and discrete $X$. 
The former is randomly drawn from $\text{Unif}[-3,3]$ and the latter is  generated by
$X = \text{round}(10 \times \text{Unif}[-3,3])/10$. That is,
$X$ is a discrete uniform random variable on the discrete support $[-3, -2.9, \ldots, 2.9, 3]$. 
The rest of the simulation design is the same as before, and 
we focus on ATT as well. 

Panels I and II in Table~\ref{tab-mc-infer} summarize the results of Monte Carlo experiments when the distribution of $X$ is discrete.
In the columns heading `Coverage', we report the Monte Carlo coverage proportion that the true value of 
ATT is included in either sample analog bounds or inference bounds. %
 In the columns heading `Non-Empty Interval', we report the Monte Carlo proportion of the cases that 
the resulting interval is non-empty. 
In the columns heading `Avg. Length', we report the average length of the interval when it is not empty.

The inference bounds are constructed by applying the method described in Section~\ref{sec:asympt}.
We first discuss the results for DGP A. In this scenario, the ATT is point-identified and the lower bound equals the upper bound; thus, the sample lower bound can be easily larger than the sample upper bound, resulting in frequent occurrence of empty intervals. However, the inference bounds are never empty and provide good coverage results. 
For DGP~B, the ATT is only partially identified.
Accordingly, the inference bounds are wide enough to cover the true value in nearly all Monte Carlo replications. Sample analog coverage can deteriorate for larger values of $Q$.

Panels III and IV in Table~\ref{tab-mc-infer} summarizes the results of Monte Carlo experiments when the distribution of $X$ is continuous.
Overall, the results are similar to the discrete $X$ case for DGP A. However, there is a rather surprising result with 
$Q=4$ for DGP B. In this case, the inference bounds include the true value only 382 out of 1,000 replications. This suggests that the clustering estimators with a large value of $Q$ may lead to severe estimation bias and size distortion, possibly due to the bias from the clustering method. 

\begin{table}[htbp]
\caption{Monte Carlo Results: Inference}\label{tab-mc-infer}
\begin{center}
\begin{tabular}{lrrrrrr}
 \hline\hline
  Q & \multicolumn{2}{c}{Coverage} & \multicolumn{2}{c}{Non-Empty Interval} & \multicolumn{2}{c}{Avg. Length} \\
  & Sample  & Inference & Sample  & Inference & Sample  & Inference \\ 
  & Analogs & Bounds & Analogs & Bounds & Analogs & Bounds \\  \hline
 \multicolumn{5}{l}{Panel I. DGP A with a Discrete Covariate} \\   
  1 & 1.000 & 1.000 & 1.000 & 1.000 & 1.000 & 1.381 \\ 
  2 & 0.114 & 0.975 & 0.451 & 1.000 & 0.018 & 0.127 \\ 
  3 & 0.058 & 0.965 & 0.438 & 1.000 & 0.009 & 0.117 \\ 
  4 & 0.031 & 0.958 & 0.419 & 1.000 & 0.006 & 0.114 \\ 
   \hline
 \multicolumn{5}{l}{Panel II. DGP B with a Discrete Covariate} \\   
  1 & 1.000 & 1.000 & 1.000 & 1.000 & 0.982 & 1.295 \\ 
  2 & 1.000 & 1.000 & 1.000 & 1.000 & 0.275 & 0.471 \\ 
  3 & 0.997 & 1.000 & 1.000 & 1.000 & 0.240 & 0.430 \\ 
  4 & 0.573 & 0.996 & 1.000 & 1.000 & 0.151 & 0.344 \\ 
   \hline
 \multicolumn{5}{l}{Panel III. DGP A with a Continuous Covariate} \\   
  1 & 1.000 & 1.000 & 1.000 & 1.000 & 0.998 & 1.249 \\ 
  2 & 0.101 & 0.983 & 0.375 & 1.000 & 0.022 & 0.141 \\ 
  3 & 0.041 & 0.977 & 0.289 & 1.000 & 0.011 & 0.128 \\ 
  4 & 0.061 & 0.979 & 0.272 & 1.000 & 0.015 & 0.140 \\ 
   \hline
 \multicolumn{5}{l}{Panel IV. DGP B with a Continuous Covariate} \\   
  1 & 1.000 & 1.000 & 1.000 & 1.000 & 0.970 & 1.182 \\ 
  2 & 0.927 & 1.000 & 1.000 & 1.000 & 0.211 & 0.380 \\ 
  3 & 0.578 & 0.990 & 1.000 & 1.000 & 0.165 & 0.335 \\ 
  4 & 0.002 & 0.382 & 0.961 & 1.000 & 0.046 & 0.241 \\ 
   \hline
\end{tabular}
\end{center}
Notes:
The nominal coverage probability is 0.95.
The sample size was $n = 1{,}000$ and the number of simulation replications was $1{,}000$.
\end{table}

\newpage
\renewcommand{\baselinestretch}{0.1}
\setlength{\bibsep}{0pt}
%

\end{document}